\documentclass{emulateapj}

\newcommand{\myemail}{quanz@astro.phys.ethz.ch}

\shorttitle{NACO PDI observations of the disk around HD100546}
\shortauthors{Quanz et al.}

\begin{document}


\title{VLT/NACO polarimetric differential imaging of HD100546 -- \\Disk structure and dust grain properties between 10--140 AU$^1$}


\author{Sascha P. Quanz$^2$, Hans Martin Schmid$^2$, Kerstin Geissler$^3$, Michael R. Meyer$^2$, Thomas Henning$^4$,Wolfgang Brandner$^4$, and Sebastian Wolf$^5$}
\email{\myemail}
%\author{Kerstin Geissler}
%\affil{...}
%\author{Hans Martin Schmid}
%\affil{Institute for Astronomy, ETH Zurich, Wolfgang-Pauli-Strasse 27, 8093 Zurich, Switzerland}    
%\author{Sebastian Wolf}
%\affil{...Kiel, Germany}
%\and
%\author{Thomas Henning, Wolfgang Brandner}
%\affil{Max Planck Institute for Astronomy, K\"onigstuhl 17, 69117 Heidelberg, Germany}
%\author{Sebastian Wolf}
%\affil{...Kiel, Germany}
%\and 
%\author{Kerstin Geissler}

\altaffiltext{1}{Based on observations collected at the European Organisation for Astronomical Research in the Southern Hemisphere, Chile (program number: 077.C-0106A).}
\altaffiltext{2}{Institute for Astronomy, ETH Zurich, Wolfgang-Pauli-Strasse 27, 8093 Zurich, Switzerland}    
\altaffiltext{3}{Physics and Astronomy Department, Stony Brook University, Stony Brook, NY 11794-3800, USA}    
\altaffiltext{4}{Max Planck Institute for Astronomy, K\"onigstuhl 17, 69117 Heidelberg, Germany}    
\altaffiltext{5}{University of Kiel, Institute of Theoretical Physics and Astrophysics, Leibnizstrasse 15, 24098 Kiel, Germany}    

%\altaffiltext{1}{Based on observations with ISO, an ESA project with instruments funded by ESA Member States (especially the PI countries: France, Germany, the Netherlands and the United Kingdom) and with the participation of ISAS and NASA. This work is based in part on observations made with the {\sc Spitzer Space Telescope}, which is operated by the Jet Prop_Ulsion Laboratory.}

%% style, abstract will outp_Ut a Received/Accepted line after the

\begin{abstract}
We present polarimetric differential imaging (PDI) data of the circumstellar disk around the Herbig Ae/Be star HD100546 obtained with VLT/NACO. We resolve the disk in polarized light in the $H$ and $K_s$ filter between $\sim$0.1--1.4$''$ (i.e., $\sim$10--140 AU). The innermost disk regions are directly imaged for the first time and the mean apparent disk inclination and position angle are derived. The surface brightness along the disk major axis drops off roughly with $S(r)\propto r^{-3}$ but has a maximum around 0.15$''$ suggesting a marginal detection of the main disk inner rim at $\sim$15 AU. %although a steeper profile is found along the south-east direction in the $H$ band. 
We find a significant brightness asymmetry along the disk minor axis in both filters with the far side of the disk appearing brighter than the near side. This enhanced backward scattering and a low total polarization degree of the scattered disk flux of 14$^{+19}_{-8}$\% suggests that the dust grains on the disk surface are larger than typical ISM grains. %Taking into account the disk inclination and a flaring angle we construct scattering functions for both filters covering scattering angles of $40^\circ\lesssim\theta\lesssim130^\circ$. 
Empirical scattering functions reveal the backward scattering peak at the largest scattering angles and a second maximum for the smallest scattering angles. This indicates a second dust grain population preferably forward scattering and smaller in size. It shows that, relatively, in the inner disk regions (40--50 AU) a higher fraction of larger grains is found compared to the outer disk regions (100--110 AU). Finally, our images reveal distinct substructures between 25--35 AU physical separation from the star and we discuss the possible origin for the two features in the context of ongoing planet formation. %Finally, we compare our findings to other circumstellar disks observed in scattered light and conclude that HD100546 is an ideal laboratory to study the formation of a future planetary system.
\end{abstract}

%% what keyword p_Unctuation is appropriate.

%% subject header.  Objects should be in the appropriate "individual"
%% headers (e.g. quasars: individual, stars: individual, etc.) with the
%% additional provision that the total number of headers, including each
%% individual object, not exceed six.  The \objectname{} macro, and its
%% alias \object{}, is used to mark each object.  The macro takes the object
%% name as its primary argument.  This name will appear in the paper
%% and serve as the link's anchor in the electronic edition if the name
%% is recognized by the data centers.  The macro also takes an optional
%% argument in parentheses in cases where the data center identification
%% differs from what is to be printed in the paper.

\objectname{HD100546} 


%%%%%%%%%%%%%%%%%%%%%%%%%%%%%%%%%%%%%%%%%%%%%%%%%%%%%%%%%%%%%%%%%%%%%
\section{Introduction}\label{intro}
The direct detection and characterization of exoplanets around young A-type stars \citep[e.g.,][]{marois2008,marois2010,kalas2008,lagrange2009a,lagrange2010,hinz2010,quanz2010} have demonstrated that planets with several Jupiter masses can form in a broad range of orbital separations providing important constraints on planet formation processes. Understanding the formation of planets in more detail requires a solid knowledge of the geometrical, physical and chemical parameters of the circumstellar disks in which planets are thought to form \citep[e.g.,][]{alibert2011}.

To study these planet-forming disks by means of direct imaging in the near-infrared requires high contrast techniques that sufficiently suppress the stellar light and, in case of ground-based observations, minimize the residual speckle noise caused by uncorrected aberrations in the wavefront. A powerful technique to achieve this is polarimetric differential imaging \citep[PDI; see, e.g.,][]{kuhn2001,potter2003,apai2004,hinkley2009} that takes advantage of the fact that the direct stellar light is essentially unpolarized whereas scattered light from dust grains in the surface layer of a circumstellar disk is polarized \citep[e.g.,][]{fischer1995,fischer1996}. In the past, this technique has been successfully applied to study the circumstellar disks around young massive stars \citep{jiang2005,jiang2008}, intermediate mass Herbig Ae/Be stars \citep[e.g.,][]{hashimoto2011,perrin2004,oppenheimer2008,hales2006}, the TTauri star TW Hya \citep{apai2004,hales2006} and some debris disks \citep[e.g.,][]{hinkley2009}. 

Polarimetry from space with the \emph{Hubble Space Telescope (HST)} has the advantage of a more stable point spread function (PSF) and a higher surface brightness sensitivity compared to ground-based facilities and the disk structure and dust grain properties have been investigated for some selected objects \citep[e.g., AU Mic and AB Aur;][]{graham2007,perrin2009}. 
Although most of the studies mentioned above had to cope with technical or observational limitations, such as the use of a coronagraph (which often reduces significantly the accessible inner working angles around the star), a rather modest telescope size or only single filter images, they provided valuable scientific insights for the observed targets.  

In this paper, we present the first PDI study of the transition disk around the well-studied Herbig Be star HD100546 (see, Table~\ref{parameters} for stellar parameters). The circumstellar environment of this object has been intensively studied by previous imaging campaigns using direct imaging from the ground \citep[][$J$ and $K$ band]{pantin2000} and from space with \emph{HST NICMOS} (1.6$\mu$m), \emph{STIS}, and \emph{ACS} (F435W, F606W, F814W) \citep{augereau2001,grady2001,ardila2007}. In particular, the scattered light images across optical and NIR wavelengths of the latter studies revealed a complex circumstellar structure interpreted as a large disk ($R\geq300$ AU) and a remnant envelope. Emission at 10$\mu$m possibly probing the warmer dust layers of the circumstellar disk has been spatially resolved on tens of AU scales by means of mid-infrared (MIR) interferometry \citep{leinert2004,liu2003_b}. MIR spectroscopy studying the emission features of heated dust grains on the disk surface hinted towards noticeable dust processing, i.e., a significant amount of crystalline grains and comparatively large grains \citep{vanboekel2005}. In particular, the resemblance of the MIR spectrum to that of the comet Hale Bopp \citep{malfait1998} led to the conclusion that dust processing similar to that in our own solar system must have taken place \citep{bouwman2003}. Furthermore, to account for a deficit in the NIR excess compared to the strong MIR excess emission, \citet{bouwman2003} suggested the presence of an inner cavity with a radius of 10 AU in order to fit the spectral energy distribution. These authors speculated that a giant planet orbiting at roughly 10 AU may have opened such a gap. An inner cavity was further supported by FUV long slit spectroscopy using \emph{HST/STIS} \citep{grady2005} and spatially resolved echelle spectroscopy of the [OI] 630 nm emission line \citep{acke2006}. The latter authors argued that temporal changes in their [OI] line profile may be another signpost for the existence of the yet unseen massive planet orbiting within the cavity. From rovibrational CO emission lines \citet{brittain2009} found further evidence for an inner cavity ($\sim$13 AU radius) existing not only in the dust but also in the gaseous component of the disk (at least in CO). Qualitatively the same was found by \citet{vanderplas2009} although these authors could only put a lower limit of 8 AU on the inner radius of the CO emission region. \citet{panic2010} used CO emission lines in the (sub-)mm to constrain the mass, kinematics and outer radius of the molecular gas and found that more than 10$^{-3}$ M$_\sun$ of molecular gas is present in a Keplerian disk with an outer radius of $\sim$400 AU. A wealth of additional molecular lines (e.g., CO, OH) as well as a 69 $\mu$m silicate feature was found in data taken with the \emph{PACS} instrument onboard the \emph{Herschel Space Observatory} \citep{sturm2010}. 

\begin{figure*}
\centering
\epsscale{.5}
\plotone{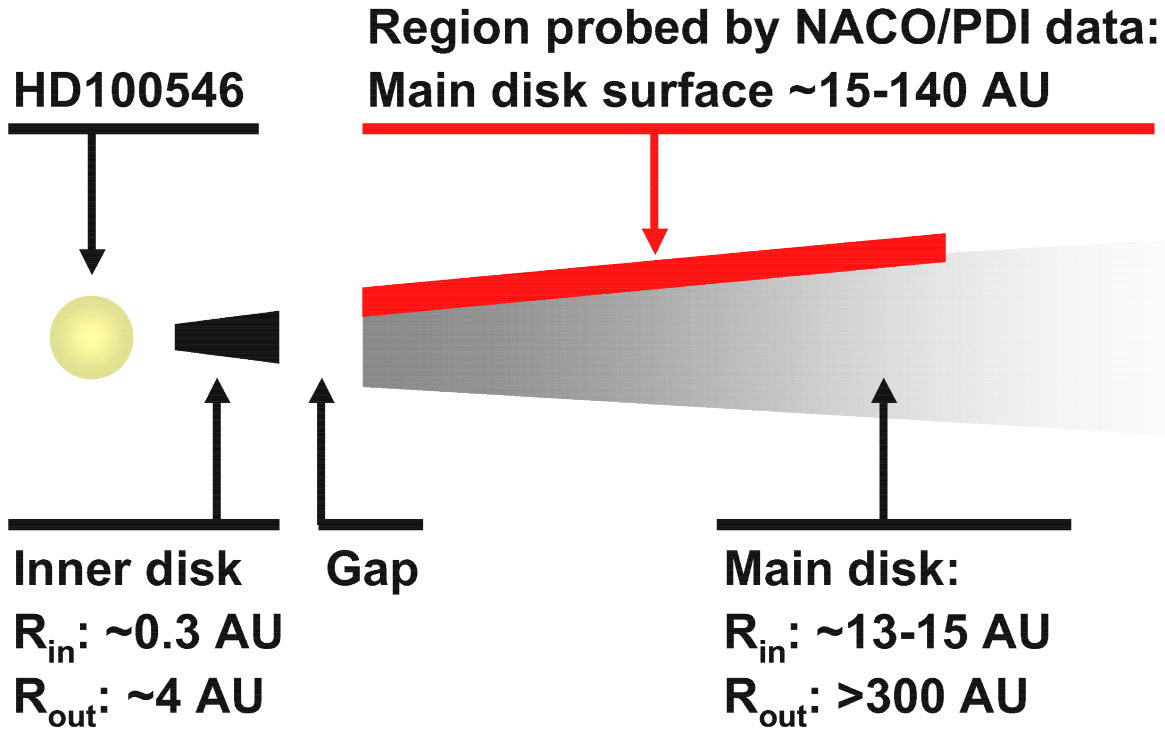}
\caption{Sketch of the HD100546 system. Our data probe the surface layer of the main disk.
\label{disk_sketch}}
\end{figure*}

Currently, the most sophisticated model of the HD100546 system was presented by \citet{benisty2010} who inferred the existence of an additional inner disk from $K$-band interferometry data. These authors model the SED and the NIR/MIR visibilities with a tenuous inner disk component (reaching from $\sim$0.3--4 AU) consisting of micron-sized grains, a gap (stretching from $\sim$4--13 AU) and a massive optically thick main disk ($\sim$13--350 AU). A sketch of the basic disk geometry is shown in Figure~\ref{disk_sketch}. In this model, most of the disk mass is contained in the main disk comprised of mostly large dust grains (1 $\mu$m to 10 mm), but a population of small grains (0.05 to 1 $\mu$m) is also required on the disk surface at a few tens of AU to account for the MIR and FIR excess emission.

Using PDI on an 8m class, AO-assisted telescope in two infrared filters ($H$ and $K_s$) without any occulting mask, our images allow us to analyze disk regions with unprecedented spatial resolution and small inner working angle ($\sim$0.1$''$ to the central star) which are difficult to probe with other direct imaging techniques. Not only are we able to probe the disk geometry and structure but we can also put new constraints on the dust grain properties on the disk surface between $\sim$15--140 AU. In section ~\ref{observationssection}, we describe the observations. The data reduction is explained in detail in section~\ref{datareductionsection}. The key results are presented in section~\ref{resultssection} and analyzed in a broader context in section~\ref{analysissection}. Finally, we discuss our findings in section~\ref{discussionsection} and summarize and conclude in section~\ref{summaryconclusionsection}.

%%%%%%%%%%%%%%%%%%%%%%%%%%%%%%%%%%%%%%%%%%%%%%%%%%%%%%%%%%%%%%%%%%%%%
\section{Observations}\label{observationssection}
The observations were carried out on April 7, 2006, with the AO-assisted, high resolution NIR camera NACO \citep{lenzen2003,rousset2003} mounted on UT4 at the VLT. The observing conditions were photometric. We used the SL27 camera with a pixel scale of 27 mas/pixel and the detector was set to {\tt HighDynamic} mode and read out in {\tt Double RdRstRd} mode. 
NACO is equipped with a polarimetric focal plane mask, 
a rotatable half-wave retarder plate, and a Wollaston prism which can be 
inserted into the light path for imaging polarimetry.
The Wollaston splits the $I_\perp$ and $I_\parallel$ polarization 
into the ordinary and extraordinary beam offset by 3.5$''$ in $y$-direction
on the detector. Overlap of the two images is avoided by the polarimetric mask
which restricts the original field of view of ~27$''\times$27$''$ to $x=27''$ long and $y=3.1''$ wide 
stripes separated by 3.5$''$ in $y$-direction. 
For linear polarization measurements different polarization 
directions can be selected with the rotatable half-wave plate.
One full polarization cycle includes observations in four
different retarder plate positions 
($0.0^\circ,-22.5^\circ,-45.0^\circ,-67.5^\circ$).
To correct for bad pixels we moved the object along the detector's 
$x$-axis between consecutive polarization cycles. 
%To split the beam into an ordinary and an extraordinary beam a Wollaston prism was inserted into the light path. Both resulting beams are imaged simultaneously on the detector. Due to the Wollaston mask the original field of view of ~27$''\times$27$''$ is split into several $\sim$27$''\times3.1''$stripes parallel to the detector's x-axis due with the two central stripes containing both images of the source separated by roughly 3.5$''$.
%To correct for bad pixels we moved the object along the stripes between consecutive exposures. On each detector position we obtained redundant datasets of the Stokes components $Q$ and $U$ by rotating the image on the detector using the retarder plate. In total, four different retarder plate positions were used (0.0$^\circ$, -22.5$^\circ$, -45.0$^\circ$, -67.5$^\circ$) yielding four different rotation angles of the object (0$^\circ$, -45$^\circ$, -90$^\circ$, -135$^\circ$) on each detector position. 
In total, we observed HD100546 in four different filters ($H$, $K_s$, $NB1.64$, $IB2.18$). While we saturated the core of the PSF in the $H$ and $K_s$ filter so that the innermost 5-6 pixels (in diameter) were no longer in the linear detector regime (i.e. $>$10000 counts), the images in the narrow band filters were never saturated and only individual frames showed count rates slightly above the linearity threshold. The FWHM of the PSF in the narrow band filters was typically 4.0-4.5 pixels wide. 

In addition to HD100546, we also observed the main sequence star HD106797 as a reference star in the two broadband filters to check the validity of our data reduction process. The same star was already used as a PSF reference star for HD100546 by \citet{augereau2001}. In the meantime, however, \citet{fujiwara2009} detected a hot debris disk around HD106797 based on mid-infrared observations with the AKARI satellite making this object no longer an ideal reference source and we would pick a different object next time.

While Table~\ref{parameters} summarizes the basic parameters of the two objects Table~\ref{observations} provides a detailed summary of the observations and the observing conditions.

%%%%%%%%%%%%%%%%%%%%%%%%%%%%%%%%%%%%%%%%%%%%%%%%%%%%%%%%%%%%%%%%%%%%%
\section{Data reduction and calibration}\label{datareductionsection}
%For each object and each filter the data were reduced in the same way using two different approaches. While the first approach yielded the polarization fraction $p_Q$ and $p_U$ the second approach resulted directly in the Stokes components $Q$ and $U$. We define the meaning of these quantities below. 

%%%%%%%%%%%%%%%%%%%%%%%%%%%%%%%%%%%%%%%%%%%%%%%%%%%%%%%%%%%%%%%%%%%%%
\subsection{Basic data reduction steps}
The data reduction was carried out using self-written IDL\,--\,scripts unless otherwise described. 
Each exposure was first dark and flatfield corrected and then cleaned for bad pixels using a 3-$\sigma$ filtering process that replaces the central pixel of a 5$\times$5 pixel box with the mean of the other pixels. Dither positions where the AO correction was poor (e.g., due to an open loop) were disregarded from any further analysis (see Table~\ref{observations}). Afterwards, the ordinary and extraordinary image in each exposure was cut out and all images of each filter (i.e., all dither positions, all rotation angles) were aligned to one reference images that was chosen to be the first image of the series. As pointed out by previous authors \citep[e.g.,][]{apai2004}, the alignment of the images has to be done with sub-pixel accuracy in order to avoid the introduction of any spurious features and to effectively suppress the unpolarized light from the central star. For the alignment, we used the IDL-based program {\tt IDP3}\footnote{http://mips.as.arizona.edu/MIPS/IDP3/} that was initially developed for \emph{HST} imaging data  \citep{schneider2002_idp3}. To determine the centroid of each individual image, a two-dimensional Gaussian was fitted to the PSF over a 20-pixel radius. The shifting onto the reference image was carried out using "bicubic sinc" interpolation. Once all images were aligned, we flagged those regions in the individual images where the count rates exceeded the linear regime, i.e., $\>$10000 counts in this detector read-out mode. This way, in the final images at the end of the data reduction process non-linear pixels are flagged and disregarded in the analyses.

%%%%%%%%%%%%%%%%%%%%%%%%%%%%%%%%%%%%%%%%%%%%%%%%%%%%%%%%%%%%%%%%%%%%%
\subsection{Computation of Stokes parameter}
We measured the fractional Stokes parameters $p_Q=Q/I$ and $p_U=U/I$ using
double ratios as described in detail in \citet{tinbergen1996} or as applied
in the polarimetric analysis of planets by \citet{schmid2006}.
%The Stokes parameters are defined as 
%$Q=I_{0^\circ}-I_{90^\circ}$ and $U=I_{45^\circ}-I_{135^\circ}$, where
%position angles refer to the instrumental zero point position of the 
%half wave plate. 

From the two exposures taken with retarder position
$0^\circ$ and $-45^\circ$ we derive $p_Q$. This rotation of the
retarder plate allows to swap the two polarization images in the
two Wollaston beams so that differential effects cancel out. First,
we derive 
%This approach is described in detail in \citet{tinbergen1996} and was used in \citet{schmid2006} for polarization data of Neptune and Uranus. The underlying idea is that by computing the following ratios, all unwanted polarization dependent and time dependent signals are cancelled out. Consider the quantity
\begin{equation}
{\rm R}_Q=\sqrt{\frac{{I^{0^{\circ}}_{\rm ord}}/{I^{0^{\circ}}_{\rm extra}}}{I^{-45^{\circ}}_{\rm ord}/I^{-45^{\circ}}_{\rm extra}}}\quad .
\end{equation}
Here, the superscript denotes the rotation angle of the retarder plate, while the subscript denotes either the ordinary or the extraordinary beam. 
By first dividing the ordinary images by the extraordinary images for a given retarder position, all time dependent factors such as atmospheric transmission and seeing are cancelled out as these images were taken simultaneously. Then, by dividing the ratios of the different retarder positions, all polarization influencing factors such as reflection coefficients of mirrors or individual pixel sensitivities are cancelled out. The fractional polarization $Q/I$ is derived via
\begin{equation}
p_Q=\frac{Q}{I}=\frac{{\rm R}_Q-1}{{\rm R}_Q+1}\quad.
\end{equation}
Since $Q$ denotes the flux measured for the Stokes Q parameter and $I$ refers to the total intensity of the object (i.e., all counts summed up: $I_{\rm ord}+I_{\rm extra}$), $p_Q$ can be expressed in units of "percent" or simply as a fraction. Using the images taken at the other retarder plate positions (-22.5$^\circ$ and -67.5$^\circ$) the fractional polarization for the $U$ component, $p_U$, is derived similarly. We computed $p_Q$ and $p_U$ for each dither position individually. Afterwards the resulting images are averaged to increase the signal-to-noise ratio (S/N) of the final image. The total fractional polarization $p_{I}$ is derived from the final $p_Q$ and $p_U$ images via
\begin{equation}
p_{I}=\sqrt{p_Q^2+p_U^2}\quad .
\end{equation}

%%%%%%%%%%%%%%%%%%%%%%%%%%%%%%%%%%%%%%%%%%%%%%%%%%%%%%%%%%%%%%%%%%%%%
%\subsection{Stokes $Q$ and $U$ using double difference}\label{double_difference}
 To compute the individual Stokes components $Q$ and $U$ for each dither position we also applied the double difference approach as presented in \citet{hinkley2009}:
\begin{equation}
+Q = I^{0^{\circ}}_{\rm ord} - I^{0^{\circ}}_{\rm extra}
\end{equation}
\begin{equation}
-Q = I^{-45^{\circ}}_{\rm ord} - I^{-45^{\circ}}_{\rm extra}
\end{equation}
\begin{equation}
\Rightarrow Q=(+Q - (-Q)) / 2. \quad\quad
\end{equation}
Again, the superscript denotes the rotation angle of the retarder plate, while the subscript denotes either the ordinary or the extraordinary beam. 
We refer the reader to the aforementioned paper for a detailed discussion of this method and its advantages. The Stokes $U$ component is obtained in the same way using the images with rotation angles of -22.5$^\circ$ and -67.5$^\circ$. Finally, the images from each dither position were averaged to increase the S/N of the final $Q$ and $U$ images. The intensity of the polarization (i.e., the total polarized flux) is obtained from the final images via
\begin{equation}
P=\sqrt{Q^2+U^2}
\end{equation}

%%%%%%%%%%%%%%%%%%%%%%%%%%%%%%%%%%%%%%%%%%%%%%%%%%%%%%%%%%%%%%%%%%%%%
\subsection{Correction for instrumental polarization}\label{instrumental_pol}
Being a Nasmyth instrument and never optimally designed for polarimetry, NACO suffers from significant instrumental polarization effects. Unfortunately, the instrumental polarization changes whenever the positions of reflecting surfaces in the light path change relative to each other (i.e., due to different telescope pointing, different rotation angle of the camera, etc.). These effects have been investigated by \citet{witzel2011} for the $K$-band. According to this study the introduced 
polarization is about $0.9~\%$ for the telescope and $0.6~\%$ for NAOS,
while the $U\rightarrow V$ cross talk between linear and circular 
polarization in NAOS can reduce the $U$-polarization flux by up to
15~\%. In this sense the measured $U$ fluxes should be regarded as lower limits. At the time of our observations the polarimetric characteristics
of VLT/NACO were not known. For this reason we have chosen a calibration
approach which uses the central source as zero point polarization reference to estimate the amount of instrumental polarization for each data set. A detailed description is provided in Appendix A.

%%%%%%%%%%%%%%%%%%%%%%%%%%%%%%%%%%%%%%%%%%%%%%%%%%%%%%%%%%%%%%%%%%%%%
\subsection{Alignment and photometric calibration}\label{calibration}
%As can be seen from Table~\ref{observations} the position angle of the camera was different between the $H$ filter and the other filters. In addition to derotating the camera to get the correct image orientation there was an offset in the encoder positions of the retarder plate. 
It turned out that for data taken before fall 2009 the retarder plate position values saved in the image headers were off by $\sim$13.2$^\circ$ compared to the actual angular positions on the sky due to a bug in the encoder positions \citep{witzel2011}. This angular offset had to be subtracted from the camera positions saved in the image headers to obtain the correct orientation of the measured polarization direction on the sky (see, Table~\ref{observations}). Since we chose to align the disk major axis as observed with \emph{HST} \citep{augereau2001} with the x-axis of the detector\footnote{Due to an error in the initial rotation command the camera was rotated by 90$^\circ$ in the $H$ band compared to the other filters so that the disk minor axis was aligned along the detector's x-axis.} the positive $Q$ component is not aligned in north-south direction in our final images as it is the usual convention.

The photometric calibration was based on the unsaturated images in the $NB1.64$ and $IB2.18$ filters because the  images in the broadband filters were saturated as described above. For all dark, flatfield and bad pixel corrected images we cut out the $\sim$19.8$''\times3.1''$ sub-images of the detector that contained either the ordinary ($I^{x^{\circ}}_{\rm ord}$)
or the extraordinary image ($I^{x^{\circ}}_{\rm extra}$) of the star. The size of the sub-images along the x-axis was chosen so that the edges of the detector, where dark and white fringes were apparent, were excluded ($\sim$3.5$''$ on both sides). The size along the y-axis was determined by the size of the Wollaston mask. We computed and subtracted the local sky background in the individual sub-images. The remaining count rates in the sub-images were then regarded as the photometric signal from HD100564. 

To estimate the total intensity of the object we summed up the contribution from ordinary and extraordinary sub-images, i.e., $I^{x^{\circ}}_{\rm tot}=I^{x^{\circ}}_{\rm ord}+I^{x^{\circ}}_{\rm extra}$, for each retarder plate and dither position and computed the mean value.
For this we found approximately $(1.446\pm0.012)$$\times$$10^6$ counts  and  $(8.210\pm0.079)$$\times$$10^6$ counts in the $NB1.64$ and $IB2.18$ filter, respectively. The errors are the standard deviation of the mean ($\sim$0.8\% and $\sim$1.0\%, in the respective filters). Using the transmission curves of the filters\footnote{http://www.eso.org/sci/facilities/paranal/instruments/naco/\newline inst/filters.html} we calculated that the $NB1.64$ filter has a transmission of only $\sim$3.4\% relative to the $H$ band filter. Between the $IB2.18$ and the $K_s$ filter this fraction amounts to $\sim$13.2\%. %These calculations assume a flat SED of the source over the respective wavelength regimes of the filters. 
Taking these factors into account as well as the different integration times between the filters (Table~\ref{observations}) we could estimate the expected count rates for the broadband filters. This intrinsically assumes that HD100546 has the same magnitude in the $NB1.64$ and $IB2.18$ filter as in the $H$ and $K_s$ filter, respectively (see, Table~\ref{parameters}). Knowing now the count rates in the $H$ and $K_s$ filters and the respective magnitudes we can derive the zero point from
\begin{equation}\label{magnitude_equation}
m_{\rm 2MASS}=-2.5\cdot{\rm log}\Big(\frac{I_{\rm {CR}}}{I_{\rm{ZP}}}\Big)
\end{equation}
with $I_{\rm {CR}}$ being the count rate and $I_{\rm{ZP}}$ being the zero point. With the zero points we can then convert the count rates in each pixel in the final images to magnitudes using again equation~(\ref{magnitude_equation}) above, and, finally, we can convert the magnitudes into surface brightness 
\begin{equation}
S=m_{\rm 2MASS}+2.5\cdot{\rm log}(A)\quad [\rm{mag\;arcsec}^{-2}]
\end{equation}
where $A$ equals the square of the pixel scale (given in arcsec pixel$^{-1}$) of the camera. 

Taking into account all the different sources of uncertainties and errors as described in this and the previous sub-section, some of which are difficult to quantify exactly, we estimate that the absolute flux calibration of the final polarized flux images $P$ is only good to 30-40\%. The main uncertainty arises from the correction of the instrumental polarization.

%%%%%%%%%%%%%%%%%%%%%%%%%%%%%%%%%%%%%%%%%%%%%%%%%%%%%%%%%%%%%%%%%%%%%
\section{Results}\label{resultssection}
\subsection{Detection of polarized flux from HD100546}\label{butterfly}
In Figures~\ref{H_images} and \ref{Ks_images} we present the final Stokes $Q$ and $U$ images and the final $p_Q$ and $p_U$ images in the $H$ and $K_s$ filter, respectively. A "butterfly" pattern\footnote{The term "butterfly" pattern arises from the fact that we see two bright lobes along one axis superimposed on two dark lobes rotated by $\sim$90$^\circ$. From equations (4)--(6) it is clear that \emph{both} the bright and the dark lobes contain a polarization signal because we are looking at a difference image.} is apparent in all the images which we interpret as polarized emission coming from light that was scattered on dust grains on the surface layer of the circumstellar disk surrounding HD100546. Such a pattern is a manifestation of tangential polarization vectors and similar images have been obtained for other sources, see e.g., \citet{apai2004} and \citet{hales2006}. In Appendix B we provide an estimate for the typical signal-to-noise ratio (S/N) of our images showing that the observed features are real. 

In our data, the corresponding $H$ and $K_s$ are qualitatively very similar and, as expected, the pattern in the $U$ images is rotated by roughly 45$^\circ$ compared to the $Q$ images. In Figure~\ref{NB_images} we show the final Stokes $Q$ images taken in the $NB1.64$ and $IB2.18$ filter. Again, the observed butterfly pattern is similar to those obtained with the broadband filters, but due to shorter total integration times and a more narrow bandwidth of the filters the S/N in these images is lower. This explains why the polarization pattern can not be traced to the same angular separation from the central star as in Figures~\ref{H_images} and~\ref{Ks_images}. Since our main analyses are based on the higher S/N broadband images we do not show and analyze the complete set of the images obtained in the $NB1.64$ and $IB2.18$ filters and mention only that the observed structures agree very well. 

Figure~\ref{reference_images} is identical with Figure~\ref{Ks_images} but shows the results for the reference star HD106797. No strong and consistent polarization pattern is apparent as most prominently seen in the $p_Q$ and $p_U$ images. As mentioned in section~\ref{observationssection}, \citet{fujiwara2009} found that HD106797 is surrounded by a hot debris disk. They estimated the inner radius  to be $\sim$14 AU and put an upper limit of 41 AU for the outer radius. Thus, theoretically, we could have resolved this disk with our observations as HD106797 and HD100546 are at the same distance. However, the HD106797 debris disk has so far only been detected at thermal infrared wavelengths. No scattered light images in the optical or NIR exist. Thus, given what is known about this source, it is impossible to estimate by what factor the polarized flux potentially coming from the debris disk is below our detection limits. However, as there are a lot of examples where bright debris disks were not detected in scattered light even with \emph{HST} \citep{krist2010} and taking into account that only a fraction of the scattered light will be polarized, it seems seems not surprising that the HD106797 debris disk is not detected in our data. Thus, we are convinced that our data reduction process works correct and does not introduce any spurious features (see also, Appendix A and B where we discuss the polarimetric calibration and the S/N estimates for both objects). 

%At the time of the observations no disk was known to exist around HD106797 which is why this object was chosen as a reference star. In the meantime, however, \citet{fujiwara2009} reported the detection of a hot debris disk around HD106797 based on mid-infrared observations. Due to the shorter total integration times we had for HD106797 the noise is higher in these images compared to HD100546 (see Appendix B for S/N estimates) and up to now, we are not aware that the HD106797 debris disk was detected in scattered light. Hence, we would not expect to see this disk in our data and Figure~\ref{reference_images} shows that our data reduction is not  

A first more quantitative analysis of the HD100546 butterfly pattern is shown in Figure~\ref{azimuthal_profile} where we plot the azimuthal profile of the polarization pattern in Stokes $Q$ and $p_Q$ for the $H$ band as well as Stokes $U$ and $p_U$ for the $K_s$ band. For four different annuli centered on the central star the mean count rate was computed as a function of position angle (averaged over 10$^\circ$ bins). The error bars are the standard deviation of the mean values computed in the individual images at each dither position divided by the square root of the number of images that were combined (see, Table~\ref{observations}).

The strongest sinusoidal variation is found between 0.1--0.25$''$ (corresponding to roughly 10--26 AU in projected separation without taking into account disk inclination), but the pattern can be traced outward to angular separations $\sim$1.4$''$ as can be directly seen in the $p_Q$ and $p_U$ images of Figures~\ref{H_images} and~\ref{Ks_images}.
The polarization pattern in our two filters is qualitatively very similar, but quantitative differences between the filters become apparent. These differences will be discussed in more detail below where the flux calibrated images are analyzed.

%%%%%%%%%%%%%%%%%%%%%%%%%%%%%%%%%%%%%%%%%%%%%%%%%%%%%%%%%%%%%%%%%%%%%
\subsection{Disk orientation}
To assess the disk orientation we fitted isophotal ellipses to the photometrically calibrated total polarized flux images ($P$) obtained in the $H$ and $K_s$ filter (upper panels Figure~\ref{P_and_pI_images}). The fit yields the apparent disk position angle and its inclination assuming the disk is intrinsically centro-symmetric and geometrically thin (i.e., no disk flaring). For the fit we used three surface brightness ranges in the inner disk region ($\le 0.5''$) going from 9.0--9.5, 9.75--10.25, and 10.5--11.0 mag/arcsec$^2$, respectively. Fitting three different brightness ranges allows us to derive a more robust estimates about the disk orientation. For the disk position angle the fit yielded  a mean position angle of 136.0$^\circ\pm$2.9$^\circ$ (east of north) in the $H$ band and 140.0$^\circ\pm$2.4$^\circ$ in the $K_s$ band. For the inclination the values were 45.9$^\circ\pm$1.7$^\circ$ in $H$ and 
48.0$^\circ\pm$2.0$^\circ$ in $K_s$. The errors are the standard deviation of the values measured in the three regions. Thus, the apparent disk orientation on these scales is comparable between the two filters and for the rest of the paper we adopt mean values of 138.0$^\circ\pm$3.9$^\circ$ and 47.0$^\circ\pm$2.7$^\circ$ for the position angle and the disk inclination, respectively. We note, however, that the true geometrical disk inclination might be somewhat different from this value as we used regions of the same brightness in polarized flux for the fit. These regions do not necessarily correspond to identical geometric regions on the disk surface as the polarization depends on the scattering properties of the underlying dust grains and the true disk flaring angle. However, most previous studies used isophotal fitting to derive estimates about the disk orientation and we compare our values to previous measurements in Table~\ref{disk_orientation}. Our results agree with most of the previously derived values although all of them were obtained at larger angular separations. It is interesting to note that we obtain significantly different values if we fit isophots between $\sim$0.7--1.0$''$ (13.0--13.5 mag/arcsec$^2$). In this case the inclination is only $\sim20^\circ$ and $\sim23^\circ$ in $H$ and $K_s$, respectively. Thus, at these separations the disk appears much more face-on but we note that the observed polarized flux is no longer symmetric with respect to the disk major axis obtained from the inner disk (see, Figure~\ref{P_and_pI_images}) and a more detailed modeling of the scattering properties of the dust grains is required.

%%%%%%%%%%%%%%%%%%%%%%%%%%%%%%%%%%%%%%%%%%%%%%%%%%%%%%%%%%%%%%%%%%%%%
\subsection{Disk surface brightness}
\subsubsection{Radial profiles and asymmetries}\label{radial_profiles}
\begin{figure*}
%\centering
\epsscale{1}
\plottwo{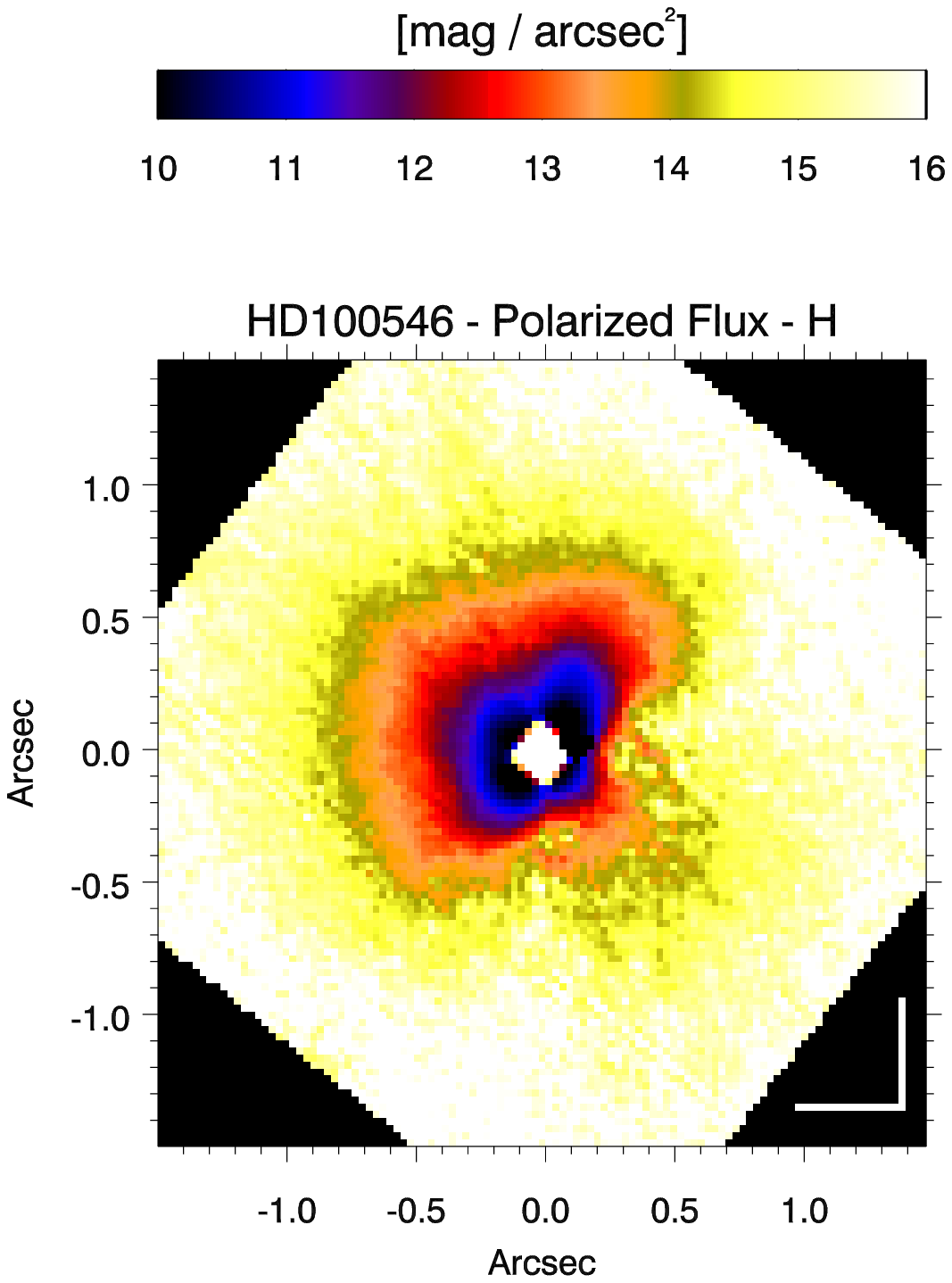}{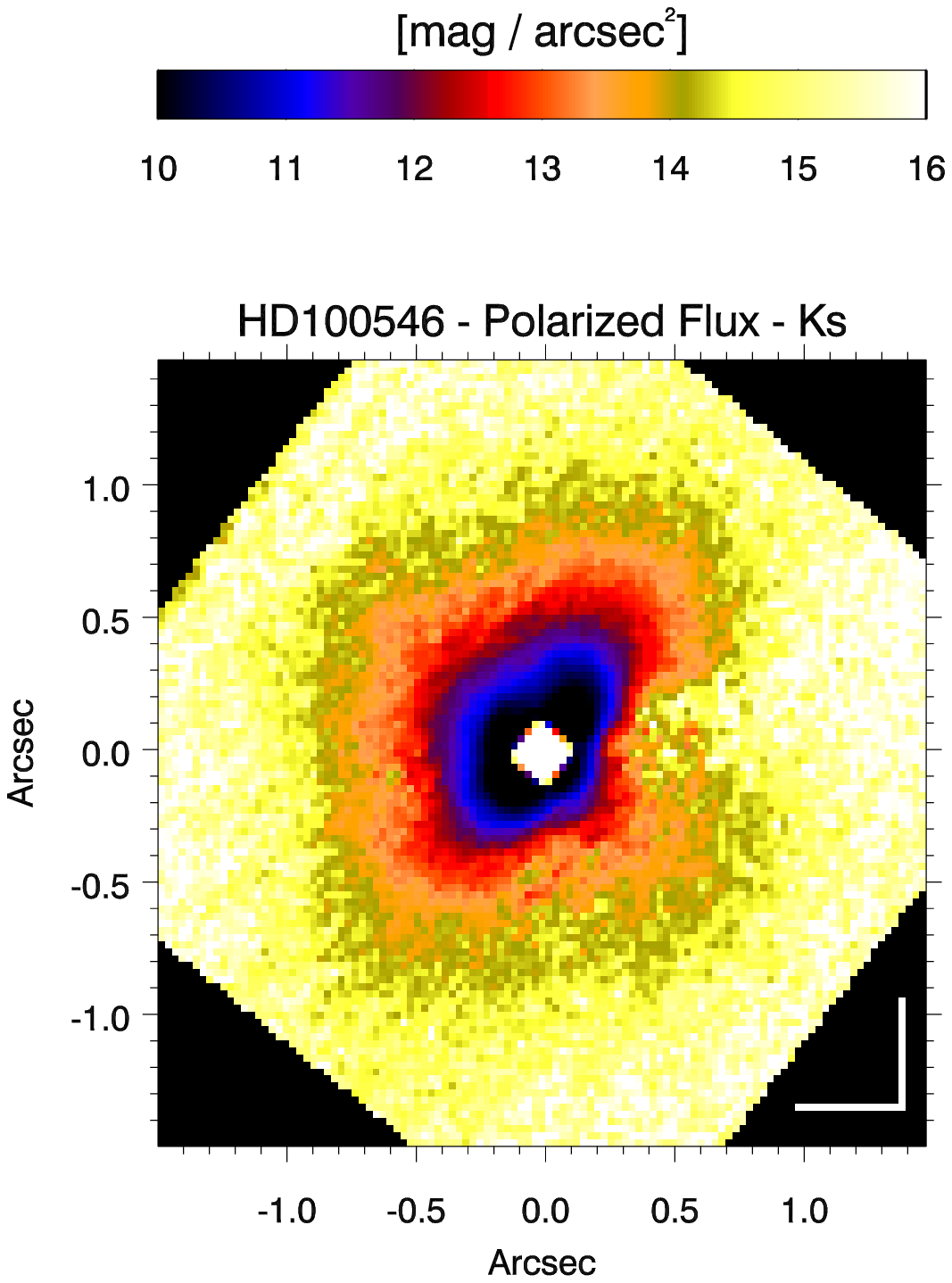}
\plottwo{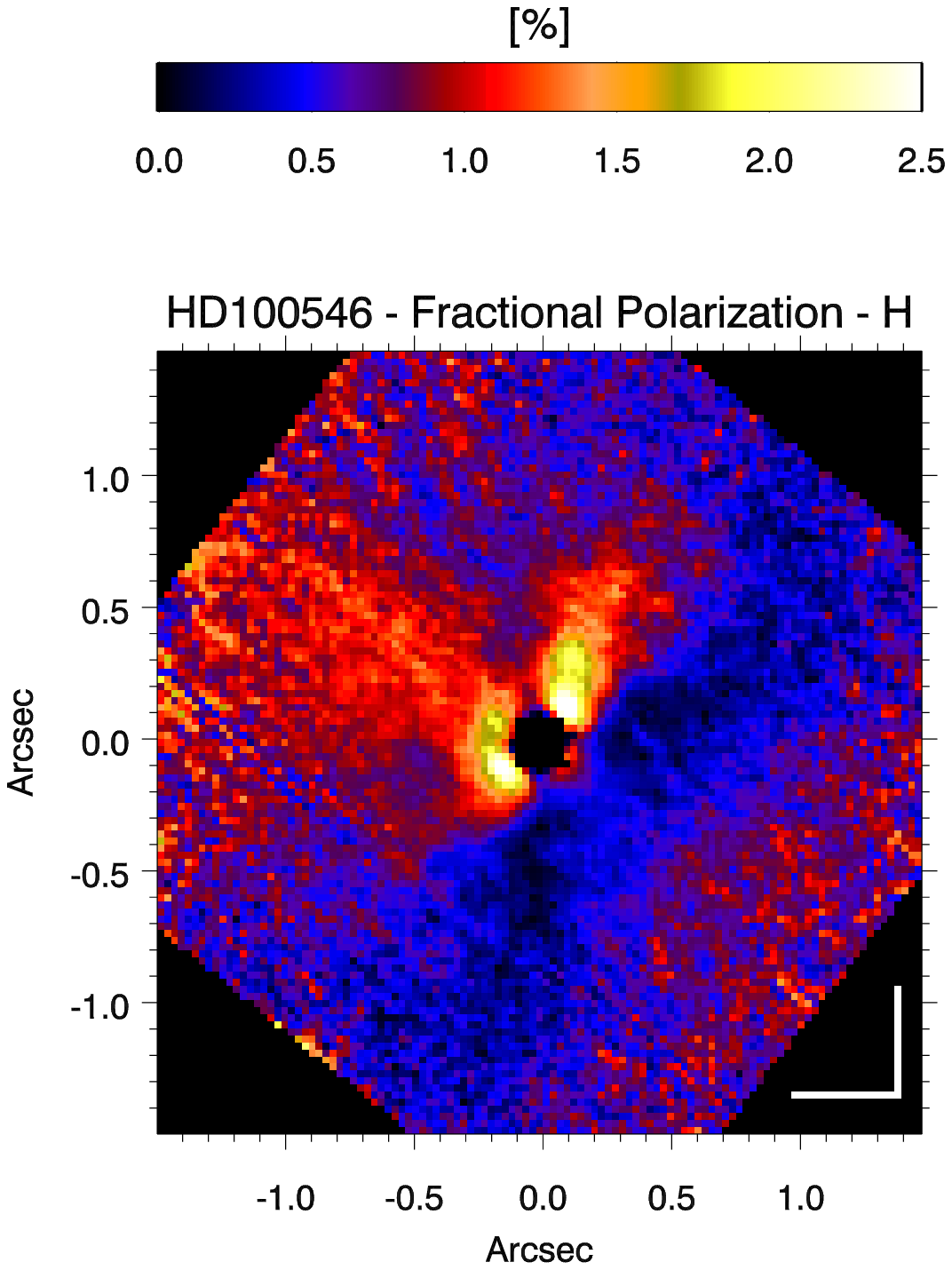}{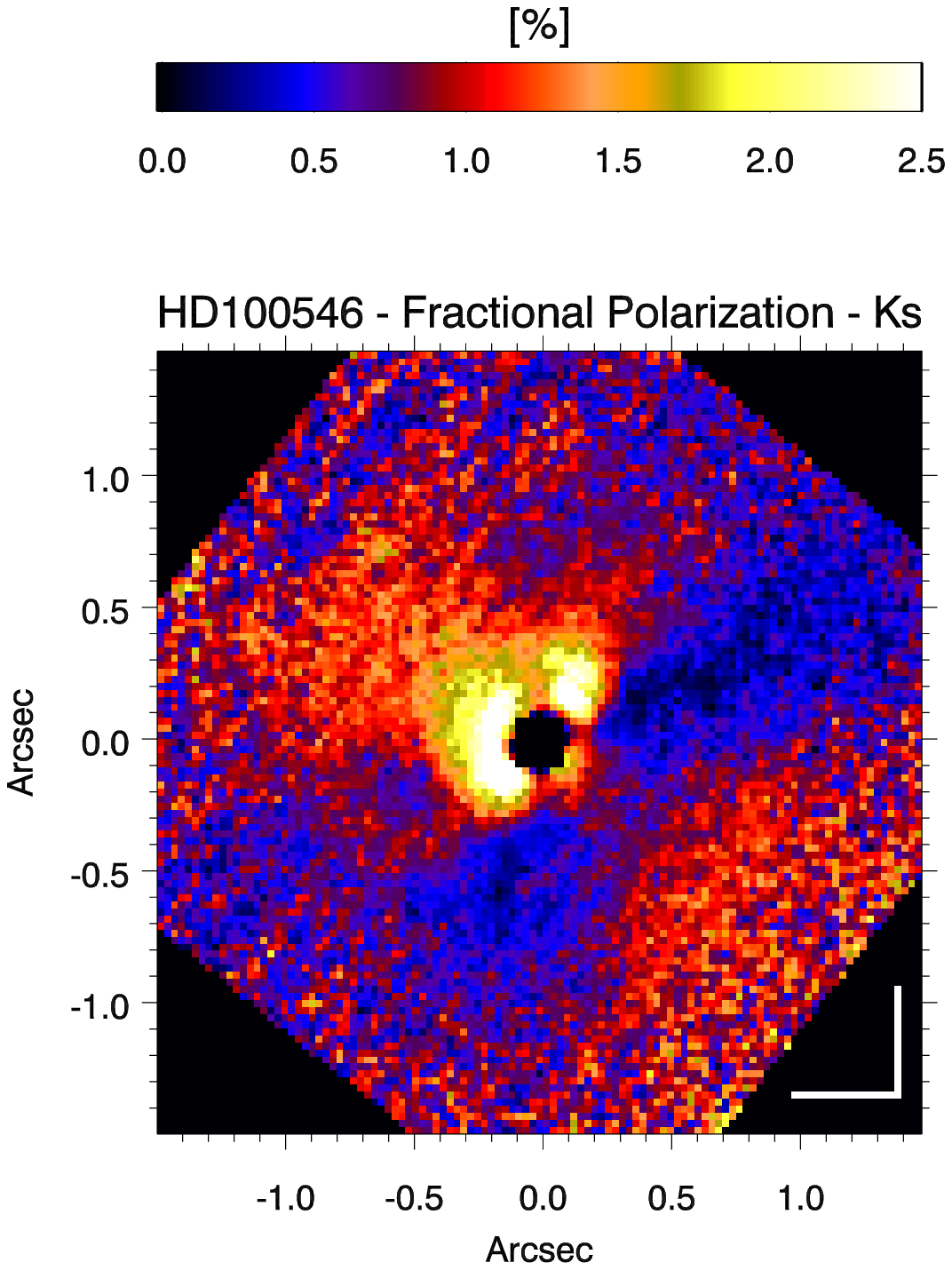}
\caption{\emph{Top:} Surface brightness of polarized flux  $P$ of the HD100546 disk in the $H$ filter (left) and $K_s$ filter (right). \emph{Bottom:} Fractional polarization $p_{I}$ in the $H$ filter (left) and $K_s$ filter (right). North is up and East is left in all images.
\label{P_and_pI_images}}
\end{figure*}

We extracted the profiles of the polarized flux $P$ and the polarization fraction $p_I$ along the semi-major and semi-minor axes of the disk in both filters (see, Figure~\ref{P_and_pI_images}). For both filters we used a 3-pixel wide slit with an orientation of 138$^\circ$ for the major axis (Table~\ref{disk_orientation}) and a perpendicular orientation for the minor axes. The resulting mean profiles are shown in Figure~\ref{brightness_profile}. The error bars are the standard deviation in the 3-pixel wide slit for a given position. The disk has an overall higher polarized flux surface brightness $S(r)$ in the $K_s$ band than in the $H$ band. However, in both filters the brightness decreases similarly fast as a function of radius. Between 0.1--1.4$''$ we find the following power-law exponents that fit the profiles best along the major axis: $-3.47\pm0.05$ ($H$) and $-2.87\pm0.04$ ($K_s$) along the SE semi-major axis, and $-2.84\pm0.04$ ($H$) and $-2.86\pm0.01$ ($K_s$) along the NW semi-major axis. 
%In Table~\ref{brightness_profile_table} we summarize the power-law exponents that fit the profiles best along the semi-major axes between 0.1--1.4$''$. 
The fits were "error-weighted" to account for the uncertainties in the measured radial profile. The errors quoted for the power-law indices are fitting errors. With one exception the profiles show the same radial dependency, roughly $S(r)\propto r^{-2.85}$. Only in the south-east the flux drops even steeper in the $H$-band with $S(r)\propto r^{-3.47}$.

%\begin{deluxetable}{lcc}
%\tablecaption{Resulting exponents from power-law fits to surface brightness profiles between 0.1$''$-1.4$''$. 
%\label{brightness_profile_table}}           % title of Table
%\tablewidth{0pt}
      % is used to refer this table in the text
%\tablehead{
%\colhead{Disk axis} & \colhead{$H$-band}  & \colhead{$K_s$-band} 
%}
%\startdata
%SE major axis & -3.47$\pm$0.05 & -2.87$\pm$0.04 \\
%NW major axis &  -2.84$\pm$0.04 & -2.86$\pm$0.01\\
%\enddata
%\end{deluxetable}

In case of a centro-symmetric disk seen face-on the brightness profile looks identical in all directions. For an inclined disk we would expect the radial profiles to be identical on both sides of the star along the disk major axis (as it is observed in the $K_s$ filter). Along the disk minor axis, which is an axis of mirror symmetry in case of an inclined but intrinsically centro-symmetric disk, the forward and backward scattering properties of the dust grains are directly imprinted and the profiles will be different for both semi-minor axes. This effect is seen in Figure~\ref{P_and_pI_images}: The polarized flux drops much more rapidly between roughly 0.1--0.8$''$ in the south-west direction compared to the north-east direction. This trend is observed in both filters. It is even more easily seen in the brightness profiles for the polarization fraction where the fractional polarization in the south-west is significantly lower than that in the north-east (bottom panels of Figure~\ref{P_and_pI_images}).

\subsubsection{Disk polarization fraction}\label{absolute_polarization}
The fractional polarization $p_I$ measured in our data does not measure the fractional polarization of the scattered light coming from the disk alone as it relates the observed polarized flux to the total intensity composed of both the flux coming from the disk and that coming from the PSF of the central star.
However, by combining the results from \citet{augereau2001} with ours we can get an idea of the polarization fraction of the light scattered from the disk surface. \citet{augereau2001} found an azimuthally averaged value of roughly 10.9$\pm$0.5 mag arcsec$^{-2}$ for the surface brightness of the disk in total scattered light using \emph{HST/NICMOS} at 1.6 $\mu$m. Averaging the polarized flux along the semi-major and semi-minor axes at the same separation in our $H$-filter image with comparable central wavelength (Figure~\ref{brightness_profile}), we find a value of roughly 13.0$\pm$0.4 mag arcsec$^{-2}$. This leads to a total polarization fraction of the disk of $\approx$14$^{+19}_{-8}$\% where the errors relate to the minimum and maximum values given the photometric uncertainties. Ideally, to reduce systematic errors, such a computation is based on data where the total disk intensity $I_{\rm Disk}$ and the polarized flux $P$ is measured at the same time as it is possible with \emph{HST} \citep[e.g.][]{perrin2009}. %In section~\ref{dust_properties} we discuss the absolute polarization fraction in more detail. 

%%%%%%%%%%%%%%%%%%%%%%%%%%%%%%%%%%%%%%%%%%%%%%%%%%%%%%%%%%%%%%%%%%%%%
\subsection{Disk color}
\begin{figure*}
%\centering
\epsscale{1.}
\plotone{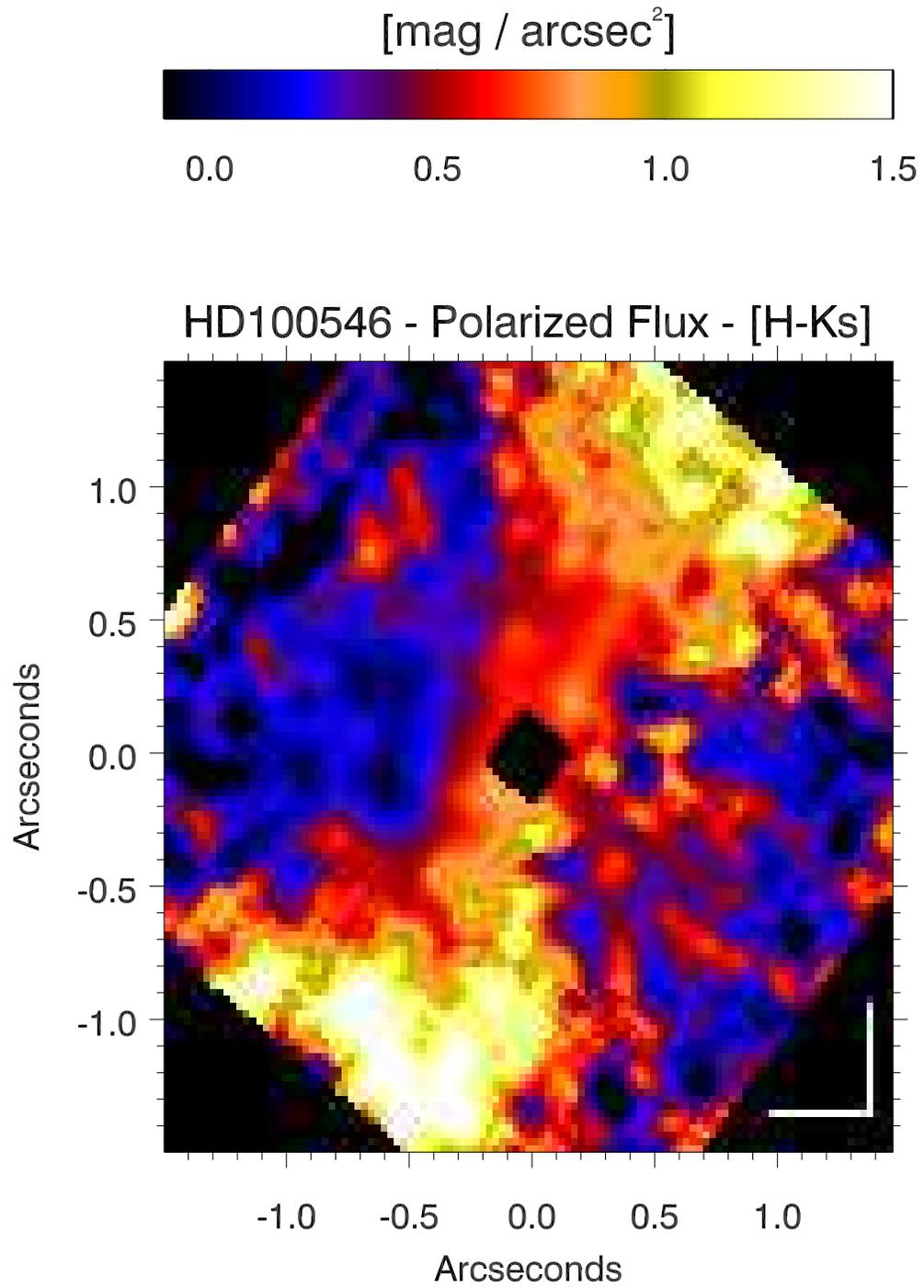}
\caption{$[H-K_s$]-color of the polarized flux computed from the images shown in the upper row in Figure~\ref{P_and_pI_images}. North is up, east to the left.
\label{disk_color}}
\end{figure*}

Having the surface brightness of the polarized flux available in two filters we can compute the $[H-K_s]$-color of the disk surface as shown in Figure~\ref{disk_color}. The image has been smoothed with a moving box of 4$\times$4 pixels so that large scale structures are enhanced. Filtering the image with a 2-D Gaussian kernel with a FWHM of 4 pixels instead of the moving box does not change the results and leads to an almost perfectly identical image. It shows that the disk surface has on average a red color where the reddest parts with $[H-K_s]>0.5$ mag arcsec$^{-2}$ run roughly along the disk major axis. The innermost 1.0$''$ (in diameter) of this area appear like a bar with still moderate colors ($[H-K_s]=0.5$...1 mag arcsec$^{-2}$), while further out the disks becomes increasingly redder over a bigger wedge of the disk surface. It is interesting to note that the position angle of this red symmetry axis is $\sim$160$^\circ$ (east of north) which is  $\sim$20$^\circ$ larger than the position angle derived from the isophot fitting in section~\ref{radial_profiles} but consistent with the disk position angle found by \citet{augereau2001} in total intensity at 1.6$\;\mu$m. Perpendicular to this axis, i.e., roughly along the disk minor axis of the disk, the color is typically $[H-K_s]<0.5$ mag arcsec$^{-2}$. In this context, we remind the reader that the intrinsic  $[H-K_s]$ color of a B9V star is $\sim$0 mag and that the NIR excess emission and the observed NIR colors of HD100546 (see Table~\ref{parameters}) have their origin most likely in the tenuous inner disk of the system: \citet{benisty2010} found from their model that only $\sim$20\% of the total observed $K$-band flux comes directly from the star, while direct thermal emission from the inner disk and thermal emission that was scattered within the inner disk account for $\sim$38\% each. This leaves $\sim$4\% of the total flux to be coming from the star and then being scattered from the inner disk. 
Hence, most of the scattered (polarized) flux that we observe here was emitted and/or scattered from the inner disk.

%%%%%%%%%%%%%%%%%%%%%%%%%%%%%%%%%%%%%%%%%%%%%%%%%%%%%%%%%%%%%%%%%%%%%
\subsection{Sub-structure in the innermost regions}\label{structures}
\subsubsection{A hole and a clump?}
\begin{figure*}
\centering
\epsscale{1}
\plottwo{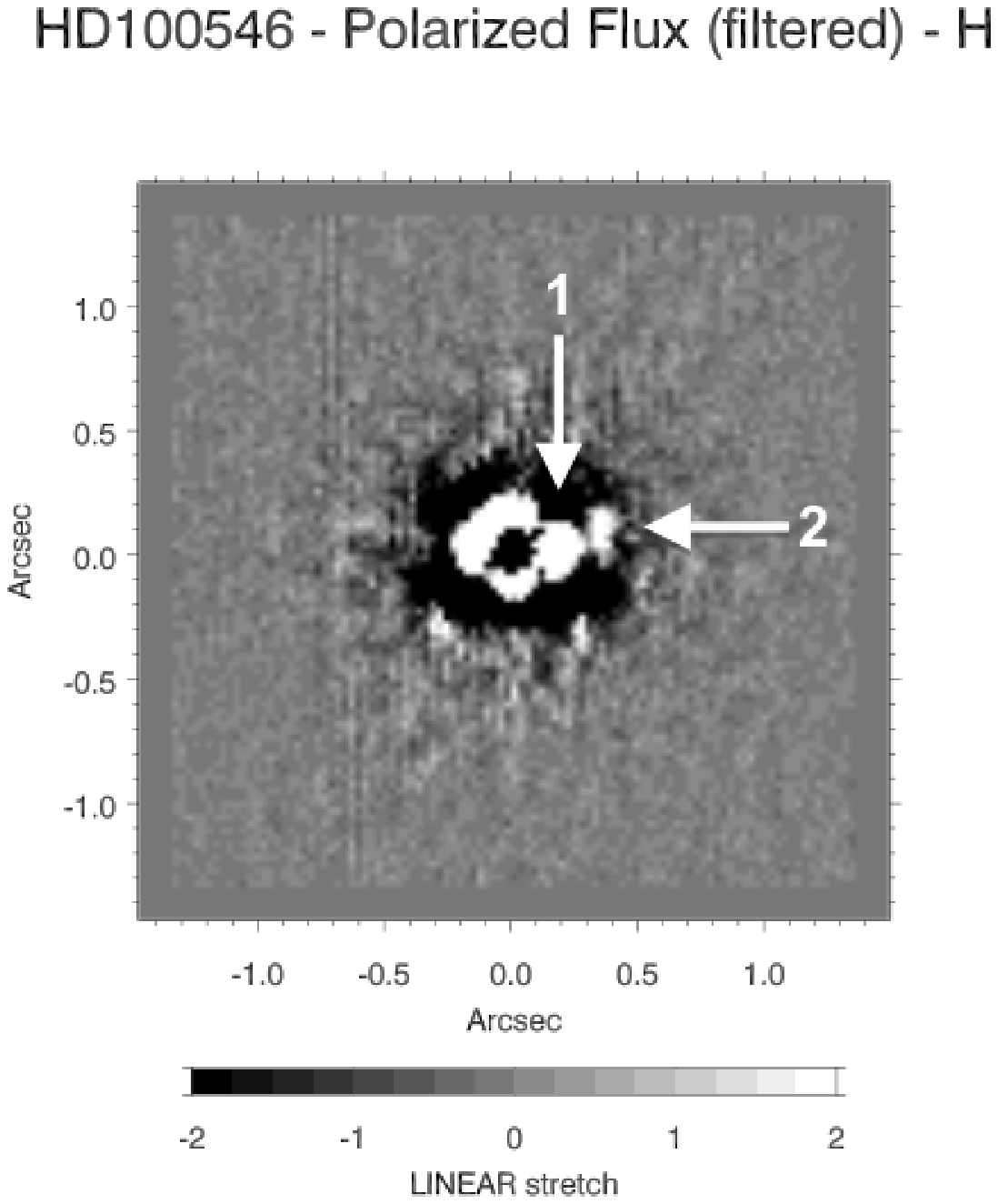}{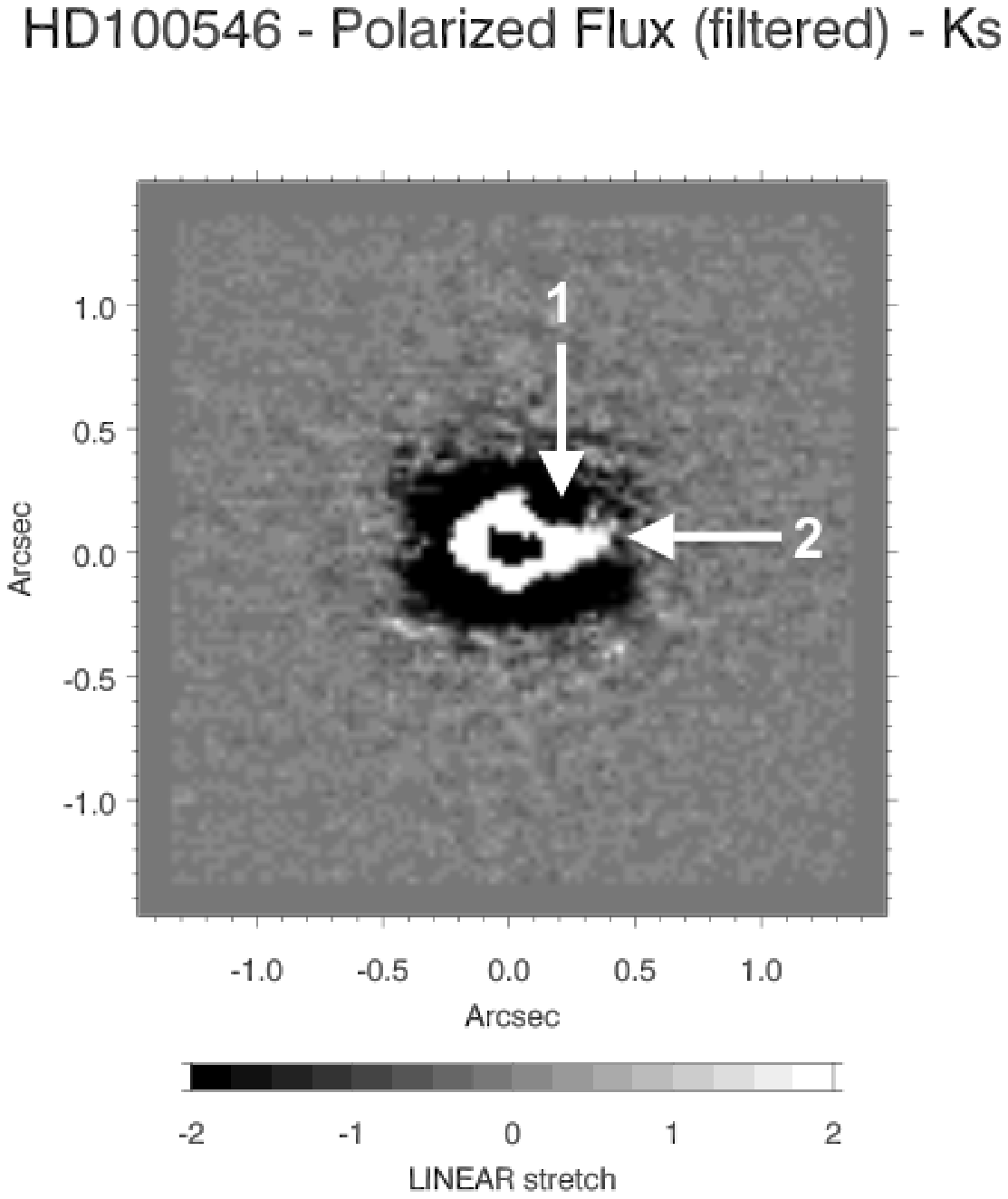}
\caption{High-pass filtered versions of the images shown in the top row of Figure~\ref{P_and_pI_images}. The images have been rotated clockwise so that the disk major axis runs horizontally. The white arrows indicate the positions of the two sub-structures: (1) a 'hole' and (2) a 'clump'. See text for more details. 
\label{disk_filter}}
\end{figure*}

In Figures~\ref{disk_filter},~\ref{disk_hole_zoom} and ~\ref{disk_subtract}  we analyze the structure of the disk in more detail. These images have been rotated clockwise so that the disk major axis runs horizontally. Image~\ref{disk_filter} shows the high-pass filtered version of the $P$ images shown in Figure~\ref{P_and_pI_images}. Here, we have subtracted a smoothed version (smoothing box size 10$\times$10 pixels) of the images from the images themselves so that large scale structures are filtered out and small scale structures are enhanced. It shows that outward of $\sim$0.5$''$ (50 AU) no sub-structures are apparent and the disk surface appears rather homogeneous and smooth. In the inner parts of the disk, however, two features appear as indicated by the arrows and numbers in Figure~\ref{disk_filter}: 

%(1) In particular in the $H$ filter image a spiral-arm-like structure is seen up-left from the center. We note that this structure shows a negative count rate, which means that it is deficient in polarized flux compare to its surroundings. In the $K_s$ image this feature is, however, less pronounced. 

(1) Up-right of the image center, there seems to be a "hole" in the disk. To explore this region further, we zoom in the innermost $\sim$40 AU of the disk and show the polarized flux images as well as the fractional polarization images in the $H$ and $K_s$ filter in Figure~\ref{disk_hole_zoom}. It shows that in the north-east of the disk (at a position angle of $\approx$12$^\circ$ E of N) we have a significant drop in polarized flux and fractional polarization which shows up most clearly in the fractional polarization images. The minimum of the polarized flux is located at the same location in all images, i.e. at $\approx$8 pixels radial separation from the central star (corresponding to $\approx$22 AU projected separation). Also, the magnitude of the flux drop is similar in all images. In case of a centro-symmetric disk, the left-hand and right-hand side of the images should be identical as the disk minor axis of the disk runs vertically through the center. Hence, we can use the mirror-symmetric location in the left-hand side of the images to quantify the flux drop on the right-hand side. It shows that we measure only $69\%\pm4\%$ of the flux / fractional polarization in an aperture with 3-pixels radius. As this figure is the mean and the standard deviation of the ratios in all 4 images shown in Figure~\ref{disk_hole_zoom} we conclude that this flux drop is real as it occurs at the same position in $H$ and $K_s$. 

(2) Extending almost exactly to the north-west along the semi-major axis of the disk we find a local flux maximum in the disk (a 'clump'). Although this feature is present in both filters the location of its peak flux is not exactly identical. The separation from the central star amounts to $\sim$0.35$''$ in both images but the position angle (E of N) is roughly $-21^\circ$ in the $H$ filter and $-32^\circ$ in the $K_s$ filter. Still, the region of enhanced polarized flux clearly overlaps in $H$ and $K_s$. In Figure~\ref{disk_subtract} we have subtracted the mirrored left-hand side of the disk from the right-hand side, to check whether this apparent flux maximum in Figure~\ref{disk_filter} is real. As mentioned above, an inclined centro-symmetric disk should have a mirror symmetry with respect to its minor axis. However, we can clearly see that the right-hand side of the disk has a flux excess in the inner $\sim$0.6$''$ compared to the left-hand side in both filters, and the 'clump' seen in Figure~\ref{disk_filter} is part of a larger structure. At larger radial distances no additional structures are seen and both sides of the disk appear very similar and cancel out in the subtracted images. %We note that the disk 'hole' is also apparent as a flux deficit in Figure~\ref{disk_subtract}.

Since the differential PDI technique compensates for instrumental effects within the CONICA camera and since we took our data with the camera rotated by 90$^\circ$ between the $H$ and $K_s$ filter we conclude that these features are real and cannot be a residual or ghost introduced by the camera, the telescope or the AO-system. Furthermore, we applied a second high-pass filtering method by taking the Fourier transform of the images and filtering all the low frequency components. Also here we recovered both features in both images. Finally, the features are located in image regions that show comparatively high S/N ratios (S/N$>$10 in the individual $Q$ and $U$ images at the location of the features; see, Appendix B). We will discuss these features in more detail in section~\ref{planet_formation}.

%%%%%%%%%%%%%%%%%%%%%%%%%%%%%%%%%%%%%%%%%%%%%%%%%%%%%%%%%%%%%%%%%%%%%
\subsubsection{The rim of the main disk}\label{rim}
\begin{figure*}
%\centering
\epsscale{1}
\plottwo{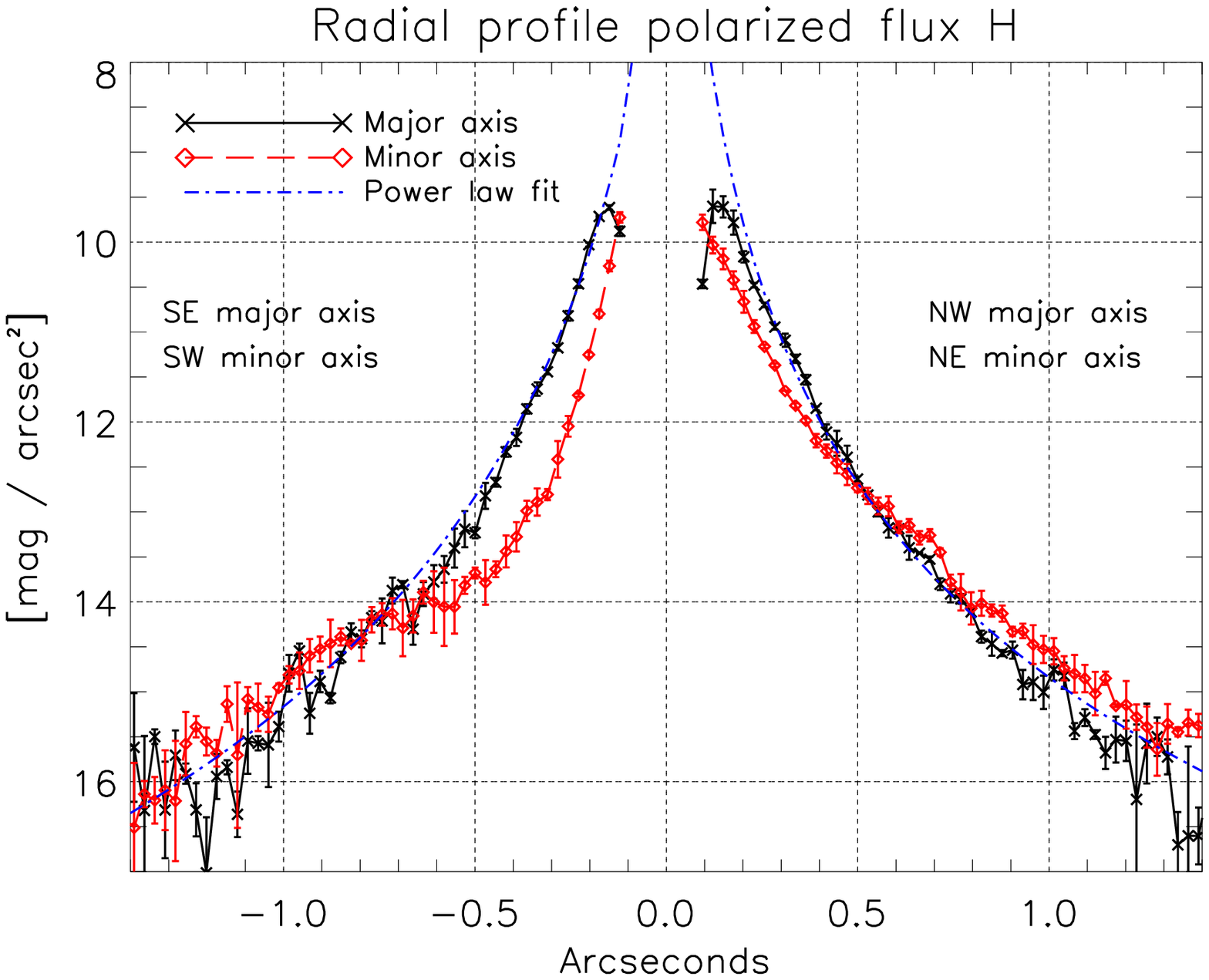}{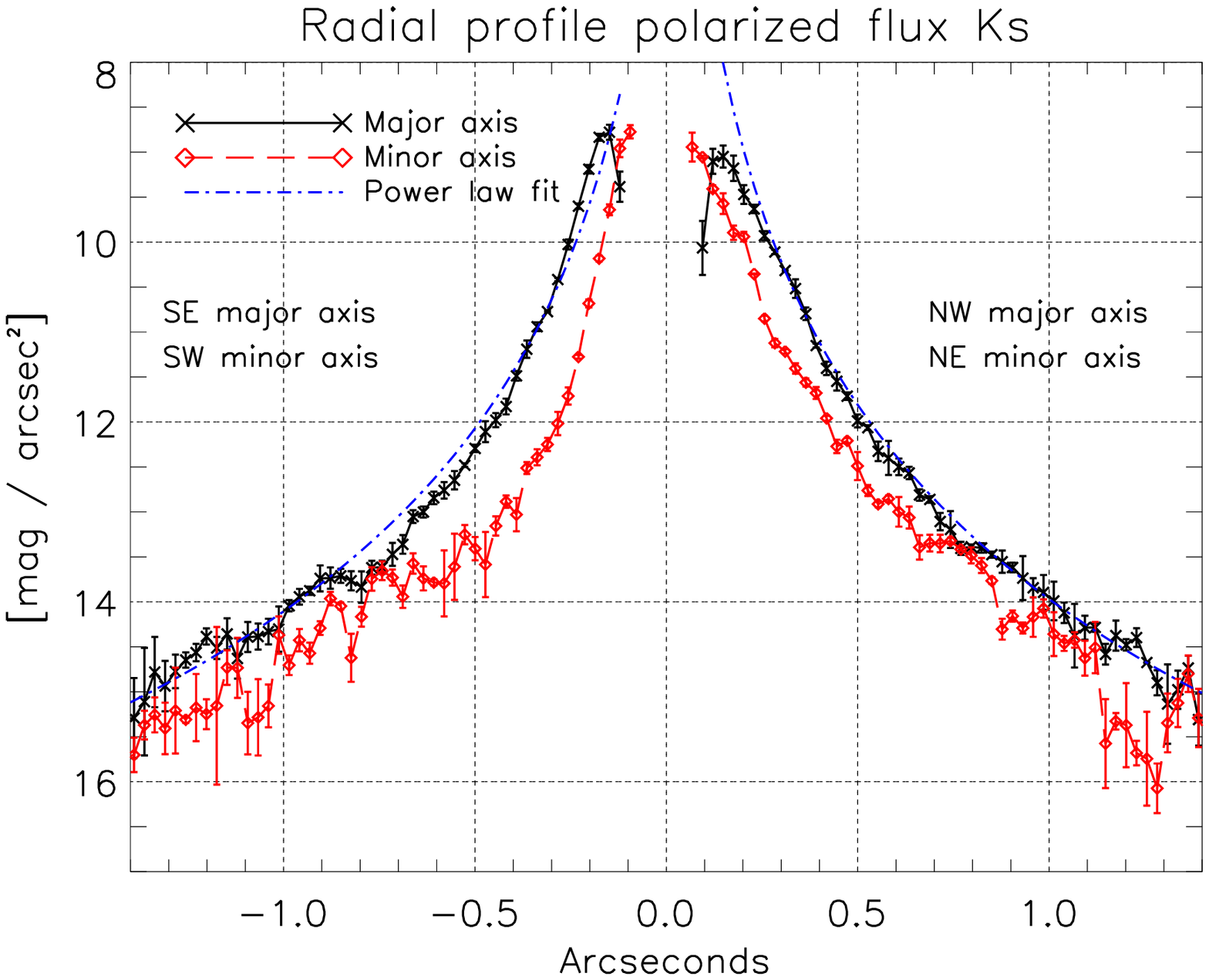}
\plottwo{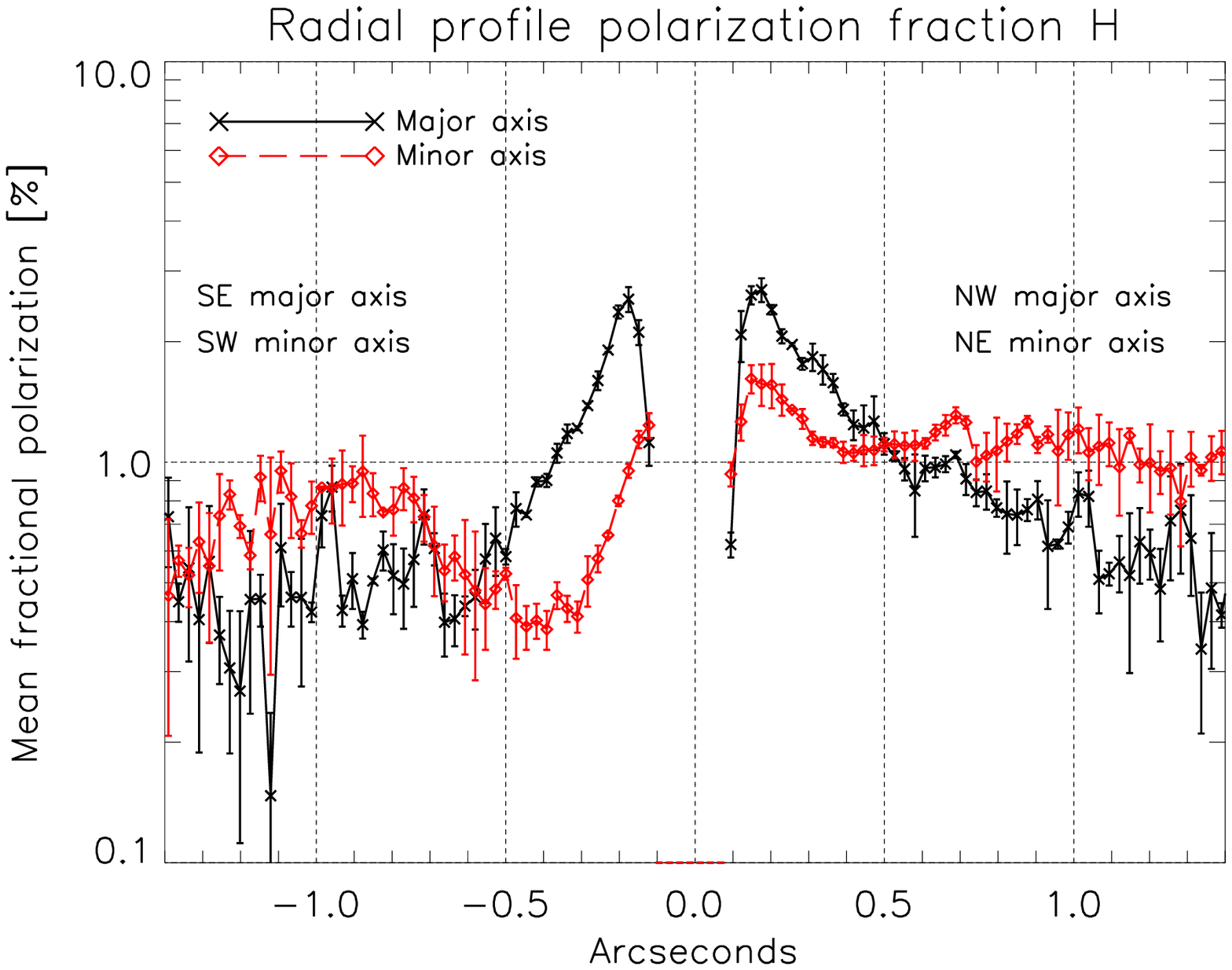}{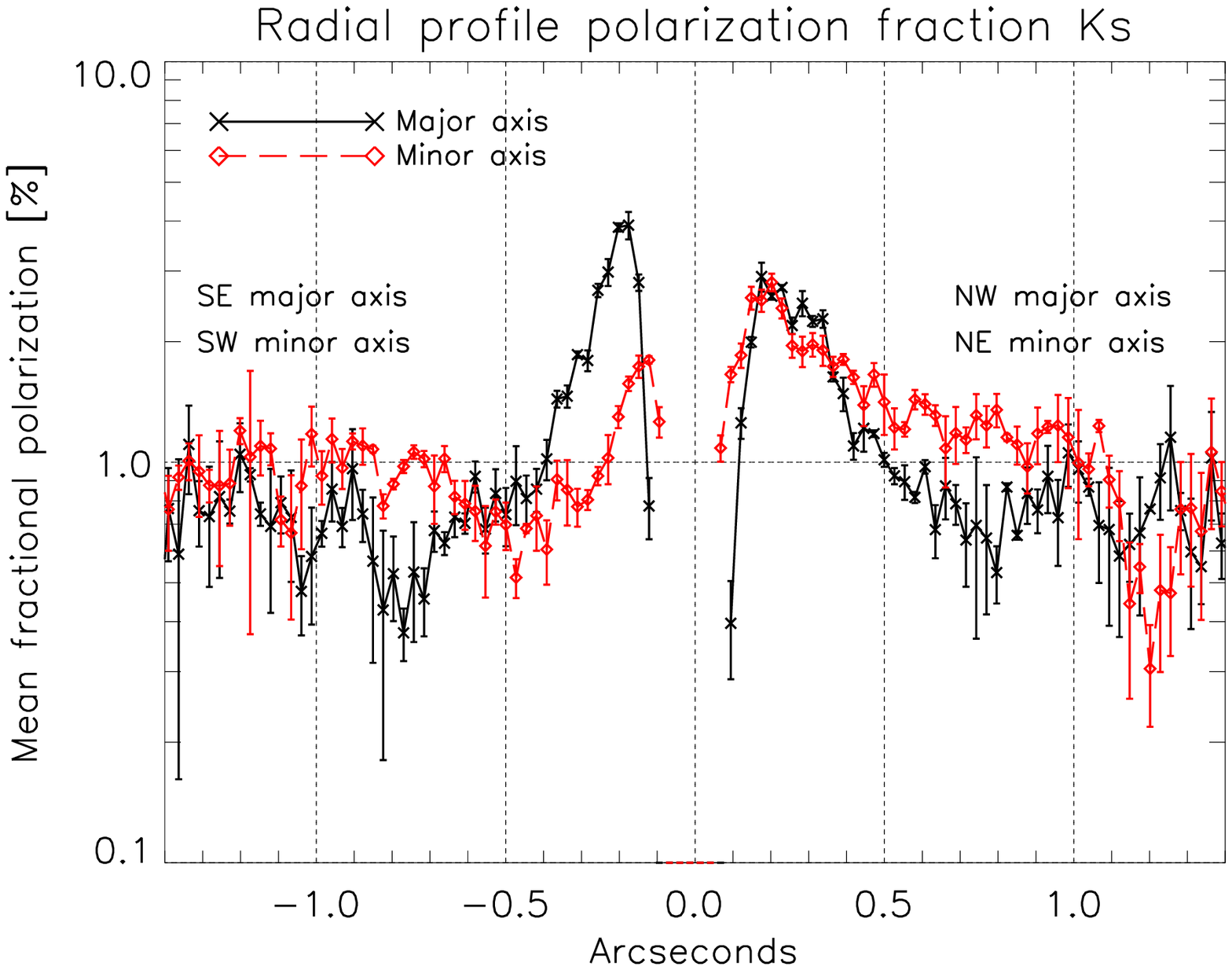}
\caption{Radial surface brightness profiles for the semi-major and semi-minor axes of the HD100546 disk derived from the images shown in Figure~\ref{P_and_pI_images}. We put a 3-pixel wide slit along the axes and computed the mean and the standard deviation across the slit for each pixel position. The black lines show the results for the semi-major axes, the red line those for the semi-minor axes. The blue line in the top row is a power-law fit to the profile of the semi-major axes as discussed in the text. The negative part of the x-axis refers to the SE semi-major axis and SW semi-minor axis, while the positive part shows the NW semi-major axis and NE semi-minor axis.
\label{brightness_profile}}
\end{figure*}

As mentioned in the introduction, the current model of the HD100546 system locates the inner rim of the main disk at $\sim$13 AU \citep{benisty2010} and there is observational support for this from different studies \citep[e.g.,][]{grady2005}. We checked in our images whether we are able to confirm this finding. Figure~\ref{inner_rim} shows a horizontal cut through the center of the images shown in the upper row of Figure~\ref{disk_hole_zoom}. This is essentially identical to the -0.4 to +0.4 regions of the plots shown in the upper row of Figure~\ref{brightness_profile}. We used a 3-pixel wide horizontal box and computed the mean and the standard deviation of the pixels for a given position. The innermost regions close to the star have been masked out if any of the pixels in the slit at this position showed count rates exceeding the linear detector regime. It shows that we find the maximum polarized flux between 0.1--0.15$''$ on \emph{both} sides of the central star in \emph{both} filters and a steep drop of flux closer to the star. The angular separation of the maximum flux corresponds to 10--15 AU at the distance of HD100546 which would support the idea that there is a jump in the density of scattering dust grains at this location. 

%The polarization fraction images show a very similar behavior with a significant decrease in the polarization fraction as we approach the star. However, we must be careful when interpreting this latter finding because the flux contribution from the stellar PSF clearly dominates the inner regions: Even if there were a similar amount of scattering particles close to the star the polarization fraction would drop due to the overwhelming flux from the central star. 

%A careful look at 
%Figure~\ref{inner_rim} reveals that the maxima in the polarization fraction plot (lower panel) do not perfectly coincide with the maxima in the polarized flux (upper panel), but are shifted 1--2 pixels further away from the star due to the effect described above. 

The interpretation of these results should be taken with caution. Although we find the maximum flux at the suspected location of the inner rim, there is only 1 radial pixel closer to the star on both sides and in both filters that  did not suffer from saturation effects. Also, only along the north-west axis (positive x-values) the flux drop appears to be significant taking into account the error bars. %If one were to believe that we are seeing the inner rim in our data, then one could take this as the basis to speculate about an asymmetry in the inner cavity. 
However, the consistent behavior we find for the polarized flux profiles close to the star  in both filters allows us to conclude that we may indeed have directly imaged the suspected inner rim of the main disk between 10--20 AU. ideally, additional new data with a non- (or less) saturated PSF core would be required to fully confirm this interpretation. 

%%%%%%%%%%%%%%%%%%%%%%%%%%%%%%%%%%%%%%%%%%%%%%%%%%%%%%%%%%%%%%%%%%%%%
\section{Analysis}\label{analysissection}
\subsection{Disk geometry - Backward scattering preferred}
In order to relate the observed polarized flux to some dust grain properties of the disk surface, it is necessary to understand which side of the inclined disk is the near side and which is  the far side. Based on the rotation signature of the [OI] gas in the inner 100 AU, \citet{acke2006} concluded that the disk is rotating counterclockwise, i.e., the north-west side is receding. This is also confirmed by the CO measurements from \citet{panic2010}.
%This is confirmed by test data images of HD100546 taken with the Atacama Large Millimeter Array (ALMA) and made public on the internet\footnote{http://www.almaobservatory.org/es/visuales/\\imagenes/main.php?g2\_itemId=2979}. 
Combining this information with the large-scale spiral structure seen by \citet{grady2001} and also \citet{ardila2007}, and assuming that these spiral structures are trailing spiral arms, the north-east side of the disk is the far side and the south-west side of the disk the near side relative to Earth. Based on this, the observed brightness asymmetry along the disk minor axis (section~\ref{radial_profiles}) tells us that \emph{the far side of the disk appears brighter in polarized flux than the near side}. This in turn means that the dust grains responsible for the polarized flux are preferentially backward scattering.

We note that the observed brightness asymmetry within the innermost $\approx$0.8$''$ along the disk minor axis was already recognized in \emph{HST/NICMOS} data by \citet{augereau2001} and anisotropic scattering properties of the dust grains was mentioned as one possible explanation. Interestingly, in the \emph{HST/ACS} data presented by \citet{ardila2007} there is also a brightness asymmetry along the minor axis but in the opposite sense: the amount of scattered light in the south-west is larger than in the north-east in all three optical bands. However, these authors were only sensitive to disk regions $\ge$1.6$''$ as subtraction residuals dominated in the innermost regions. Still, this result needs to be considered if one tries to derive dust grain properties from a combination of all scattered light images. We discuss this in more detail below in section~\ref{dust_properties}.

\subsection{Polarized scattering functions}
To further quantify this brightness asymmetry we computed the polarized scattering function for the disk in both filters (Figure~\ref{scat_func}). For this purpose we used again the polarized flux images shown in the upper panels of Figure~\ref{P_and_pI_images}. Given the structures we found in the north-west side of the disk (see, section~\ref{structures}), the scattering functions were computed for the opposite disk side. The scattering angle $\theta$ is a function of  the azimuthal angle $\phi$ (starting from the semi-minor axis in the north-east  going counterclockwise to the semi-minor axis in the south-west), the disk inclination $i$ and the disk flaring angle $\gamma$ and is given by
\begin{equation}
\begin{array}{ll}
\theta= 90^\circ+(i-\gamma)\cdot{\rm cos}\;\phi & \quad ({\rm for}\;\phi\le90^\circ)\\
\theta= 90^\circ+(i+\gamma)\cdot{\rm cos}\;\phi & \quad ({\rm for}\;\phi>90^\circ)
\end{array}
\end{equation}
For the disk inclination we adopted the mean value of 47$^\circ$ (see Table~\ref{disk_orientation}) and for the flaring angle 7$^\circ$ which corresponds roughly to the flaring index $\beta$ used in the disk model from \citet{benisty2010}. For two different annuli (40--50 AU and 100--110 AU) we computed the mean and the standard deviation of polarized flux  in wedges of 10$^\circ$ as a function of $\theta$ and plotted the results in Figure~\ref{scat_func}. For the annuli we also had to take into account the disk inclination to make sure we were analyzing the same disk regions in terms of physical and not projected separation from the star. The curves were normalized to unity at the brightest disk region which is the far side of the disk as described above. The maximum scattering angle is roughly 130$^\circ$ while on the near side of the disk it is a bit less than 40$^\circ$. 

In these scattering functions a few points points are noteworthy: (1) All four curves are qualitatively very similar: Going from the backward scattering peak at the largest scattering angles to smaller $\theta$, all curves drop, show a minimum between $\theta=40^\circ$--70$^\circ$ (i.e., a scattering angle with minimum scattering efficiency) before they rise again for even smaller angles. (2) Comparing the curves for the $H$ filter with those for the $K_s$ filter, the flux drop seems less pronounced in the $K_s$ filter in both annuli. (3) Comparing the curves for the two annuli, in both filters the black curves (probing 40--50 AU) seem to fall off less steep than the red curves (probing 100--110 AU) going from the backward scattering peak to smaller values for $\theta$. Also, the black curves fall more or less steadily until they reach their minimum and the rise at the smallest scattering angles is not as high as for the red curves. In contrast, the red curves, after their sharp drop-off coming from the backward scattering peak, remain almost flat until they increase again at angles $<$50$^\circ$.

%Looking at the NIR color of the disk (Figure~\ref{disk_color}) we find that the the far side, with scattering angles between 110--130$^\circ$, shows modest $[H-K_s]$ color almost regardless of distance from the star. Interestingly, a lot of Kuiper Belt Objects in our own Solar System show also very little $[H-K_s]$ color \citep{delsanti2004} even if the scattering angles for these objects may have been even closer to 180$^\circ$. 

%Finally, we remind the reader that the MIR spectrum of HD100546 is remarkably similar to that of comet Hale Bopp \citep{malfait1998,bouwman2003}. Taking all these findings together it appears as if in the HD100546 system very similar processes like in our own Solar System took place and left their imprints on the disk and dust grain properties. 

%WORK IN PROGRESS
%\begin{itemize}
%\item{WHAT ELSE CAN WE LEARN FROM THE SCATTERING FUNCTIONS AND THE COLOR OF THE DISK?}
%\item{CAN DIFFERENT GRAIN SIZES EXPLAIN THE INCREASINGLY REDDER COLOR OF THE DISK AS WE GO FURTHER OUT ALONG THE SEMI MAJOR AXIS?}
%\item{DO WE WANT TO BUILT A COMPLETE DISK MODEL OR WOULD A DUST MODEL EXPLAINING THE DISK COLORS AND SCATTERING FUNCTIONS BE SUFFICIENT? ALL MODELS I KNOW (FROM THE AMSTERDAM AND FROM THE GRENOBLE GROUP) DO NOT MATCH THE SED AND THE SCATTERED LIGHT IMAGES SIMULTANEOUSLY YET AND THEY INVESTED A LOT OF TIME!}
%\end{itemize}

%%%%%%%%%%%%%%%%%%%%%%%%%%%%%%%%%%%%%%%%%%%%%%%%%%%%%%%%%%%%%%%%%%%%%
\section{Discussion}\label{discussionsection}
\subsection{Brightness profile and comparison to other disks}
The measured profiles show that the scattered surface brightness along the semi-major axes in the sky plane drops off roughly $S\propto r^{-3}$. 
%Such a profile is expected for an optically thick, geometrically thin irradiated disk \citep[e.g.,][]{hartmann1998}. 
If one includes disk flaring then the expected profile becomes less steep and fits even better to our observed values. Taking into account the $f(r)\propto r^{-2}$ drop in stellar radiation this implies that the surface density of the scattering dust grains drops off roughly $\Sigma\propto r^{-1}$ if we assume constant scattering properties of the dust grains over the observed radial range.

Our values for the inner $\sim$140 AU of the disk are in good agreement with the results from \citet{augereau2001} who found an azimuthally averaged surface brightness profile $S(r)\propto r^{-2.9}$ between $\sim$55--250 AU in their \emph{HST/NICMOS} 1.6$\mu$m data. Further out, the power law seems to become steeper: \citet{augereau2001} measured $S(r)\propto r^{-5.5}$ between 250--350 AU while \citet{ardila2007} found $S(r)\propto r^{-3.8}$ between 160-500 AU in their optical \emph{HST/ACS} data. This suggests either a discontinuity in the surface density of the dust disk, different dust grain properties at larger disk radii or a different disk geometry (e.g., changing flaring angle). We come back to this point below in the next section.

Comparing the observed surface brightness profile to other directly imaged objects we find that also the Herbig Ae star HD169142 shows $S(r)\propto r^{-3}$ between 80--190 AU \citep[][; \emph{HST/NICMOS}]{grady2007}. For lower mass TTauri stars \citet{apai2004} found $S(r)\propto r^{-3.1}$ for TW Hya between 50--80 AU using NACO/PDI\footnote{We note that based on \emph{HST/NICMOS} data \citet{weinberger2002} found a more shallow profile for TW Hya of  $S(r)\propto r^{-2.6}$ between 45--150 AU.} and \citet{schneider2003} found $S(r)\propto r^{-3...-3.5}$ for GM Aurigae for distances $\ge$85 AU using \emph{HST/NICMOS} data. Irrespective of stellar mass, many circumstellar disks show a very similar surface brightness profile in scattered light images for disk radii $\geq$50 AU consistent with optically thick (and geometrically thin or only moderately flared) disks. 

Interesting is the comparison of HD100546 with TW Hya as both objects are transition disks and both were observed with NACO/PDI. For HD100546 the strongest polarimetric signal is found in the innermost regions of the disk (i.e., in the annulus from 0.1--0.25$''$ corresponding to roughly 10--26 AU in projected separation without taking into account disk inclination) while \citet{apai2004} found the stronger signal not close in but further out in the TW Hya disk (0.76--1.03$''$ corresponding to roughly 43--58 AU). Thus, apparently those regions in the TW Hya transition disk that are optically thin and depleted of dust reach further out compared to HD100546. And since TW Hya lies at only $\sim$54 pc distance, the data from \citet{apai2004} seem to resolve these regions of reduced dust surface density. As we have seen in section~\ref{rim} our data tentatively resolve the inner rim of the HD100546 main disk but probing even further in (i.e, closer than 10 AU) is beyond the capabilities of our observations. %To investigate in more detail why the TW Hya disk appears more evolved than the HD100546 disk despite the fact that both objects have a comparable age and TW Hya is less massive and thus less luminous (i.e., the disk is suffering less evaporating radiation) is beyond the scope of this work. 

\subsection{Dust grain properties}\label{dust_properties}
\begin{figure*}
%\centering
\epsscale{1.}
\plottwo{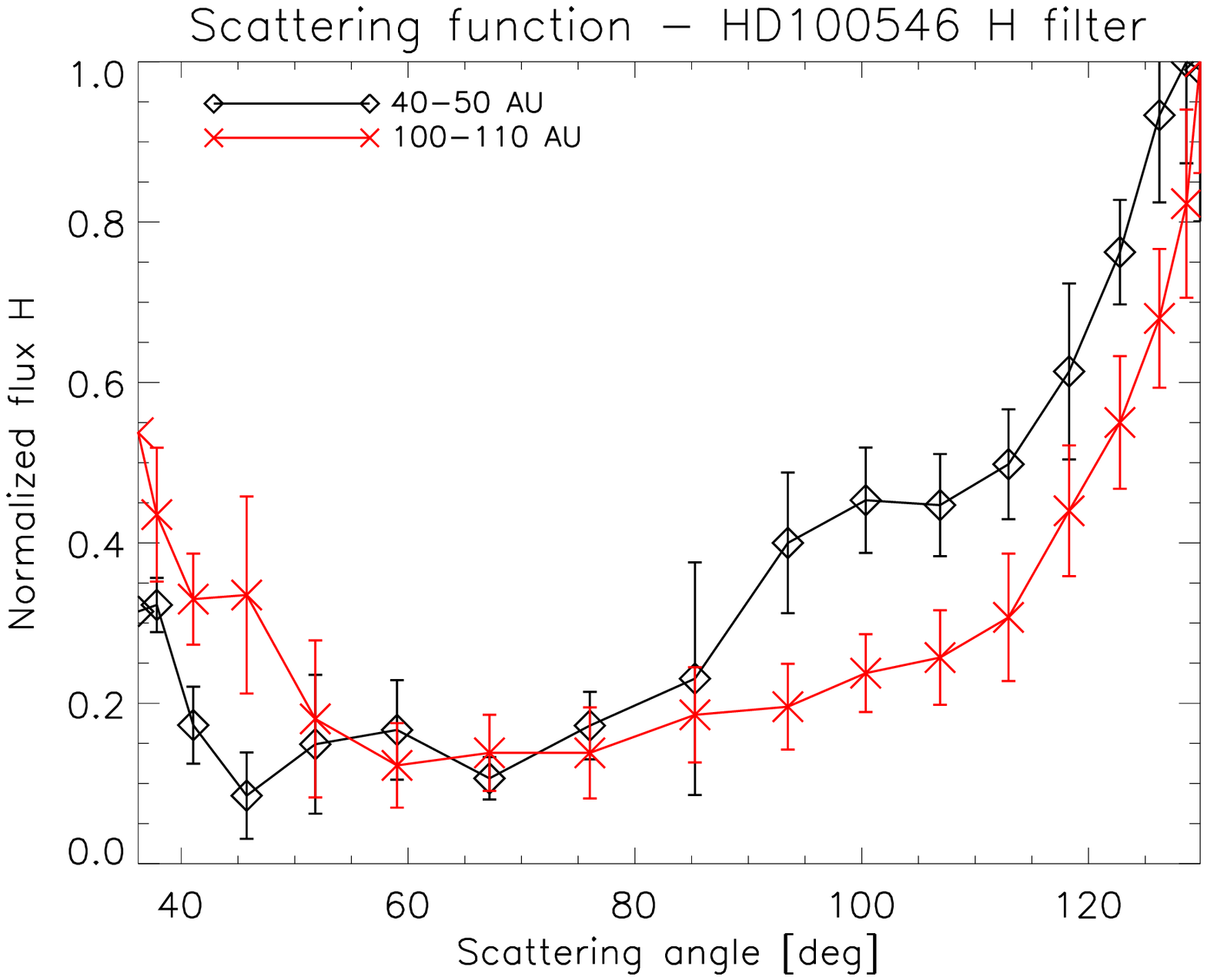}{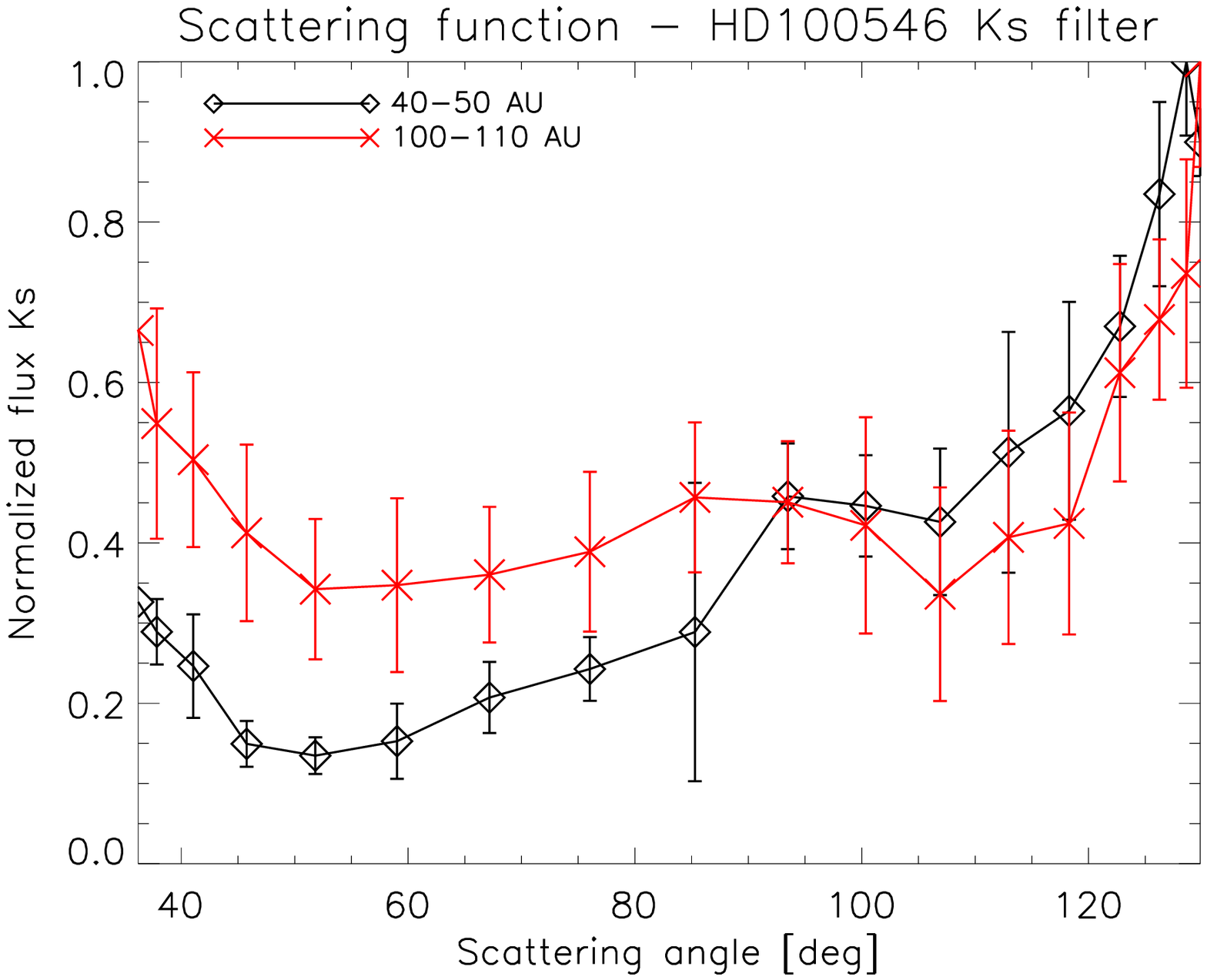}
\caption{Scattering function based on the images shown in the upper panels of Figure~\ref{P_and_pI_images} (left: $H$ filter; right: $K_s$ filter). The mean polarized flux was computed in two annuli between 40--50 AU (black curves) and 100--110 AU (red curves) in wedges of 10$^\circ$ as a function of the scattering angle (see text). The error bars denote the standard deviation within each wedge. The flux was normalized to the maximum flux measured in the direction of the north-east semi-minor axis.
\label{scat_func}}
\end{figure*}

Having disk images in polarized flux one can constrain dust grain properties much better than based on disk intensity image alone \citep[e.g.,][]{graham2007,perrin2009}. However, in order to interpret the polarized scattering functions shown in Figure~\ref{scat_func} one has to keep in mind that they are the product of two functions which depend both on dust grain properties such as grain sizes, composition and porosity: The "scattering function" describing the dust scattering efficiency as a function of scattering angle, and the "polarization function" describing the introduced scattering polarization as a function of scattering angle. 

As mentioned above and shown in Figure~\ref{scat_func}, we find that the far side of the disk appears brighter than the near side indicating that backward scattering if preferred over forward scattering. This, however, requires scattering particles that are typically at least as large as the wavelength of the incoming light. The strong flux detected in mm-observations \citep{henning1998,wilner2003} already indicated the existence of dust grains that are considerably larger than those found typically in the ISM (i.e., sub-micron sized). While most flux observed in the mm-domain is in general attributed to the mid-plane of a circumstellar disk, our scattered light observations probe grain properties of the disk surface. 

To our knowledge there are only two other prominent examples of disks where backward scattering dominates: The first one is Saturn's Rings which are made up of icy particles that are predominantly cm-sized or larger \citep[see,][and references therein]{dones1993}. The second one is the debris ring around Fomalhaut. \citet{min2010} showed that the scattering function observed by \citet{kalas2005} could be matched with grains several tens of micrometer in size. Comparing our scattering functions to those published by \citet{min2010} it seems as if grains of similar size could match our observations. With this interpretation one has to be cautious, though: First, the HD100546 is very likely optically thick which means that optical depths effects and multiple scattering will influence the scattering function compared to the optically thin, single scattering case. Secondly, we only observe the \emph{polarized} component of the scattered light. It is important to consider that the polarization efficiency typically peaks for scattering angles close to 90$^\circ$ and is always zero in exact forward or backward scattering direction. Thus, a direct quantitative comparison with the dust grains sizes \citep{min2010} and dust composition of other circumstellar disks \citep{graham2007,perrin2009} requires detailed modeling which is beyond the scope of the present paper. Qualitatively, however, it is clear that rather large, micron-sized grains are required to explain the strong backscattering peak we observe. 

In addition, the rather low polarization degree of the scattered light of $\sim$14\% (measured at a separation of 0.5$''$; see,~\ref{absolute_polarization}) also supports the existence of large grains on the disk surface as the polarization efficiency decreases with increasing particle size \citep{graham2007}.  The value we find for HD100546 is lower than that of AB Aur where \citet{perrin2009} found roughly 40\% of the scattered light to be polarized (on average) in an annulus of 0.7--1.5$''$ ($\sim$100--215 AU). For the TW Hya disk, \citet{hales2006} found the absolute polarization to increase from $\sim$10\% at 0.5$''$ to $\sim$30\% at 1.2$''$ in both $J$ and $H$ band. Further out, at roughly 2$''$, the value increased up to $\sim$40\% in $H$ but falls back to $\sim$10\% in $J$ what the authors attributed to a steeper surface brightness profile of polarized flux in $J$. 
For HD100546, we can expected that, since our surface brightness profile along the disk major axis and that of \citet{augereau2001} are similar, the ratio of polarized light to total scattered light does not change significantly and hence the polarization degree between 0.5--1.4$''$ should not change significantly either. \citet{perrin2009} used a disk model to fit their observations and the best fit model uses the typical grain size distribution with a power low index of -3.5 and a maximum grain size of 1$\mu$m. The same maximum grain size is currently used in the latest disk models for the surface layer of the HD100546 main disk \citep{benisty2010} but our data suggests that a significant fraction of large grains is required. 

Further, we can speculate that the differences we see in the scattering function between the inner disk and the outer disk (see, Figure~\ref{scat_func}) could hint toward slightly different dust grain populations. Larger grains tend to have their scattering minimum at smaller scattering angles \citep{min2010}. Perhaps the grains in the inner disk regions are slightly larger than those in the outer disk regions. This hypothesis is further supported by the existence of the second scattering peak we observe for the smallest scattering angles: This peak is indicative of an additional dust grain population of smaller grains that are preferably forward scattering. Figure~\ref{scat_func} shows that this peak is stronger for the outer disk regions (100--110 AU) compared to the inner disk regions (40--50 AU). In this context it is interesting to note that \citet{augereau2001} derived a blow-out size for the dust grains of $\sim$13 $\mu$m, which would mean that smaller dust grains should be blown out to larger radii and eventually out of the system. As these computations took only radiation and gravitational forces into account it is unclear how the total gas-to-dust ratio on the disk surface is and how potential pressure for the remaining gas (see, section~\ref{intro}) affects this number. However, having a higher fraction of small grains further out on the scattering surface could also explain the brightness asymmetry that \citet{ardila2007} found along the disk minor axis in their optical \emph{HST/ACS} images. In their data forward scattering, possibly coming from smaller particles, was enhanced. This brightness asymmetry was most prominent for disk radii $>$2.0$''$ ($>$200 AU). Inside of 1.6$''$ the authors were limited by the \emph{HST/ACS} coronagraph and subtraction residuals and no information could be derived. Thus, our data are not probing the same disk region and maybe different dust grain populations are involved: smaller dust grains (blown out?) at large separations, and an increasing fraction of larger particles closer to the star. We note that, similarly, another abrupt change in the color and brightness in their scattered light images for radial distances $>8''$ led \citet{ardila2007} to the conclusion that they were probing yet again a different regime of the circumstellar matter - the authors referred to it as an extended envelope - possibly featuring again a different dust grain population.

\subsection{Disk sub-structures as signposts of ongoing planet formation?}\label{planet_formation}
The small inner working angle of our observations revealed structures in the inner disk regions of HD100546 that have not been detected previously (see~\ref{structures}). Similar structures have recently been observed in the innermost regions of the AB Aurigae disk also using ground-based polarimetric imaging \citep{hashimoto2011}. It is difficult to identify the underlying physical reasons for these structures as different mechanisms can leave an observational imprint on the disk surface density ranging from disk intrinsic phenomena such as gravitational instabilities or magneto-rotational instabilities to interactions between the disk and an embedded body such as a young planet. 

Both features, i.e., 'the hole' and 'the clump', were detected in both filters in the $p_I$ and $P$ images so that we can be sure that they are features of the disk (see~\ref{structures}). However, while interpreting these features we have to keep in mind that our observations are probing the dust component of the disk \emph{surface}
which is presumably optically thick at NIR wavelengths. We can not say whether the features are caused by a deficiency/overabundance of dust grains at these locations (i.e., a physical 'hole'/'clump' in the disk) or whether different scattering properties of the dust grains are responsible for the observed features. Since both features are very localized   and no additional large scale disk structures are present in our data\footnote{As mentioned above, there is evidence for larger scale structures such as spiral arms at greater radial distances in \emph{HST} images.} we will discuss 'the hole' and 'the clump' in the context of interactions with embedded bodies and will not consider possible mechanisms driven by the disk alone or very localized changes in the grains' scattering properties.  

Numerical simulations have demonstrated that embedded planets can rapidly create gaps and sub-structures in the dust component of a circumstellar disk which depend on the mass of the planet but also on the size of the dust grains with more massive planets creating deeper and wider gaps and larger grains being more affected \citep[e.g.,][]{fouchet2010,paardekooper2006}. These features are most prominent in the disk midplane, i.e., where the planet is formed, and are most easily seen in (sub-mm) observations. This in turn means, that a lower mass body orbiting in a disk is causing only little signatures and is more challenging to see in scattered light images \citep[see, e.g.,][]{wolf2008}. However, the gravitational influence of the planet could remove dust and gas particles locally and create small scale depressions on the disk surface. This could create planet-induced disk 'shadows' in the NIR which have been modeled by \citet{jang-condell2009}. Even if the parameter space studied in the aforementioned paper does not cover the setup we observe here, it provides at least qualitatively an interesting explanation for 'the hole': Even bodies with masses of $10 \rm{M}_\earth \le \rm{M} \le 50 \rm{M}_\earth$ can leave an observational imprint on the disk surface on the order of the flux drop we observe here \citep{jang-condell2009}. 

Since the apparent distance between 'the hole' and 'the clump' is rather small ($\sim$9 pixels) it seems interesting to investigate whether they may be somehow related. Taking into account the rotation sense of the disk, 'the clump' is trailing behind 'the hole'. If we de-project the apparent separations from central source ($r_{\rm clump}\approx 13$ px; $r_{\rm hole}\approx 8$ px) considering the disk inclination angle and the position angle of the features we find that 'the clump' has a physical separation of $\sim$35 AU while for 'the hole' the separation amounts to $\sim$27 AU. In the above mentioned numerical simulations from \citet{fouchet2010} it is shown that on the outer edge of the planets orbital gaps the dust surface density is enhanced. Also, in the \citet{jang-condell2009} simulations, in addition to the planet-induced disk shadow roughly at the planet's location, the disk will appear brighter radially behind the planet compared to the surroundings. It would be interesting to see whether the static radiative transfer models by \citet{jang-condell2009}, when combined with dynamical SPH simulations, would predict a brighter spot trailing behind the 'shadow'. This would match with our observations and support a physical link between 'the hole' and 'the clump'. 

%An alternative explanation for 'the clump' could be the fact that dust particles (if decoupled from the gas in the disk) can become temporarily trapped into orbital resonance with a planet. For instance, \citet{reach2010}, using data from the \emph{Spitzer Space Telescope}, showed that the Earth is followed by a cloud of dust particles along its orbit. In this context, local overdensities in the disk result from gravitational resonances and are a direct indication for additional orbiting bodies. 

We emphasize that our observations and the structures we see in our data provide no immediate proof that planet formation is taking place in the disk around HD100546. The observed features are, however, consistent with the predictions of certain theoretical models dealing with the observational signatures of planet-disk interactions. A good observational test for our structures will be to re-observe HD100546 in the future. If 'the hole' and 'the clump' are caused by (an) unseen planetary companion(s) then a Keplerian orbital motion with a baseline of 10 years will lead to a rotation of roughly 30$^\circ$ (assuming circular orbits), enough to be detected even if the estimated disk inclination is taken into account.

\subsection{The big picture - HD100546 a remarkable transitional disk}
In combination with previous findings, the results presented here emphasize that the HD100546 circumstellar disk seems to be an ideal laboratory to study the formation of planets:
(1) The inner regions between 4--13 AU are mostly devoid of dust and gas \citep[e.g.,][]{bouwman2003,grady2005,acke2006,vanderplas2009,benisty2010} possibly cleared out by an unseen (planetary mass?) companion \citep[e.g.][]{acke2006}; (2) the MIR spectrum is remarkable similar to that of comet Hale-Bopp \citep{malfait1998,bouwman2003} indicating that some fraction of the HD100546 dust component is similar in composition to dust found in our Solar System; and (3) our observations revealed structures in the outer parts of the disk (at $\sim$30 AU separation) that could be the signpost of an/additional body/bodies forming at these locations. 

At the same time, HD100546 seems to be still accreting \citep{vieira1999} from its inner disk and the main disk contains a reservoir of molecular gas that is still massive enough ($\ge$10$^{-3}$ M$_\sun$) to form additional planetary bodies \citep{panic2010}. It seems somewhat surprising that HD100546 still has such a significant gas reservoir, as gaseous disks around early type stars typically disperse on timescales $\lesssim$3 Myr \citep{kennedy2009}. Even if recent estimates of the surface gravity of HD100546 indicate an age between 5--10 Myr \citep{guimaraes2006} instead of the more frequently used age of $>$10 Myr \citep{vandenancker1997}, the measured disk mass is exceptional. Finally, it should be kept in mind that at larger separations ($>$2$''$) the disk becomes highly structured again as seen in the scattered light images of \emph{HST} \citep{grady2001,augereau2001,ardila2007}. \citet{quillen2006} suggested that the spiral-arm like features and brightness variations detected in these images result from a warped and twisted disk geometry outside of $\sim$200 AU. An interesting mechanism to produce such a tilt in the disk would be precession induced by of a swarm of planetesimal-like bodies embedded in the disk at larger separations (50--200 AU) \citep{quillen2006}. The presence of such bodies would also support the existence of larger dust grains at these separations (e.g., created through collisions) as suggested by our data.

Overall HD100546 may be a great showcase example of a young star where we can directly witness the formation of a new planetary system with at least one planet already orbiting within the disk gap and additional planets and planetesimals still forming in the disk in which they are embedded.

\section{Summary and conclusions}\label{summaryconclusionsection}
In this paper we presented polarimetric differential imaging (PDI) observations of the circumstellar disk around the Herbig Be star HD100546 using the AO assisted high-resolution camera NACO at the VLT. We detected and clearly resolved the disk from $\sim$0.10--1.4$''$ (corresponding to the inner 10--140 AU) in polarized flux in the $H$ and $K_s$ filter. The innermost disk regions $<$50 AU are directly imaged for the first time emphasizing the power of PDI. 
Our main findings can be summarized as follows:
\begin{itemize}
\item{The disk inclination measured from isophot fitting in $H$ and $K_s$ between 0.2--0.5$''$ is 47.0$^\circ\pm$2.7$^\circ$ in agreement with previous estimates from direct imaging. For the disk position angle we find 138.0$^\circ\pm$3.9$^\circ$ again in general agreement with previous work.}
\item{Fits to the surface brightness $S(r)$ along the semi-major axes between 0.1--1.4$''$ yield power laws with $S(r) \propto r^{-2.8...-2.9}$ in both filters. Only in the $H$ band the south-east semi-major axis shows a steeper profile with $S(r) \propto r^{-3.5}$.} %The observed radial profiles are consistent with those expected from an optically thick, geometrically irradiated disk.} %A variety of disks resolved in scattered light with \emph{HST} show very similar profiles.} 
\item{The brightness profile of the disk in polarized flux is asymmetric along the disk minor axis. In both filters, the \emph{far side} of the disk appears \emph{brighter} than the near side indicating that the dust grains on the disk surface are \emph{preferably backward scattering}.}% To our knowledge the only other disk with preferably backward scattering dust grains is the debris ring around Fomalhaut.} 
\item{The above mentioned brightness asymmetry in combination with a low polarization fraction of the scattered light of only 14$^{+19}_{-8}$\% indicates that the dust grains probed with our observations have probably sizes on the order of micrometer, i.e., significantly larger than the size of pristine ISM dust grains. Grains of this size on the disk surface challenge current theoretical models of the HD100546 disk that typically assume a population of smaller grains.}
\item{The polarized scattering functions we measure in $H$ and $K_s$ cover scattering angles $\theta\approx40^\circ$--130$^\circ$ (taking into account the disk inclination and a theoretically motivated flaring angle) and are qualitatively very similar in both filters. While the maximum is found for the the largest scattering angles (the manifestation of preferred backward scattering) the minimum typically lies between 50$^\circ$--70$^\circ$. For even smaller scattering angles the functions do rise again which could be caused by an additional population of smaller dust grains which scattered preferably in the forward direction. Comparing the exact shape of the scattering function (i.e., position of the minimum, intensity of the forward scattering peak) for disk regions between 40--50 AU to that between 100--110 AU suggests that the relative fraction of small grains is higher in the outer disk regions. }
%A detailed modeling of these scattering functions taking into account the optical depth of the disk as well as the raw scattering function and the polarization function of different dust grain populations (with varying sizes and compositions) is beyond the scope of the current paper.  } 
\item{The small inner working angles of our imaging data reveal sub-structures in the innermost regions of the main disk that have never been observed before: a 'hole' and a 'clump' are detected in both filters. Since these two features appear in the final polarized flux images ($P$) as well as in the fractional polarization images ($p_I$) they are real and have their origin in the disk. Although these features are not a definite proof that planet formation is going on in the HD100546 disk, there is qualitative agreement between our observations and models predicting the observational features of the interaction between low-mass planets and a disk. If the observed features were caused by orbiting planetary bodies in the HD100546 disk, they should show a rotation of roughly 30$^\circ$ by 2016 (assuming Keplerian rotation and circular orbits).}
\item{Previous observations and models have suggested that a gap stretching from $\sim$4--13 AU divides the circumstellar disk into an inner disk and an main disk. Our data  support the existence of an 'inner rim' of the main disk around $\sim$15 AU. At this location the polarized flux measured along the disk major axis has its maximum in both the $H$ and $K_s$ filter and it drops as we get closer to the star.}
\end{itemize}

Our data demonstrate the immense power of PDI as a high-contrast imaging technique at large ground-based telescopes to spatially resolve dusty circumstellar disks at small inner working angles. PDI helps to put constraints on the properties of the dust grains on the surface of circumstellar disks and reveals structures within the disk that could arise from the interaction between the disk and larger bodies, such as planets, forming within the disk. When combined with scattered light images from \emph{HST} (giving access to the absolute intensity of the scattered light and probing disk regions farther out due to its superior surface brightness sensitivity) a comprehensive picture of the surface properties of circumstellar disk can be derived. 

In the near future, two new high-contrast instruments will become available both equipped with a dual-beam polarimeter: SPHERE at the VLT \citep{dohlen2006,beuzit2006} and GPI at GEMINI south \citep{macintosh2006}. Thanks to their extreme AO systems, these instruments will provide a much more stable PSF and will be able to probe polarized flux at smaller inner working angles with higher sensitivity than current instruments. The hope is that these instruments will mark a new area in resolving and analyzing the surfaces of circumstellar disks from the ground. 

Finally, ALMA will become available as of this summer. Once it has reached its final configuration it will provide valuable complementary information about the dust in the mid-plane of circumstellar disks and also about the gaseous component of disks with somewhat comparable spatial resolution. New insights - but also new questions - about the formation processes of planets lie in the not so distant future and PDI has the potential to play an important role.

%Something about the power of PDI and the future prospects of this technique in the advent of SPHERE and GPI. Use ALMA to resolve sub-mm continuum and search fro structures.

%%%%%%%%%%%%%%%%%%%%%%%%%%%%%%%%%%%%%%%%%%%%%%%%%%%%%%%%%%%%%%%%%%%%%
\acknowledgments
We are very grateful to the people who supported us during the observations, in particular Nancy Ageorges, Markus Hartung, and Nuria Huelamo. We also thank Daniel Apai, and Esther B\"unzli and Andreas Bazzon for helpful discussion during the proposal preparations and data reduction, respectively. Finally, we thank an anonymous referee for carefully reading the manuscript and providing helpful and constructive comments. This research has made use of the SIMBAD database, operated at CDS, Strasbourg, France. %% To help institutions obtain information on the effectiveness of their

%% copy editing.  Individual instruments can be provided in parentheses,
%% after the keyword, but they will not be verified.

{\it Facilities:}  \facility{VLT: Yepun (NACO)}


%\bibliographystyle{apj}
%\bibliography{mybib.bib}

\begin{thebibliography}{74}
\expandafter\ifx\csname natexlab\endcsname\relax\def\natexlab#1{#1}\fi

\bibitem[{{Acke} \& {van den Ancker}(2006)}]{acke2006}
{Acke}, B. \& {van den Ancker}, M.~E. 2006, \aap, 449, 267

\bibitem[{{Alibert} {et~al.}(2011){Alibert}, {Mordasini}, \&
  {Benz}}]{alibert2011}
{Alibert}, Y., {Mordasini}, C., \& {Benz}, W. 2011, \aap, 526, A63+

\bibitem[{{Apai} {et~al.}(2004){Apai}, {Pascucci}, {Brandner}, {Henning},
  {Lenzen}, {Potter}, {Lagrange}, \& {Rousset}}]{apai2004}
{Apai}, D., {Pascucci}, I., {Brandner}, W., {Henning}, T., {Lenzen}, R.,
  {Potter}, D.~E., {Lagrange}, A., \& {Rousset}, G. 2004, \aap, 415, 671

\bibitem[{{Ardila} {et~al.}(2007){Ardila}, {Golimowski}, {Krist}, {Clampin},
  {Ford}, \& {Illingworth}}]{ardila2007}
{Ardila}, D.~R., {Golimowski}, D.~A., {Krist}, J.~E., {Clampin}, M., {Ford},
  H.~C., \& {Illingworth}, G.~D. 2007, \apj, 665, 512

\bibitem[{{Augereau} {et~al.}(2001){Augereau}, {Lagrange}, {Mouillet}, \&
  {M{\'e}nard}}]{augereau2001}
{Augereau}, J.~C., {Lagrange}, A.~M., {Mouillet}, D., \& {M{\'e}nard}, F. 2001,
  \aap, 365, 78

\bibitem[{{Benisty} {et~al.}(2010){Benisty}, {Tatulli}, {M{\'e}nard}, \&
  {Swain}}]{benisty2010}
{Benisty}, M., {Tatulli}, E., {M{\'e}nard}, F., \& {Swain}, M.~R. 2010, \aap,
  511, A75+

\bibitem[{{Beuzit} {et~al.}(2006){Beuzit}, {Feldt}, {Dohlen}, {Mouillet},
  {Puget}, {Antichi}, {Baruffolo}, {Baudoz}, {Berton}, {Boccaletti},
  {Carbillet}, {Charton}, {Claudi}, {Downing}, {Feautrier}, {Fedrigo}, {Fusco},
  {Gratton}, {Hubin}, {Kasper}, {Langlois}, {Moutou}, {Mugnier}, {Pragt},
  {Rabou}, {Saisse}, {Schmid}, {Stadler}, {Turrato}, {Udry}, {Waters}, \&
  {Wildi}}]{beuzit2006}
{Beuzit}, J., {Feldt}, M., {Dohlen}, K., {Mouillet}, D., {Puget}, P.,
  {Antichi}, J., {Baruffolo}, A., {Baudoz}, P., {Berton}, A., {Boccaletti}, A.,
  {Carbillet}, M., {Charton}, J., {Claudi}, R., {Downing}, M., {Feautrier}, P.,
  {Fedrigo}, E., {Fusco}, T., {Gratton}, R., {Hubin}, N., {Kasper}, M.,
  {Langlois}, M., {Moutou}, C., {Mugnier}, L., {Pragt}, J., {Rabou}, P.,
  {Saisse}, M., {Schmid}, H.~M., {Stadler}, E., {Turrato}, M., {Udry}, S.,
  {Waters}, R., \& {Wildi}, F. 2006, The Messenger, 125, 29

\bibitem[{{Bouwman} {et~al.}(2003){Bouwman}, {de Koter}, {Dominik}, \&
  {Waters}}]{bouwman2003}
{Bouwman}, J., {de Koter}, A., {Dominik}, C., \& {Waters}, L.~B.~F.~M. 2003,
  \aap, 401, 577

\bibitem[{{Brittain} {et~al.}(2009){Brittain}, {Najita}, \&
  {Carr}}]{brittain2009}
{Brittain}, S.~D., {Najita}, J.~R., \& {Carr}, J.~S. 2009, \apj, 702, 85

\bibitem[{{Clarke} {et~al.}(1999){Clarke}, {Smith}, \& {Yudin}}]{clarke1999}
{Clarke}, D., {Smith}, R.~A., \& {Yudin}, R.~V. 1999, \aap, 347, 590

\bibitem[{{Cutri} {et~al.}(2003){Cutri}, {Skrutskie}, {van Dyk}, {Beichman},
  {Carpenter}, {Chester}, {Cambresy}, {Evans}, {Fowler}, {Gizis}, {Howard},
  {Huchra}, {Jarrett}, {Kopan}, {Kirkpatrick}, {Light}, {Marsh}, {McCallon},
  {Schneider}, {Stiening}, {Sykes}, {Weinberg}, {Wheaton}, {Wheelock}, \&
  {Zacarias}}]{cutri2003}
{Cutri}, R.~M., {Skrutskie}, M.~F., {van Dyk}, S., {Beichman}, C.~A.,
  {Carpenter}, J.~M., {Chester}, T., {Cambresy}, L., {Evans}, T., {Fowler}, J.,
  {Gizis}, J., {Howard}, E., {Huchra}, J., {Jarrett}, T., {Kopan}, E.~L.,
  {Kirkpatrick}, J.~D., {Light}, R.~M., {Marsh}, K.~A., {McCallon}, H.,
  {Schneider}, S., {Stiening}, R., {Sykes}, M., {Weinberg}, M., {Wheaton},
  W.~A., {Wheelock}, S., \& {Zacarias}, N. 2003, {2MASS All Sky Catalog of
  point sources.} (The IRSA 2MASS All-Sky Point Source Catalog, NASA/IPAC
  Infrared Science Archive.~http://irsa.ipac.caltech.edu/applications/Gator/)

\bibitem[{{de Zeeuw} {et~al.}(1999){de Zeeuw}, {Hoogerwerf}, {de Bruijne},
  {Brown}, \& {Blaauw}}]{dezeeuw1999}
{de Zeeuw}, P.~T., {Hoogerwerf}, R., {de Bruijne}, J.~H.~J., {Brown}, A.~G.~A.,
  \& {Blaauw}, A. 1999, \aj, 117, 354

\bibitem[{{Dohlen} {et~al.}(2006){Dohlen}, {Beuzit}, {Feldt}, {Mouillet},
  {Puget}, {Antichi}, {Baruffolo}, {Baudoz}, {Berton}, {Boccaletti},
  {Carbillet}, {Charton}, {Claudi}, {Downing}, {Fabron}, {Feautrier},
  {Fedrigo}, {Fusco}, {Gach}, {Gratton}, {Hubin}, {Kasper}, {Langlois},
  {Longmore}, {Moutou}, {Petit}, {Pragt}, {Rabou}, {Rousset}, {Saisse},
  {Schmid}, {Stadler}, {Stamm}, {Turatto}, {Waters}, \& {Wildi}}]{dohlen2006}
{Dohlen}, K., {Beuzit}, J., {Feldt}, M., {Mouillet}, D., {Puget}, P.,
  {Antichi}, J., {Baruffolo}, A., {Baudoz}, P., {Berton}, A., {Boccaletti}, A.,
  {Carbillet}, M., {Charton}, J., {Claudi}, R., {Downing}, M., {Fabron}, C.,
  {Feautrier}, P., {Fedrigo}, E., {Fusco}, T., {Gach}, J., {Gratton}, R.,
  {Hubin}, N., {Kasper}, M., {Langlois}, M., {Longmore}, A., {Moutou}, C.,
  {Petit}, C., {Pragt}, J., {Rabou}, P., {Rousset}, G., {Saisse}, M., {Schmid},
  H., {Stadler}, E., {Stamm}, D., {Turatto}, M., {Waters}, R., \& {Wildi}, F.
  2006, in Society of Photo-Optical Instrumentation Engineers (SPIE) Conference
  Series, Vol. 6269, Society of Photo-Optical Instrumentation Engineers (SPIE)
  Conference Series

\bibitem[{{Dones} {et~al.}(1993){Dones}, {Cuzzi}, \& {Showalter}}]{dones1993}
{Dones}, L., {Cuzzi}, J.~N., \& {Showalter}, M.~R. 1993, Icarus, 105, 184

\bibitem[{{Fischer} \& {Henning}(1995)}]{fischer1995}
{Fischer}, O. \& {Henning}, T. 1995, \apss, 223, 154

\bibitem[{{Fischer} {et~al.}(1996){Fischer}, {Henning}, \&
  {Yorke}}]{fischer1996}
{Fischer}, O., {Henning}, T., \& {Yorke}, H.~W. 1996, \aap, 308, 863

\bibitem[{{Fouchet} {et~al.}(2010){Fouchet}, {Gonzalez}, \&
  {Maddison}}]{fouchet2010}
{Fouchet}, L., {Gonzalez}, J., \& {Maddison}, S.~T. 2010, \aap, 518, A16+

\bibitem[{{Fujiwara} {et~al.}(2009){Fujiwara}, {Yamashita}, {Ishihara},
  {Onaka}, {Kataza}, {Ootsubo}, {Fukagawa}, {Marshall}, {Murakami}, {Nakagawa},
  {Hirao}, {Enya}, \& {White}}]{fujiwara2009}
{Fujiwara}, H., {Yamashita}, T., {Ishihara}, D., {Onaka}, T., {Kataza}, H.,
  {Ootsubo}, T., {Fukagawa}, M., {Marshall}, J.~P., {Murakami}, H., {Nakagawa},
  T., {Hirao}, T., {Enya}, K., \& {White}, G.~J. 2009, \apjl, 695, L88

\bibitem[{{Grady} {et~al.}(2001){Grady}, {Polomski}, {Henning}, {Stecklum},
  {Woodgate}, {Telesco}, {Pi{\~n}a}, {Gull}, {Boggess}, {Bowers}, {Bruhweiler},
  {Clampin}, {Danks}, {Green}, {Heap}, {Hutchings}, {Jenkins}, {Joseph},
  {Kaiser}, {Kimble}, {Kraemer}, {Lindler}, {Linsky}, {Maran}, {Moos}, {Plait},
  {Roesler}, {Timothy}, \& {Weistrop}}]{grady2001}
{Grady}, C.~A., {Polomski}, E.~F., {Henning}, T., {Stecklum}, B., {Woodgate},
  B.~E., {Telesco}, C.~M., {Pi{\~n}a}, R.~K., {Gull}, T.~R., {Boggess}, A.,
  {Bowers}, C.~W., {Bruhweiler}, F.~C., {Clampin}, M., {Danks}, A.~C., {Green},
  R.~F., {Heap}, S.~R., {Hutchings}, J.~B., {Jenkins}, E.~B., {Joseph}, C.,
  {Kaiser}, M.~E., {Kimble}, R.~A., {Kraemer}, S., {Lindler}, D., {Linsky},
  J.~L., {Maran}, S.~P., {Moos}, H.~W., {Plait}, P., {Roesler}, F., {Timothy},
  J.~G., \& {Weistrop}, D. 2001, \aj, 122, 3396

\bibitem[{{Grady} {et~al.}(2007){Grady}, {Schneider}, {Hamaguchi}, {Sitko},
  {Carpenter}, {Hines}, {Collins}, {Williger}, {Woodgate}, {Henning},
  {M{\'e}nard}, {Wilner}, {Petre}, {Palunas}, {Quirrenbach}, {Nuth},
  {Silverstone}, \& {Kim}}]{grady2007}
{Grady}, C.~A., {Schneider}, G., {Hamaguchi}, K., {Sitko}, M.~L., {Carpenter},
  W.~J., {Hines}, D., {Collins}, K.~A., {Williger}, G.~M., {Woodgate}, B.~E.,
  {Henning}, T., {M{\'e}nard}, F., {Wilner}, D., {Petre}, R., {Palunas}, P.,
  {Quirrenbach}, A., {Nuth}, III, J.~A., {Silverstone}, M.~D., \& {Kim}, J.~S.
  2007, \apj, 665, 1391

\bibitem[{{Grady} {et~al.}(2005){Grady}, {Woodgate}, {Heap}, {Bowers}, {Nuth},
  {Herczeg}, \& {Hill}}]{grady2005}
{Grady}, C.~A., {Woodgate}, B., {Heap}, S.~R., {Bowers}, C., {Nuth}, III,
  J.~A., {Herczeg}, G.~J., \& {Hill}, H.~G.~M. 2005, \apj, 620, 470

\bibitem[{{Graham} {et~al.}(2007){Graham}, {Kalas}, \& {Matthews}}]{graham2007}
{Graham}, J.~R., {Kalas}, P.~G., \& {Matthews}, B.~C. 2007, \apj, 654, 595

\bibitem[{{Guimar{\~a}es} {et~al.}(2006){Guimar{\~a}es}, {Alencar}, {Corradi},
  \& {Vieira}}]{guimaraes2006}
{Guimar{\~a}es}, M.~M., {Alencar}, S.~H.~P., {Corradi}, W.~J.~B., \& {Vieira},
  S.~L.~A. 2006, \aap, 457, 581

\bibitem[{{Hales} {et~al.}(2006){Hales}, {Gledhill}, {Barlow}, \&
  {Lowe}}]{hales2006}
{Hales}, A.~S., {Gledhill}, T.~M., {Barlow}, M.~J., \& {Lowe}, K.~T.~E. 2006,
  \mnras, 365, 1348

\bibitem[{{Hashimoto} {et~al.}(2011){Hashimoto}, {Tamura}, {Muto}, {Kudo},
  {Fukagawa}, {Fukue}, {Goto}, {Grady}, {Henning}, {Hodapp}, {Honda},
  {Inutsuka}, {Kokubo}, {Knapp}, {McElwain}, {Momose}, {Ohashi}, {Okamoto},
  {Takami}, {Turner}, {Wisniewski}, {Janson}, {Abe}, {Brandner}, {Carson},
  {Egner}, {Feldt}, {Golota}, {Guyon}, {Hayano}, {Hayashi}, {Hayashi}, {Ishii},
  {Kandori}, {Kusakabe}, {Matsuo}, {Mayama}, {Miyama}, {Morino}, {Moro-Martin},
  {Nishimura}, {Pyo}, {Suto}, {Suzuki}, {Takato}, {Terada}, {Thalmann},
  {Tomono}, {Watanabe}, {Yamada}, {Takami}, \& {Usuda}}]{hashimoto2011}
{Hashimoto}, J., {Tamura}, M., {Muto}, T., {Kudo}, T., {Fukagawa}, M., {Fukue},
  T., {Goto}, M., {Grady}, C.~A., {Henning}, T., {Hodapp}, K., {Honda}, M.,
  {Inutsuka}, S., {Kokubo}, E., {Knapp}, G., {McElwain}, M.~W., {Momose}, M.,
  {Ohashi}, N., {Okamoto}, Y.~K., {Takami}, M., {Turner}, E.~L., {Wisniewski},
  J., {Janson}, M., {Abe}, L., {Brandner}, W., {Carson}, J., {Egner}, S.,
  {Feldt}, M., {Golota}, T., {Guyon}, O., {Hayano}, Y., {Hayashi}, M.,
  {Hayashi}, S., {Ishii}, M., {Kandori}, R., {Kusakabe}, N., {Matsuo}, T.,
  {Mayama}, S., {Miyama}, S., {Morino}, J., {Moro-Martin}, A., {Nishimura}, T.,
  {Pyo}, T., {Suto}, H., {Suzuki}, R., {Takato}, N., {Terada}, H., {Thalmann},
  C., {Tomono}, D., {Watanabe}, M., {Yamada}, T., {Takami}, H., \& {Usuda}, T.
  2011, \apjl, 729, L17+

\bibitem[{{Henning} {et~al.}(1998){Henning}, {Burkert}, {Launhardt}, {Leinert},
  \& {Stecklum}}]{henning1998}
{Henning}, T., {Burkert}, A., {Launhardt}, R., {Leinert}, C., \& {Stecklum}, B.
  1998, \aap, 336, 565

\bibitem[{{Hinkley} {et~al.}(2009){Hinkley}, {Oppenheimer}, {Soummer},
  {Brenner}, {Graham}, {Perrin}, {Sivaramakrishnan}, {Lloyd}, {Roberts}, \&
  {Kuhn}}]{hinkley2009}
{Hinkley}, S., {Oppenheimer}, B.~R., {Soummer}, R., {Brenner}, D., {Graham},
  J.~R., {Perrin}, M.~D., {Sivaramakrishnan}, A., {Lloyd}, J.~P., {Roberts},
  L.~C., \& {Kuhn}, J. 2009, \apj, 701, 804

\bibitem[{{Hinz} {et~al.}(2010){Hinz}, {Rodigas}, {Kenworthy}, {Sivanandam},
  {Heinze}, {Mamajek}, \& {Meyer}}]{hinz2010}
{Hinz}, P.~M., {Rodigas}, T.~J., {Kenworthy}, M.~A., {Sivanandam}, S.,
  {Heinze}, A.~N., {Mamajek}, E.~E., \& {Meyer}, M.~R. 2010, \apj, 716, 417

\bibitem[{{Houk} \& {Cowley}(1975)}]{houk1975}
{Houk}, N. \& {Cowley}, A.~P. 1975, {University of Michigan Catalogue of
  two-dimensional spectral types for the HD stars. Volume I. Declinations -90\_
  to -53\_$^\circ$}, ed. {Houk, N.~\& Cowley, A.~P.}

\bibitem[{{Jang-Condell}(2009)}]{jang-condell2009}
{Jang-Condell}, H. 2009, \apj, 700, 820

\bibitem[{{Jiang} {et~al.}(2005){Jiang}, {Tamura}, {Fukagawa}, {Hough},
  {Lucas}, {Suto}, {Ishii}, \& {Yang}}]{jiang2005}
{Jiang}, Z., {Tamura}, M., {Fukagawa}, M., {Hough}, J., {Lucas}, P., {Suto},
  H., {Ishii}, M., \& {Yang}, J. 2005, \nat, 437, 112

\bibitem[{{Jiang} {et~al.}(2008){Jiang}, {Tamura}, {Hoare}, {Yao}, {Ishii},
  {Fang}, \& {Yang}}]{jiang2008}
{Jiang}, Z., {Tamura}, M., {Hoare}, M.~G., {Yao}, Y., {Ishii}, M., {Fang}, M.,
  \& {Yang}, J. 2008, \apjl, 673, L175

\bibitem[{{Kalas} {et~al.}(2008){Kalas}, {Graham}, {Chiang}, {Fitzgerald},
  {Clampin}, {Kite}, {Stapelfeldt}, {Marois}, \& {Krist}}]{kalas2008}
{Kalas}, P., {Graham}, J.~R., {Chiang}, E., {Fitzgerald}, M.~P., {Clampin}, M.,
  {Kite}, E.~S., {Stapelfeldt}, K., {Marois}, C., \& {Krist}, J. 2008, Science,
  322, 1345

\bibitem[{{Kalas} {et~al.}(2005){Kalas}, {Graham}, \& {Clampin}}]{kalas2005}
{Kalas}, P., {Graham}, J.~R., \& {Clampin}, M. 2005, \nat, 435, 1067

\bibitem[{{Kennedy} \& {Kenyon}(2009)}]{kennedy2009}
{Kennedy}, G.~M. \& {Kenyon}, S.~J. 2009, \apj, 695, 1210

\bibitem[{{Krist} {et~al.}(2010){Krist}, {Stapelfeldt}, {Bryden}, {Rieke},
  {Su}, {Chen}, {Beichman}, {Hines}, {Rebull}, {Tanner}, {Trilling}, {Clampin},
  \& {G{\'a}sp{\'a}r}}]{krist2010}
{Krist}, J.~E., {Stapelfeldt}, K.~R., {Bryden}, G., {Rieke}, G.~H., {Su},
  K.~Y.~L., {Chen}, C.~C., {Beichman}, C.~A., {Hines}, D.~C., {Rebull}, L.~M.,
  {Tanner}, A., {Trilling}, D.~E., {Clampin}, M., \& {G{\'a}sp{\'a}r}, A. 2010,
  \aj, 140, 1051

\bibitem[{{Kuhn} {et~al.}(2001){Kuhn}, {Potter}, \& {Parise}}]{kuhn2001}
{Kuhn}, J.~R., {Potter}, D., \& {Parise}, B. 2001, \apjl, 553, L189

\bibitem[{{Lagrange} {et~al.}(2009){Lagrange}, {Gratadour}, {Chauvin}, {Fusco},
  {Ehrenreich}, {Mouillet}, {Rousset}, {Rouan}, {Allard}, {Gendron}, {Charton},
  {Mugnier}, {Rabou}, {Montri}, \& {Lacombe}}]{lagrange2009a}
{Lagrange}, A., {Gratadour}, D., {Chauvin}, G., {Fusco}, T., {Ehrenreich}, D.,
  {Mouillet}, D., {Rousset}, G., {Rouan}, D., {Allard}, F., {Gendron}, {\'E}.,
  {Charton}, J., {Mugnier}, L., {Rabou}, P., {Montri}, J., \& {Lacombe}, F.
  2009, \aap, 493, L21

\bibitem[{Lagrange {et~al.}(2010)Lagrange, Bonnefoy, Chauvin, Apai, Ehrenreich,
  Boccaletti, Gratadour, Rouan, Mouillet, Lacour, \& Kasper}]{lagrange2010}
Lagrange, A.-M., Bonnefoy, M., Chauvin, G., Apai, D., Ehrenreich, D.,
  Boccaletti, A., Gratadour, D., Rouan, D., Mouillet, D., Lacour, S., \&
  Kasper, M. 2010, Science, science.1187187

\bibitem[{{Leinert} {et~al.}(2004){Leinert}, {van Boekel}, {Waters},
  {Chesneau}, {Malbet}, {K{\"o}hler}, {Jaffe}, {Ratzka}, {Dutrey}, {Preibisch},
  {Graser}, {Bakker}, {Chagnon}, {Cotton}, {Dominik}, {Dullemond},
  {Glazenborg-Kluttig}, {Glindemann}, {Henning}, {Hofmann}, {de Jong},
  {Lenzen}, {Ligori}, {Lopez}, {Meisner}, {Morel}, {Paresce}, {Pel},
  {Percheron}, {Perrin}, {Przygodda}, {Richichi}, {Sch{\"o}ller}, {Schuller},
  {Stecklum}, {van den Ancker}, {von der L{\"u}he}, \& {Weigelt}}]{leinert2004}
{Leinert}, C., {van Boekel}, R., {Waters}, L.~B.~F.~M., {Chesneau}, O.,
  {Malbet}, F., {K{\"o}hler}, R., {Jaffe}, W., {Ratzka}, T., {Dutrey}, A.,
  {Preibisch}, T., {Graser}, U., {Bakker}, E., {Chagnon}, G., {Cotton}, W.~D.,
  {Dominik}, C., {Dullemond}, C.~P., {Glazenborg-Kluttig}, A.~W., {Glindemann},
  A., {Henning}, T., {Hofmann}, K.-H., {de Jong}, J., {Lenzen}, R., {Ligori},
  S., {Lopez}, B., {Meisner}, J., {Morel}, S., {Paresce}, F., {Pel}, J.-W.,
  {Percheron}, I., {Perrin}, G., {Przygodda}, F., {Richichi}, A.,
  {Sch{\"o}ller}, M., {Schuller}, P., {Stecklum}, B., {van den Ancker}, M.~E.,
  {von der L{\"u}he}, O., \& {Weigelt}, G. 2004, \aap, 423, 537

\bibitem[{{Lenzen} {et~al.}(2003){Lenzen}, {Hartung}, {Brandner}, {Finger},
  {Hubin}, {Lacombe}, {Lagrange}, {Lehnert}, {Moorwood}, \&
  {Mouillet}}]{lenzen2003}
{Lenzen}, R., {Hartung}, M., {Brandner}, W., {Finger}, G., {Hubin}, N.~N.,
  {Lacombe}, F., {Lagrange}, A., {Lehnert}, M.~D., {Moorwood}, A.~F.~M., \&
  {Mouillet}, D. 2003, in Society of Photo-Optical Instrumentation Engineers
  (SPIE) Conference Series, Vol. 4841, Society of Photo-Optical Instrumentation
  Engineers (SPIE) Conference Series, ed. {M.~Iye \& A.~F.~M.~Moorwood},
  944--952

\bibitem[{{Liu} {et~al.}(2003){Liu}, {Hinz}, {Meyer}, {Mamajek}, {Hoffmann}, \&
  {Hora}}]{liu2003_b}
{Liu}, W.~M., {Hinz}, P.~M., {Meyer}, M.~R., {Mamajek}, E.~E., {Hoffmann},
  W.~F., \& {Hora}, J.~L. 2003, \apjl, 598, L111

\bibitem[{{Macintosh} {et~al.}(2006){Macintosh}, {Graham}, {Palmer}, {Doyon},
  {Gavel}, {Larkin}, {Oppenheimer}, {Saddlemyer}, {Wallace}, {Bauman}, {Evans},
  {Erikson}, {Morzinski}, {Phillion}, {Poyneer}, {Sivaramakrishnan}, {Soummer},
  {Thibault}, \& {Veran}}]{macintosh2006}
{Macintosh}, B., {Graham}, J., {Palmer}, D., {Doyon}, R., {Gavel}, D.,
  {Larkin}, J., {Oppenheimer}, B., {Saddlemyer}, L., {Wallace}, J.~K.,
  {Bauman}, B., {Evans}, J., {Erikson}, D., {Morzinski}, K., {Phillion}, D.,
  {Poyneer}, L., {Sivaramakrishnan}, A., {Soummer}, R., {Thibault}, S., \&
  {Veran}, J. 2006, in Society of Photo-Optical Instrumentation Engineers
  (SPIE) Conference Series, Vol. 6272, Society of Photo-Optical Instrumentation
  Engineers (SPIE) Conference Series

\bibitem[{{Malfait} {et~al.}(1998){Malfait}, {Waelkens}, {Waters},
  {Vandenbussche}, {Huygen}, \& {de Graauw}}]{malfait1998}
{Malfait}, K., {Waelkens}, C., {Waters}, L.~B.~F.~M., {Vandenbussche}, B.,
  {Huygen}, E., \& {de Graauw}, M.~S. 1998, \aap, 332, L25

\bibitem[{{Marois} {et~al.}(2008){Marois}, {Macintosh}, {Barman}, {Zuckerman},
  {Song}, {Patience}, {Lafreni{\`e}re}, \& {Doyon}}]{marois2008}
{Marois}, C., {Macintosh}, B., {Barman}, T., {Zuckerman}, B., {Song}, I.,
  {Patience}, J., {Lafreni{\`e}re}, D., \& {Doyon}, R. 2008, Science, 322, 1348

\bibitem[{{Marois} {et~al.}(2010){Marois}, {Zuckerman}, {Konopacky},
  {Macintosh}, \& {Barman}}]{marois2010}
{Marois}, C., {Zuckerman}, B., {Konopacky}, Q.~M., {Macintosh}, B., \&
  {Barman}, T. 2010, ArXiv e-prints

\bibitem[{{Min} {et~al.}(2010){Min}, {Kama}, {Dominik}, \& {Waters}}]{min2010}
{Min}, M., {Kama}, M., {Dominik}, C., \& {Waters}, L.~B.~F.~M. 2010, \aap, 509,
  L6+

\bibitem[{{Oppenheimer} {et~al.}(2008){Oppenheimer}, {Brenner}, {Hinkley},
  {Zimmerman}, {Sivaramakrishnan}, {Soummer}, {Kuhn}, {Graham}, {Perrin},
  {Lloyd}, {Roberts}, \& {Harrington}}]{oppenheimer2008}
{Oppenheimer}, B.~R., {Brenner}, D., {Hinkley}, S., {Zimmerman}, N.,
  {Sivaramakrishnan}, A., {Soummer}, R., {Kuhn}, J., {Graham}, J.~R., {Perrin},
  M., {Lloyd}, J.~P., {Roberts}, Jr., L.~C., \& {Harrington}, D.~M. 2008, \apj,
  679, 1574

\bibitem[{{Paardekooper} \& {Mellema}(2006)}]{paardekooper2006}
{Paardekooper}, S. \& {Mellema}, G. 2006, \aap, 453, 1129

\bibitem[{{Pani{\'c}} {et~al.}(2010){Pani{\'c}}, {van Dishoeck}, {Hogerheijde},
  {Belloche}, {G{\"u}sten}, {Boland}, \& {Baryshev}}]{panic2010}
{Pani{\'c}}, O., {van Dishoeck}, E.~F., {Hogerheijde}, M.~R., {Belloche}, A.,
  {G{\"u}sten}, R., {Boland}, W., \& {Baryshev}, A. 2010, \aap, 519, A110+

\bibitem[{{Pantin} {et~al.}(2000){Pantin}, {Waelkens}, \&
  {Lagage}}]{pantin2000}
{Pantin}, E., {Waelkens}, C., \& {Lagage}, P.~O. 2000, \aap, 361, L9

\bibitem[{{Perrin} {et~al.}(2004){Perrin}, {Graham}, {Kalas}, {Lloyd}, {Max},
  {Gavel}, {Pennington}, \& {Gates}}]{perrin2004}
{Perrin}, M.~D., {Graham}, J.~R., {Kalas}, P., {Lloyd}, J.~P., {Max}, C.~E.,
  {Gavel}, D.~T., {Pennington}, D.~M., \& {Gates}, E.~L. 2004, Science, 303,
  1345

\bibitem[{{Perrin} {et~al.}(2009){Perrin}, {Schneider}, {Duchene}, {Pinte},
  {Grady}, {Wisniewski}, \& {Hines}}]{perrin2009}
{Perrin}, M.~D., {Schneider}, G., {Duchene}, G., {Pinte}, C., {Grady}, C.~A.,
  {Wisniewski}, J.~P., \& {Hines}, D.~C. 2009, \apjl, 707, L132

\bibitem[{{Perryman} {et~al.}(1997){Perryman}, {Lindegren}, {Kovalevsky},
  {Hoeg}, {Bastian}, {Bernacca}, {Cr{\'e}z{\'e}}, {Donati}, {Grenon}, {van
  Leeuwen}, {van der Marel}, {Mignard}, {Murray}, {Le Poole}, {Schrijver},
  {Turon}, {Arenou}, {Froeschl{\'e}}, \& {Petersen}}]{perryman1997}
{Perryman}, M.~A.~C., {Lindegren}, L., {Kovalevsky}, J., {Hoeg}, E., {Bastian},
  U., {Bernacca}, P.~L., {Cr{\'e}z{\'e}}, M., {Donati}, F., {Grenon}, M., {van
  Leeuwen}, F., {van der Marel}, H., {Mignard}, F., {Murray}, C.~A., {Le
  Poole}, R.~S., {Schrijver}, H., {Turon}, C., {Arenou}, F., {Froeschl{\'e}},
  M., \& {Petersen}, C.~S. 1997, \aap, 323, L49

\bibitem[{{Potter}(2003)}]{potter2003}
{Potter}, D.~E. 2003, PhD thesis, UNIVERSITY OF HAWAI'I

\bibitem[{{Quanz} {et~al.}(2010){Quanz}, {Meyer}, {Kenworthy}, {Girard},
  {Kasper}, {Lagrange}, {Apai}, {Boccaletti}, {Bonnefoy}, {Chauvin}, {Hinz}, \&
  {Lenzen}}]{quanz2010}
{Quanz}, S.~P., {Meyer}, M.~R., {Kenworthy}, M.~A., {Girard}, J.~H.~V.,
  {Kasper}, M., {Lagrange}, A., {Apai}, D., {Boccaletti}, A., {Bonnefoy}, M.,
  {Chauvin}, G., {Hinz}, P.~M., \& {Lenzen}, R. 2010, \apjl, 722, L49

\bibitem[{{Quillen}(2006)}]{quillen2006}
{Quillen}, A.~C. 2006, \apj, 640, 1078

\bibitem[{{Rodrigues} {et~al.}(2009){Rodrigues}, {Sartori}, {Gregorio-Hetem},
  \& {Magalh{\~a}es}}]{rodrigues2009}
{Rodrigues}, C.~V., {Sartori}, M.~J., {Gregorio-Hetem}, J., \& {Magalh{\~a}es},
  A.~M. 2009, \apj, 698, 2031

\bibitem[{{Rousset} {et~al.}(2003){Rousset}, {Lacombe}, {Puget}, {Hubin},
  {Gendron}, {Fusco}, {Arsenault}, {Charton}, {Feautrier}, {Gigan}, {Kern},
  {Lagrange}, {Madec}, {Mouillet}, {Rabaud}, {Rabou}, {Stadler}, \&
  {Zins}}]{rousset2003}
{Rousset}, G., {Lacombe}, F., {Puget}, P., {Hubin}, N.~N., {Gendron}, E.,
  {Fusco}, T., {Arsenault}, R., {Charton}, J., {Feautrier}, P., {Gigan}, P.,
  {Kern}, P.~Y., {Lagrange}, A., {Madec}, P., {Mouillet}, D., {Rabaud}, D.,
  {Rabou}, P., {Stadler}, E., \& {Zins}, G. 2003, in Society of Photo-Optical
  Instrumentation Engineers (SPIE) Conference Series, Vol. 4839, Society of
  Photo-Optical Instrumentation Engineers (SPIE) Conference Series, ed.
  {P.~L.~Wizinowich \& D.~Bonaccini}, 140--149

\bibitem[{{Schmid} {et~al.}(2006){Schmid}, {Joos}, \& {Tschan}}]{schmid2006}
{Schmid}, H.~M., {Joos}, F., \& {Tschan}, D. 2006, \aap, 452, 657

\bibitem[{{Schneider} \& {Stobie}(2002)}]{schneider2002_idp3}
{Schneider}, G. \& {Stobie}, E. 2002, in Astronomical Society of the Pacific
  Conference Series, Vol. 281, Astronomical Data Analysis Software and Systems
  XI, ed. {D.~A.~Bohlender, D.~Durand, \& T.~H.~Handley}, 382--+

\bibitem[{{Schneider} {et~al.}(2003){Schneider}, {Wood}, {Silverstone},
  {Hines}, {Koerner}, {Whitney}, {Bjorkman}, \& {Lowrance}}]{schneider2003}
{Schneider}, G., {Wood}, K., {Silverstone}, M.~D., {Hines}, D.~C., {Koerner},
  D.~W., {Whitney}, B.~A., {Bjorkman}, J.~E., \& {Lowrance}, P.~J. 2003, \aj,
  125, 1467

\bibitem[{{Sturm} {et~al.}(2010){Sturm}, {Bouwman}, {Henning}, {Evans}, {Acke},
  {Mulders}, {Waters}, {van Dishoeck}, {Meeus}, {Green}, {Augereau},
  {Olofsson}, {Salyk}, {Najita}, {Herczeg}, {van Kempen}, {Kristensen},
  {Dominik}, {Carr}, {Waelkens}, {Bergin}, {Blake}, {Brown}, {Chen}, {Cieza},
  {Dunham}, {Glassgold}, {G{\"u}del}, {Harvey}, {Hogerheijde}, {Jaffe},
  {J{\o}rgensen}, {Kim}, {Knez}, {Lacy}, {Lee}, {Maret}, {Meijerink},
  {Mer{\'{\i}}n}, {Mundy}, {Pontoppidan}, {Visser}, \&
  {Y{\i}ld{\i}z}}]{sturm2010}
{Sturm}, B., {Bouwman}, J., {Henning}, T., {Evans}, N.~J., {Acke}, B.,
  {Mulders}, G.~D., {Waters}, L.~B.~F.~M., {van Dishoeck}, E.~F., {Meeus}, G.,
  {Green}, J.~D., {Augereau}, J.~C., {Olofsson}, J., {Salyk}, C., {Najita}, J.,
  {Herczeg}, G.~J., {van Kempen}, T.~A., {Kristensen}, L.~E., {Dominik}, C.,
  {Carr}, J.~S., {Waelkens}, C., {Bergin}, E., {Blake}, G.~A., {Brown}, J.~M.,
  {Chen}, J., {Cieza}, L., {Dunham}, M.~M., {Glassgold}, A., {G{\"u}del}, M.,
  {Harvey}, P.~M., {Hogerheijde}, M.~R., {Jaffe}, D., {J{\o}rgensen}, J.~K.,
  {Kim}, H.~J., {Knez}, C., {Lacy}, J.~H., {Lee}, J., {Maret}, S., {Meijerink},
  R., {Mer{\'{\i}}n}, B., {Mundy}, L., {Pontoppidan}, K.~M., {Visser}, R., \&
  {Y{\i}ld{\i}z}, U.~A. 2010, \aap, 518, L129+

\bibitem[{{Tinbergen}(1996)}]{tinbergen1996}
{Tinbergen}, J. 1996, {Astronomical Polarimetry}, ed. {Tinbergen, J.}

\bibitem[{{van Boekel} {et~al.}(2005){van Boekel}, {Min}, {Waters}, {de Koter},
  {Dominik}, {van den Ancker}, \& {Bouwman}}]{vanboekel2005}
{van Boekel}, R., {Min}, M., {Waters}, L.~B.~F.~M., {de Koter}, A., {Dominik},
  C., {van den Ancker}, M.~E., \& {Bouwman}, J. 2005, \aap, 437, 189

\bibitem[{{van den Ancker} {et~al.}(1997){van den Ancker}, {The}, {Tjin A
  Djie}, {Catala}, {de Winter}, {Blondel}, \& {Waters}}]{vandenancker1997}
{van den Ancker}, M.~E., {The}, P.~S., {Tjin A Djie}, H.~R.~E., {Catala}, C.,
  {de Winter}, D., {Blondel}, P.~F.~C., \& {Waters}, L.~B.~F.~M. 1997, \aap,
  324, L33

\bibitem[{{van der Plas} {et~al.}(2009){van der Plas}, {van den Ancker},
  {Acke}, {Carmona}, {Dominik}, {Fedele}, \& {Waters}}]{vanderplas2009}
{van der Plas}, G., {van den Ancker}, M.~E., {Acke}, B., {Carmona}, A.,
  {Dominik}, C., {Fedele}, D., \& {Waters}, L.~B.~F.~M. 2009, \aap, 500, 1137

\bibitem[{{van Leeuwen}(2007)}]{vanleeuwen2007}
{van Leeuwen}, F. 2007, \aap, 474, 653

\bibitem[{{Vieira} {et~al.}(1999){Vieira}, {Pogodin}, \& {Franco}}]{vieira1999}
{Vieira}, S.~L.~A., {Pogodin}, M.~A., \& {Franco}, G.~A.~P. 1999, \aap, 345,
  559

\bibitem[{{Weinberger} {et~al.}(2002){Weinberger}, {Becklin}, {Schneider},
  {Chiang}, {Lowrance}, {Silverstone}, {Zuckerman}, {Hines}, \&
  {Smith}}]{weinberger2002}
{Weinberger}, A.~J., {Becklin}, E.~E., {Schneider}, G., {Chiang}, E.~I.,
  {Lowrance}, P.~J., {Silverstone}, M., {Zuckerman}, B., {Hines}, D.~C., \&
  {Smith}, B.~A. 2002, \apj, 566, 409

\bibitem[{{Wilner} {et~al.}(2003){Wilner}, {Bourke}, {Wright}, {J{\o}rgensen},
  {van Dishoeck}, \& {Wong}}]{wilner2003}
{Wilner}, D.~J., {Bourke}, T.~L., {Wright}, C.~M., {J{\o}rgensen}, J.~K., {van
  Dishoeck}, E.~F., \& {Wong}, T. 2003, \apj, 596, 597

\bibitem[{{Witzel} {et~al.}(2011){Witzel}, {Eckart}, {Buchholz}, {Zamaninasab},
  {Lenzen}, {Sch{\"o}del}, {Araujo}, {Sabha}, {Bremer}, {Karas}, {Straubmeier},
  \& {Muzic}}]{witzel2011}
{Witzel}, G., {Eckart}, A., {Buchholz}, R.~M., {Zamaninasab}, M., {Lenzen}, R.,
  {Sch{\"o}del}, R., {Araujo}, C., {Sabha}, N., {Bremer}, M., {Karas}, V.,
  {Straubmeier}, C., \& {Muzic}, K. 2011, \aap, 525, A130+

\bibitem[{{Wolf}(2008)}]{wolf2008}
{Wolf}, S. 2008, Physica Scripta Volume T, 130, 014025

\bibitem[{{Yudin} \& {Evans}(1998)}]{yudin1998}
{Yudin}, R.~V. \& {Evans}, A. 1998, \aaps, 131, 401

\end{thebibliography}

%\begin{thebibliography}{}

%\end{thebibliography}{}

\clearpage

\clearpage

\begin{figure*}
%\centering
\epsscale{0.8}
\plottwo{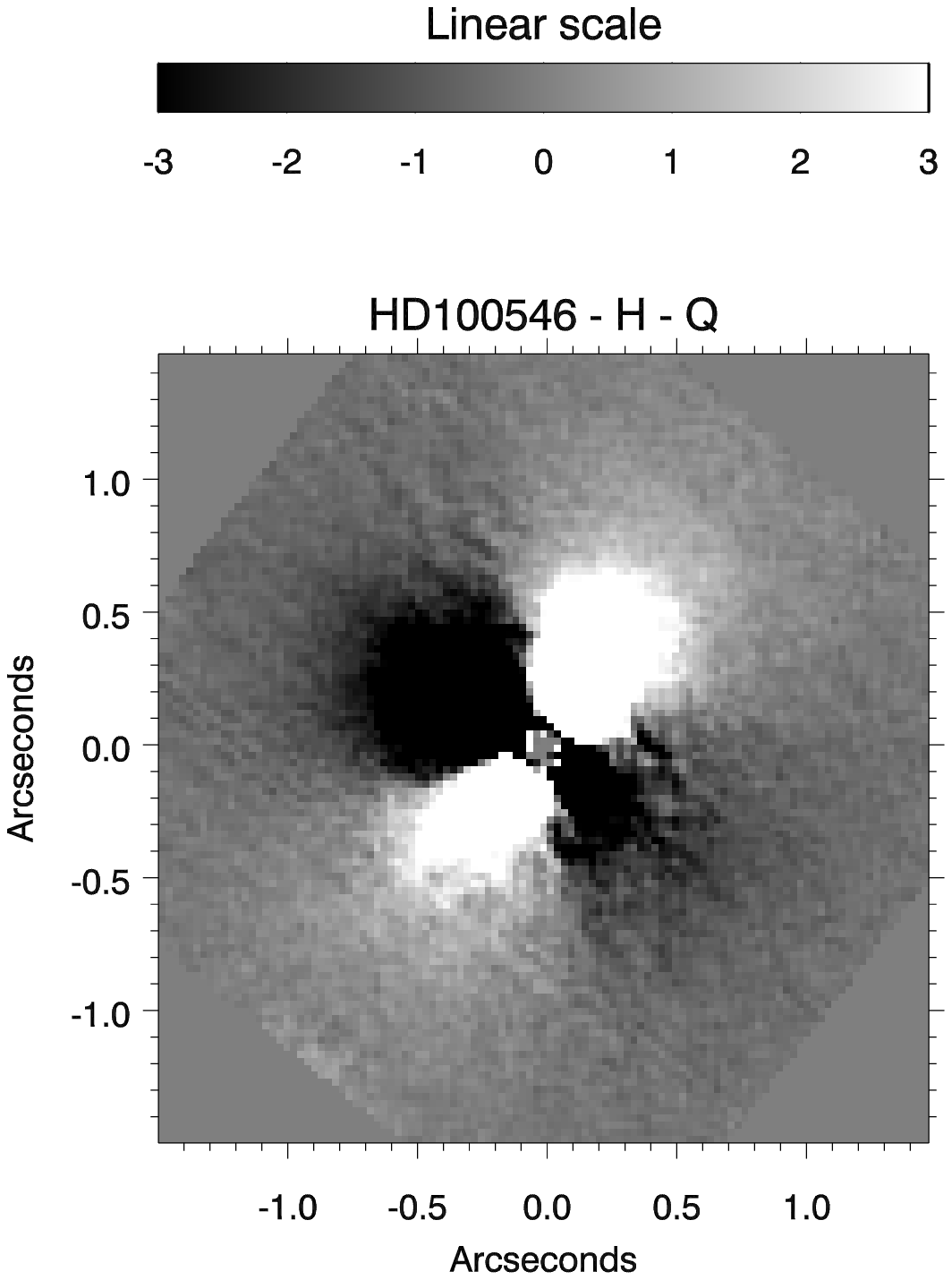}{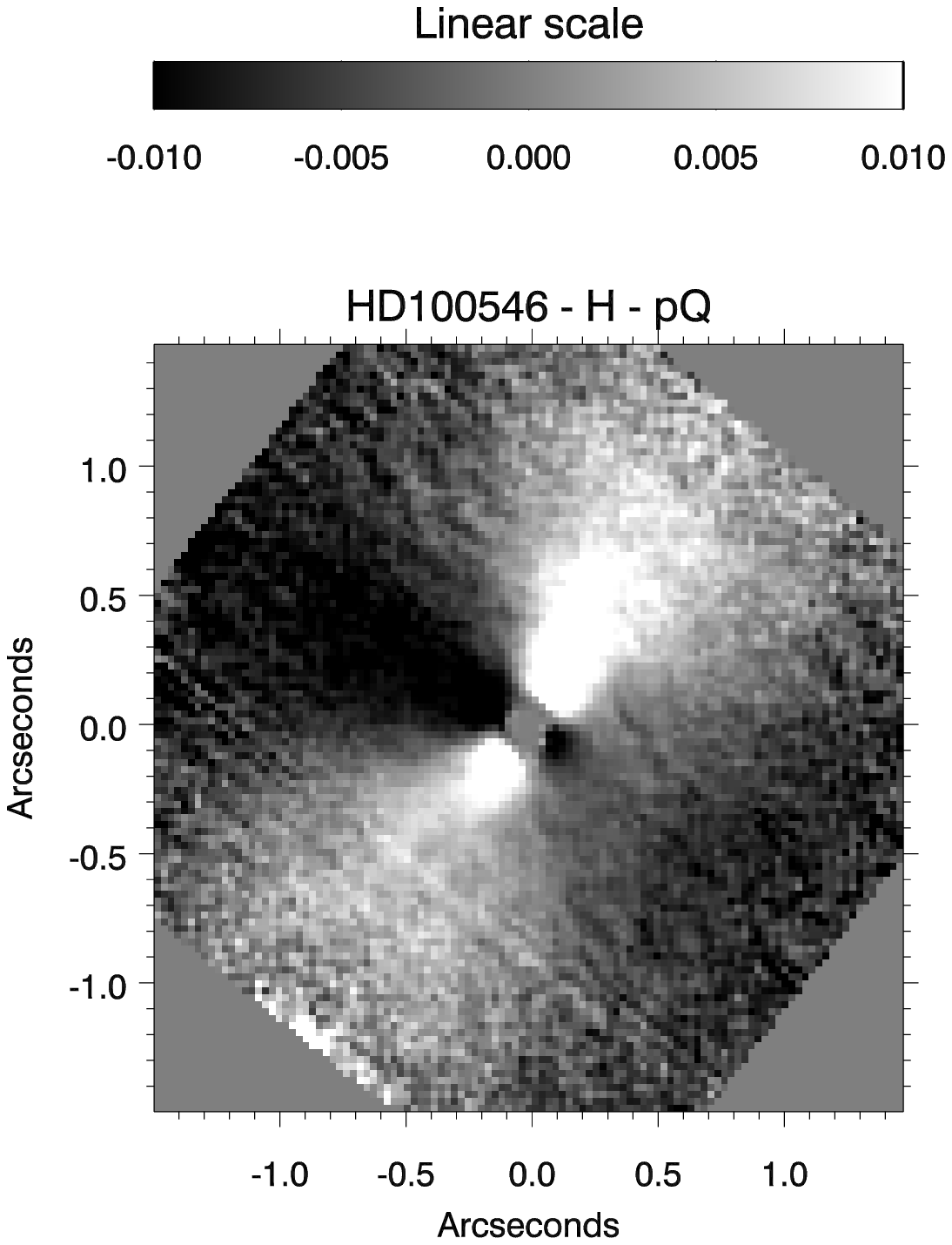}
\plottwo{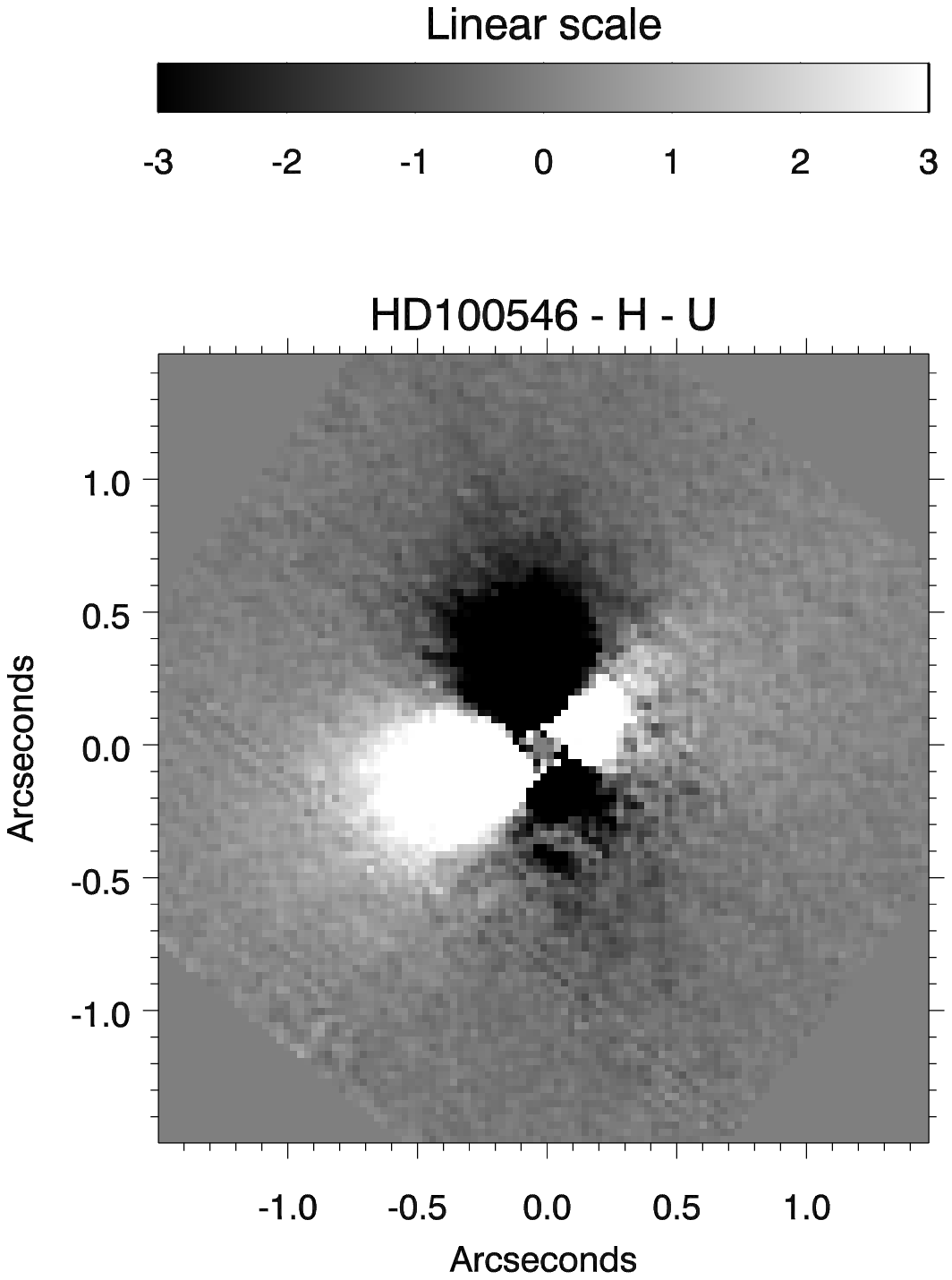}{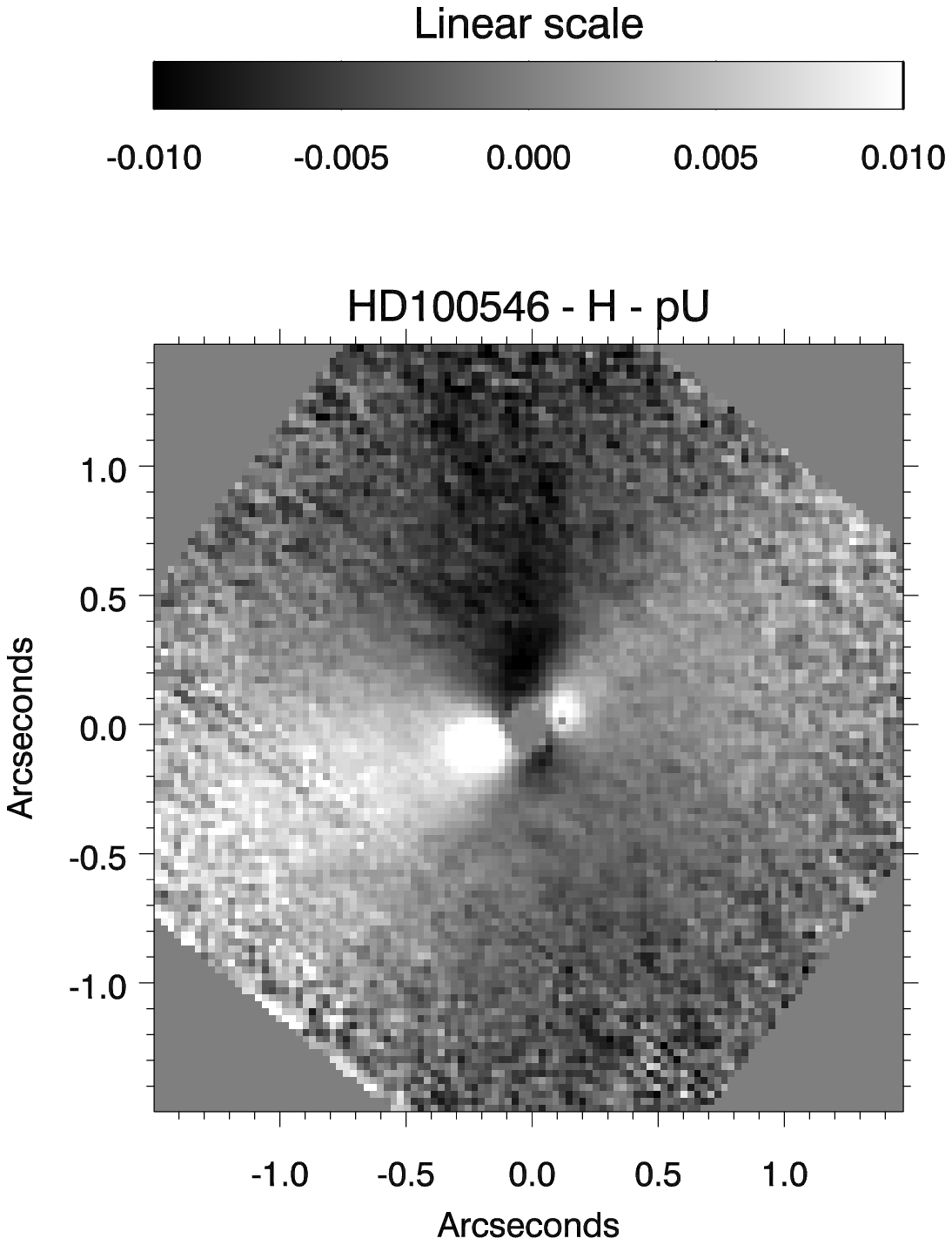}
%\plottwo{H_P_new.eps}{H_pI.eps}
\caption{Final Stokes $Q$ and $U$ images (count rates in arbitrary units, left column) and fractional polarization $p_Q$ and $p_U$ (right column) of HD100546 in the $H$ filter. The core of the PSF, where the count rate was no longer in the linear detector regime in at least one of the individual raw images before they were combined, has been masked out. All images have been multiplied by -1 so that ordinary and extraordinary beam are the same as for the other filters to compensate for a 90$^\circ$ difference in the position angle of the camera (see, Table~\ref{observations}). North is up and east to the left in all images (since we aligned the disk major axis with the detector's x-axis the positive $Q$ component is not aligned in North-South direction in our final images as it is the usual convention). 
\label{H_images}}
\end{figure*}

\clearpage

\begin{figure*}
%\centering
\epsscale{0.85}
\plottwo{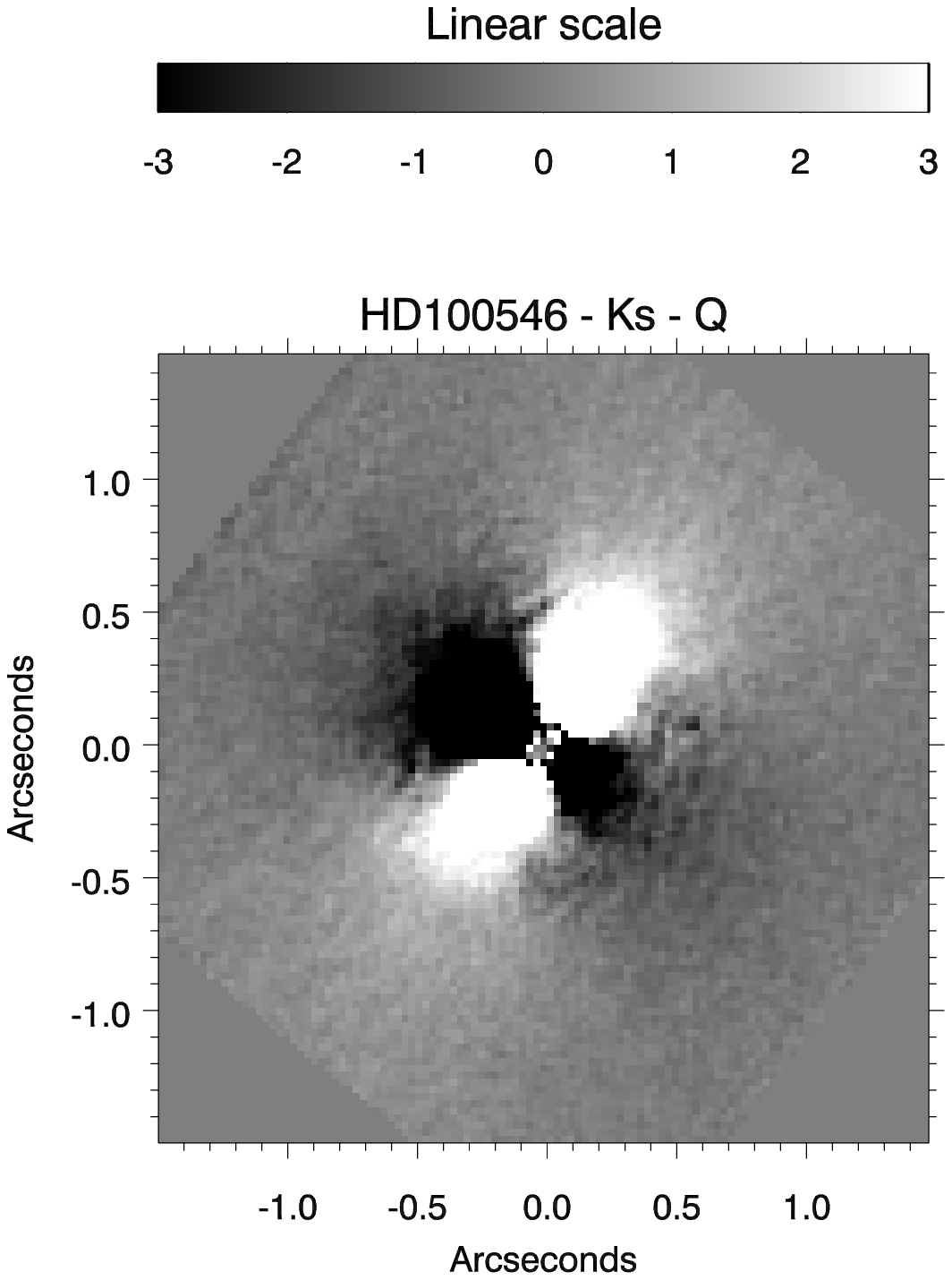}{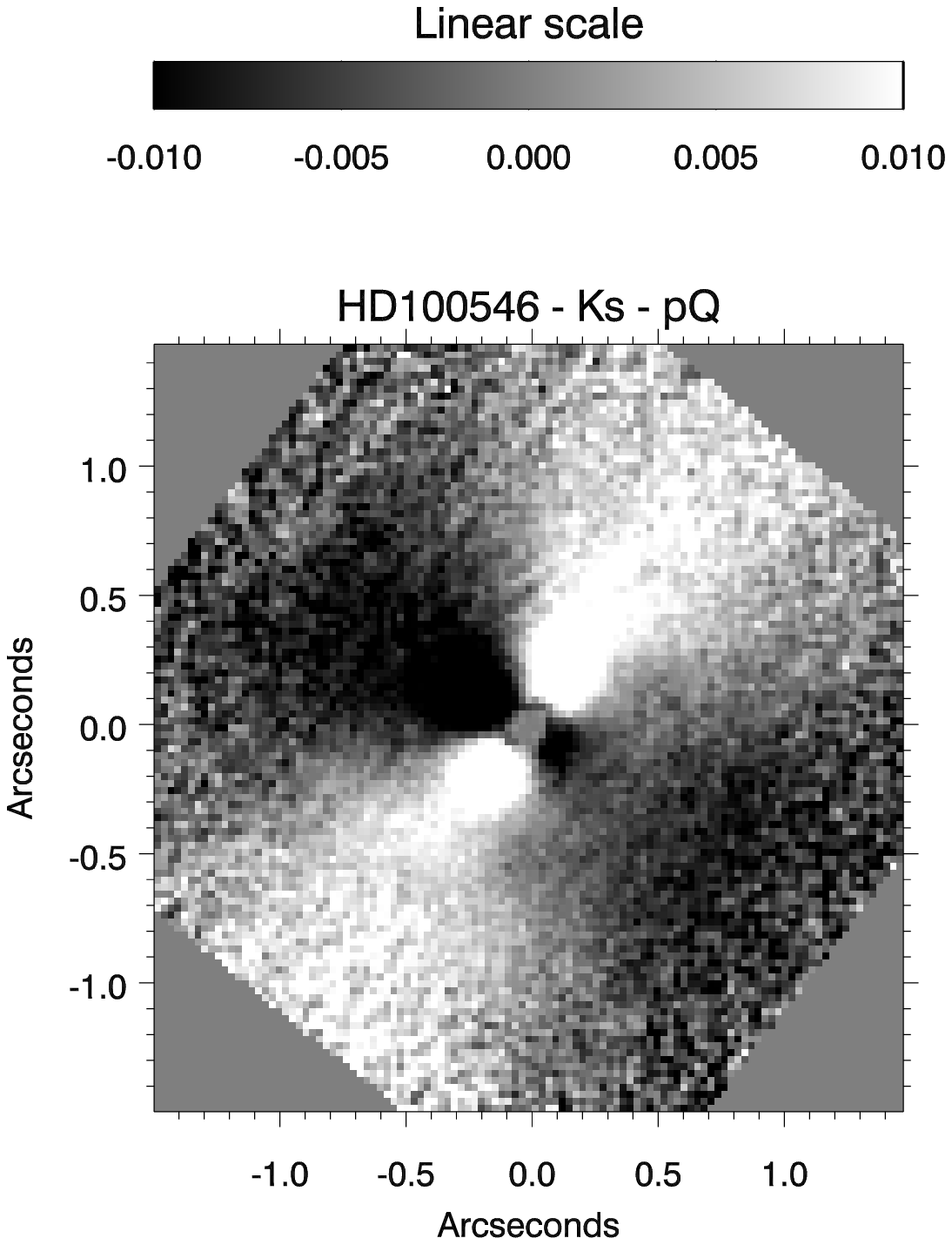}
\plottwo{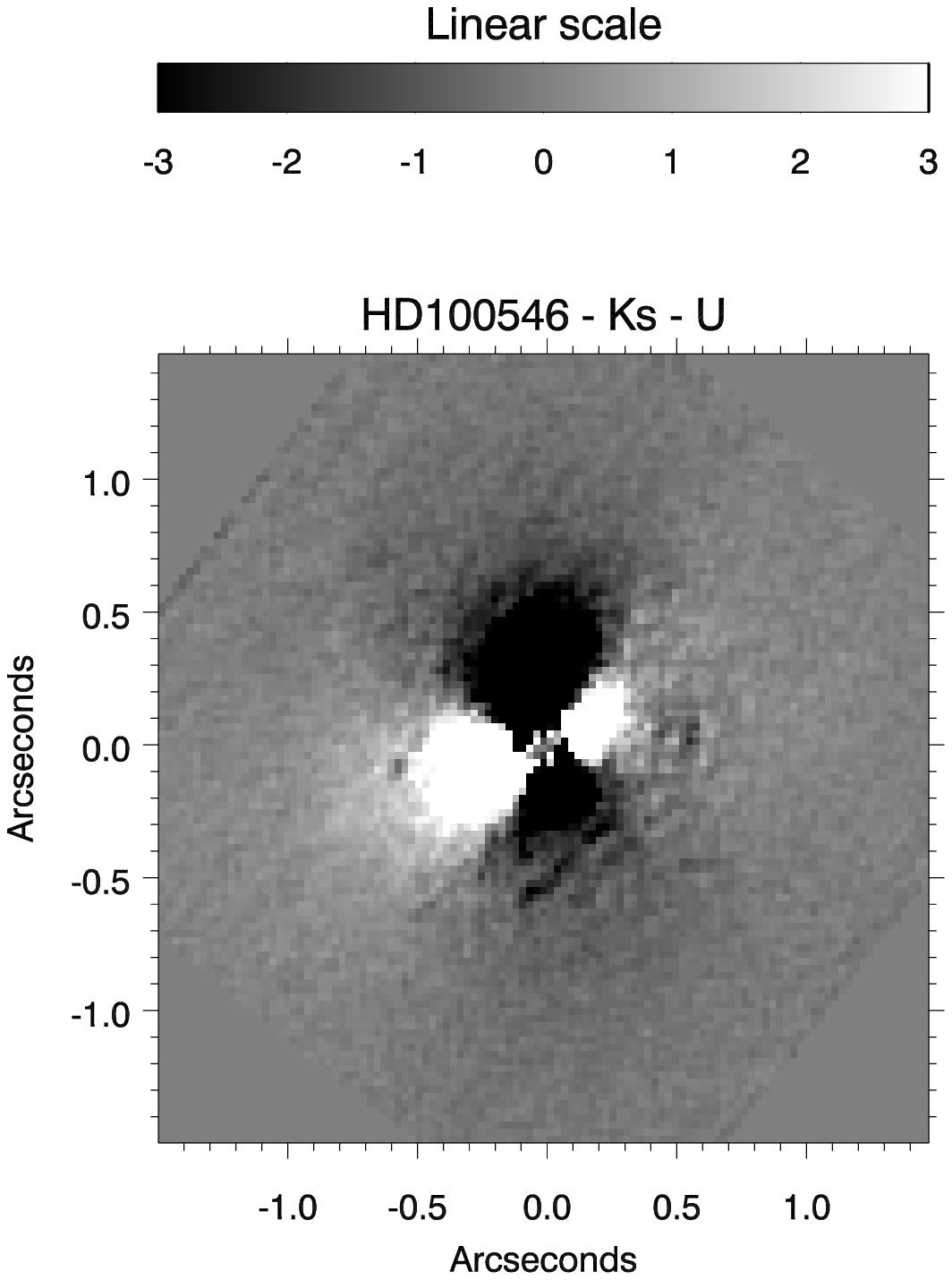}{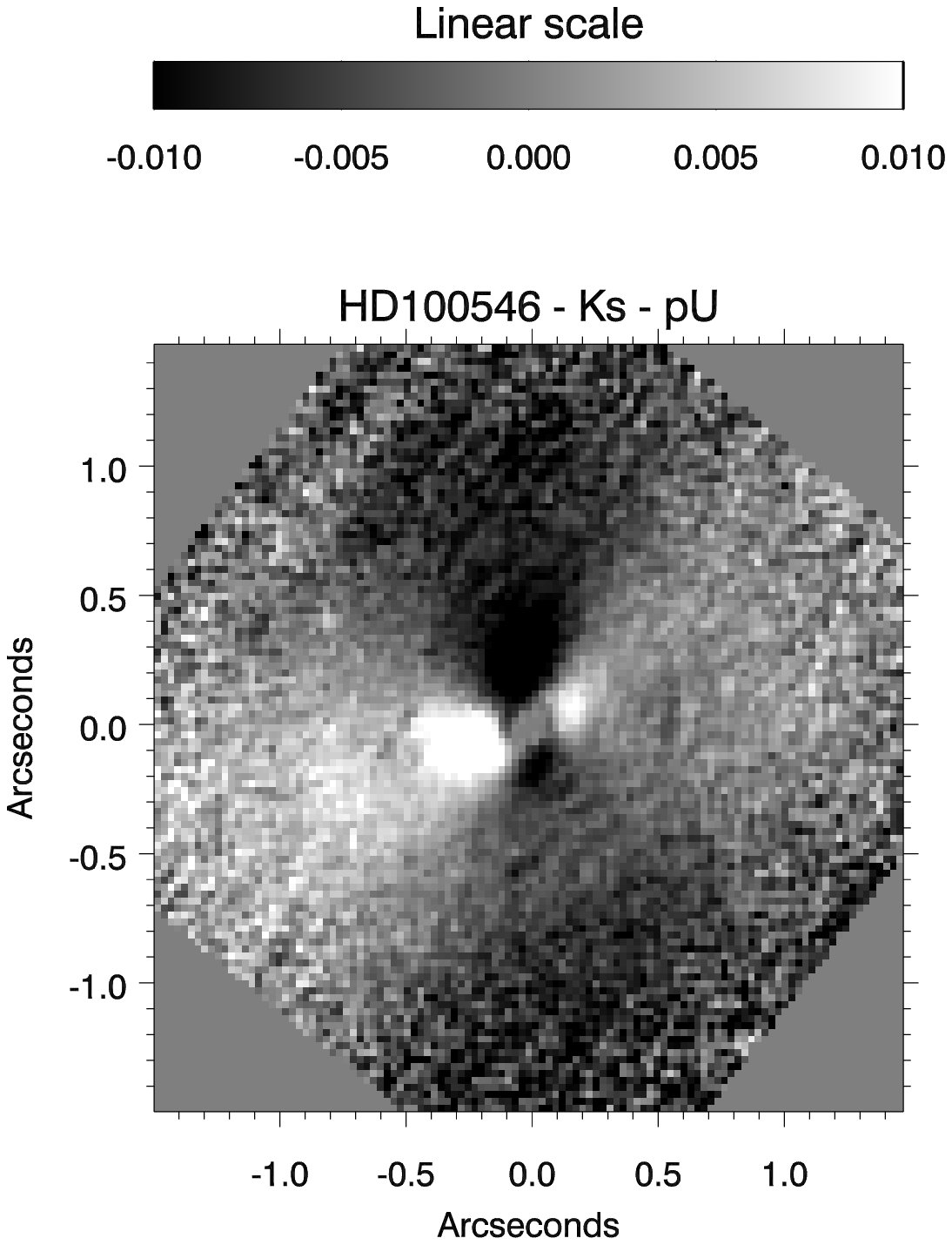}
%\plottwo{Ks_P_new.eps}{Ks_pI.eps}
\caption{Same as Figure~\ref{H_images} but for the $K_s$ filter
\label{Ks_images}}
\end{figure*}

\clearpage

\begin{figure*}
\centering
\epsscale{1}
\plottwo{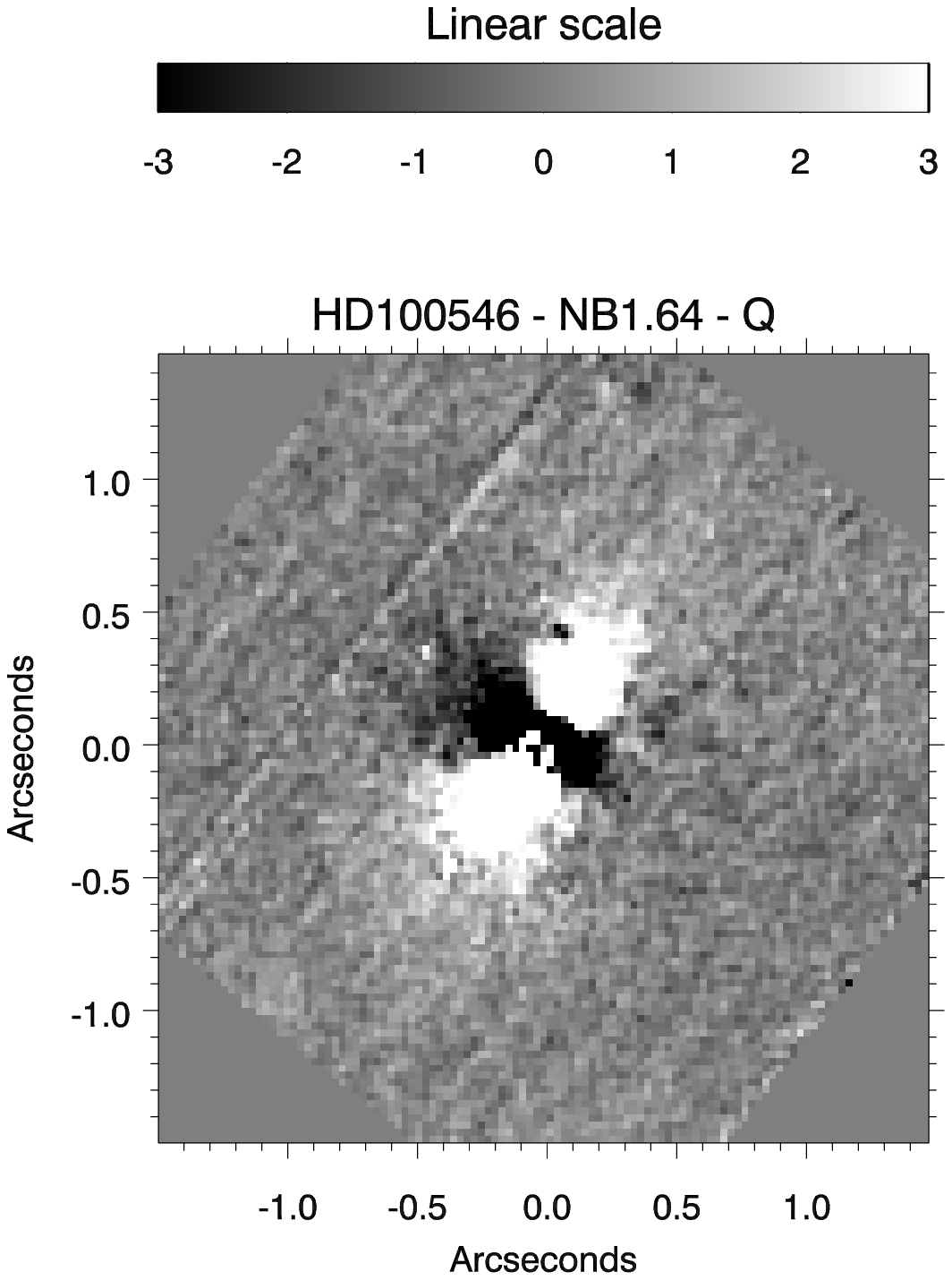}{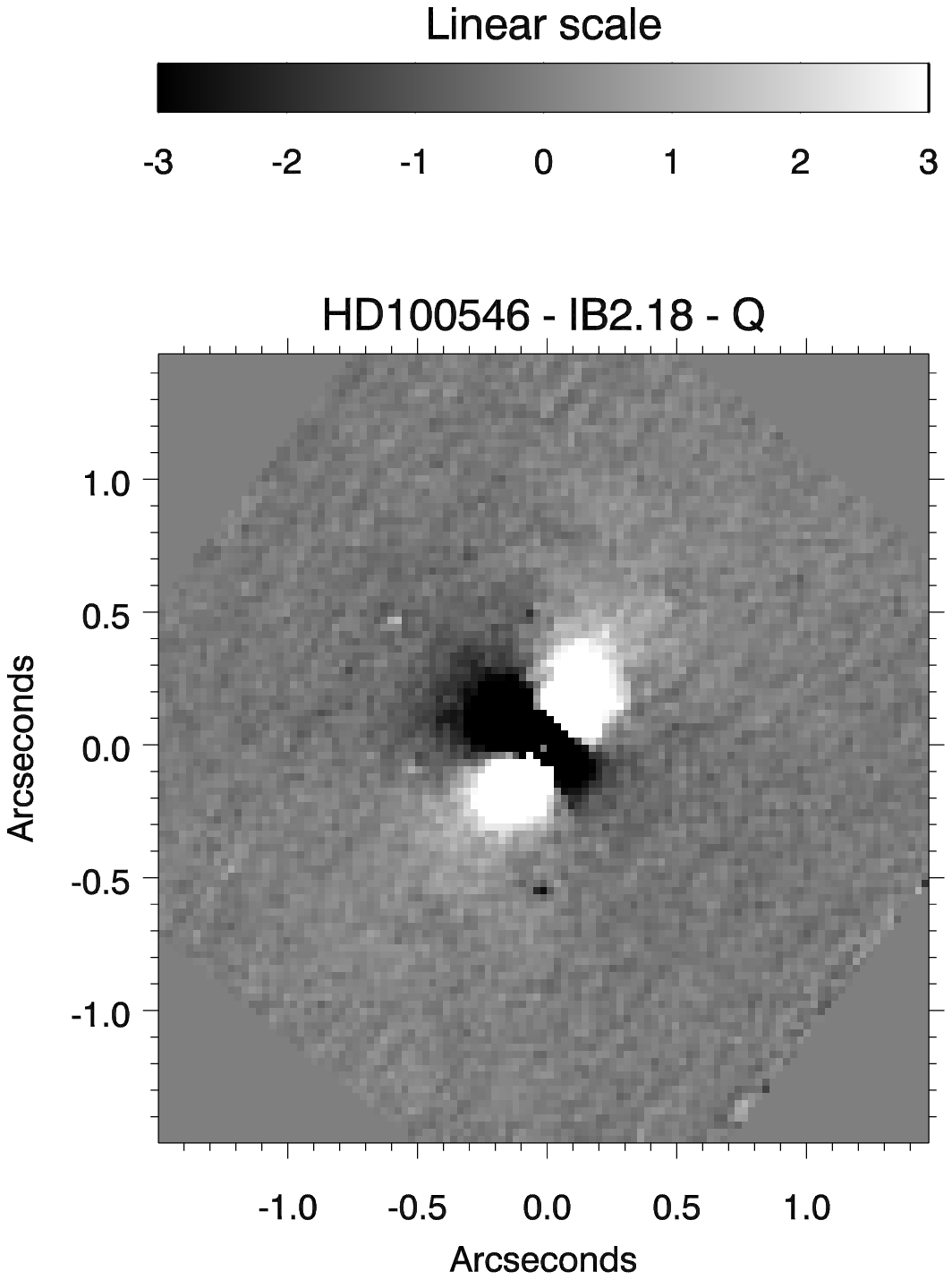}
\caption{Final Stokes $Q$ images for HD100546 in the $NB1.64$ filter (left) and $IB2.16$ filter (right). The scaling and orientation is the same as in Figures~\ref{H_images} and~\ref{Ks_images}.
\label{NB_images}}
\end{figure*}

\clearpage

\begin{figure*}
\centering
\epsscale{1}
\plottwo{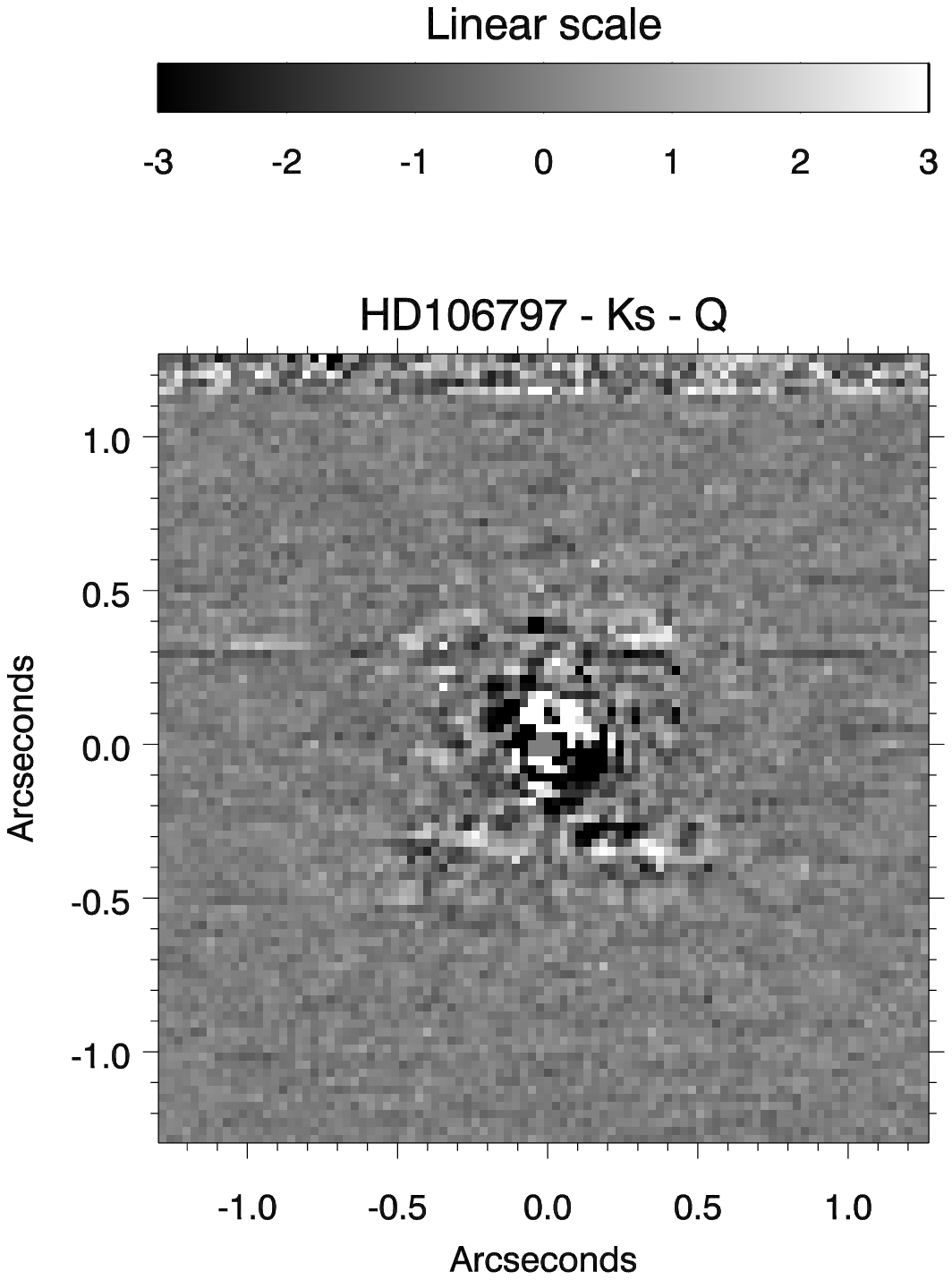}{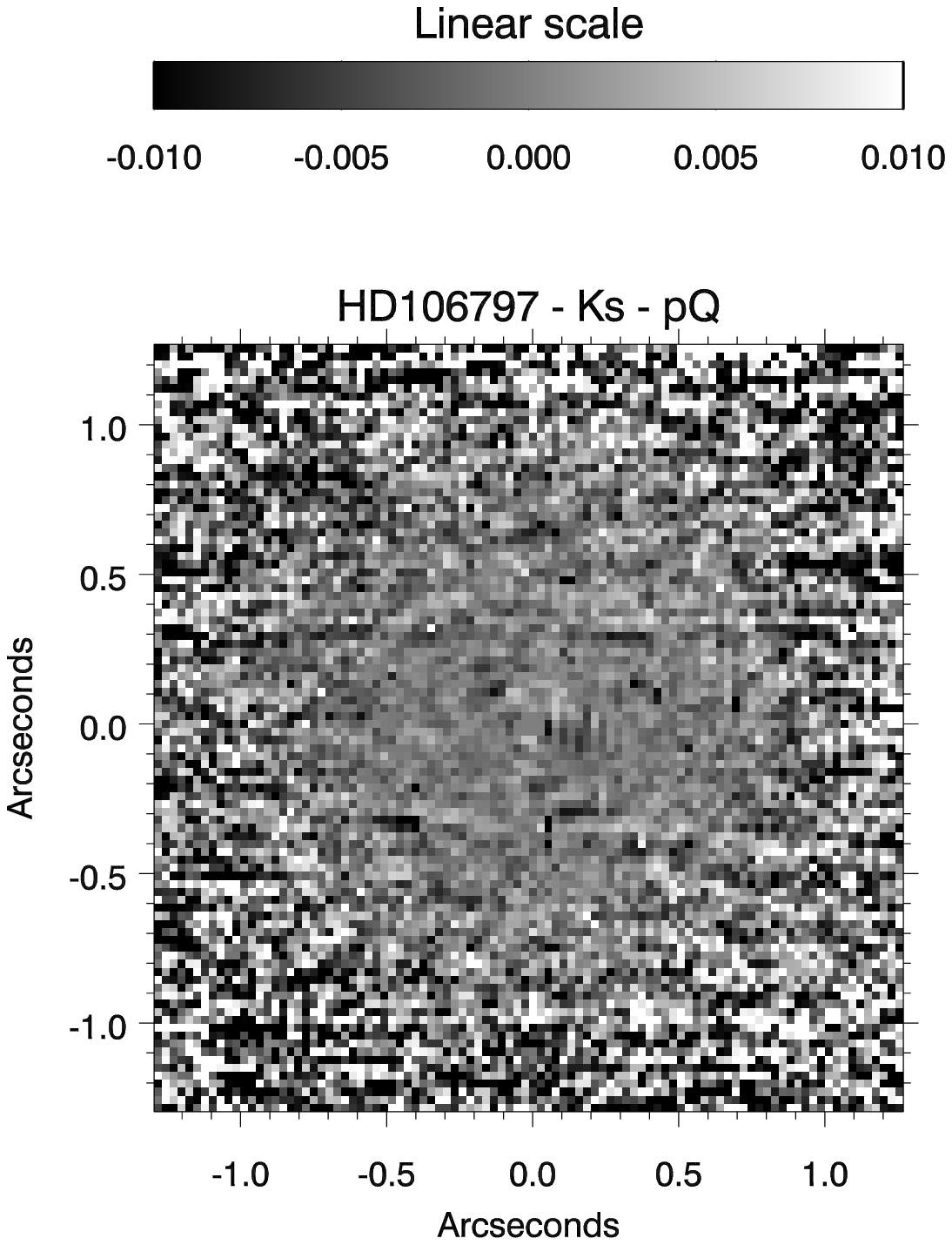}
\plottwo{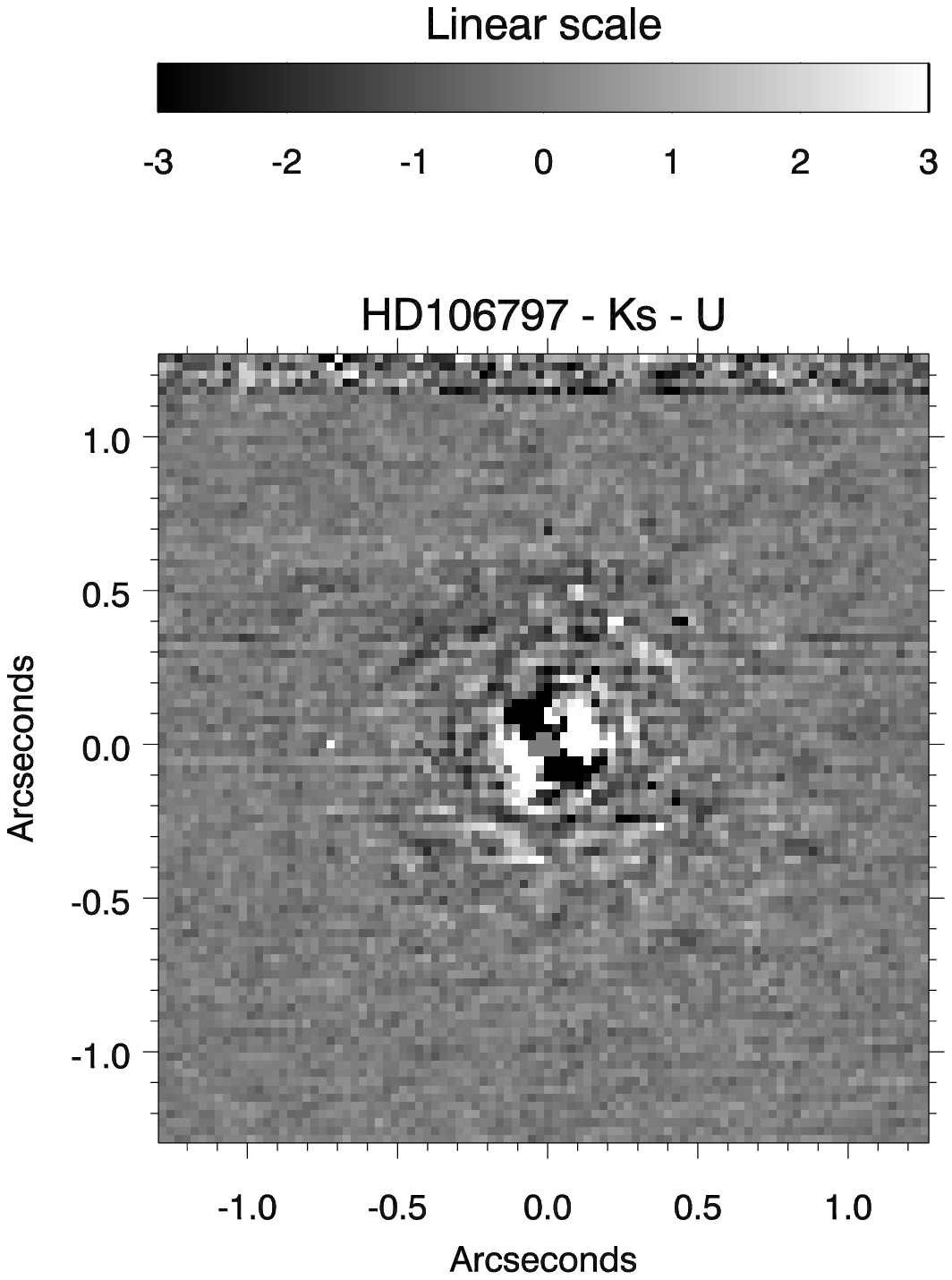}{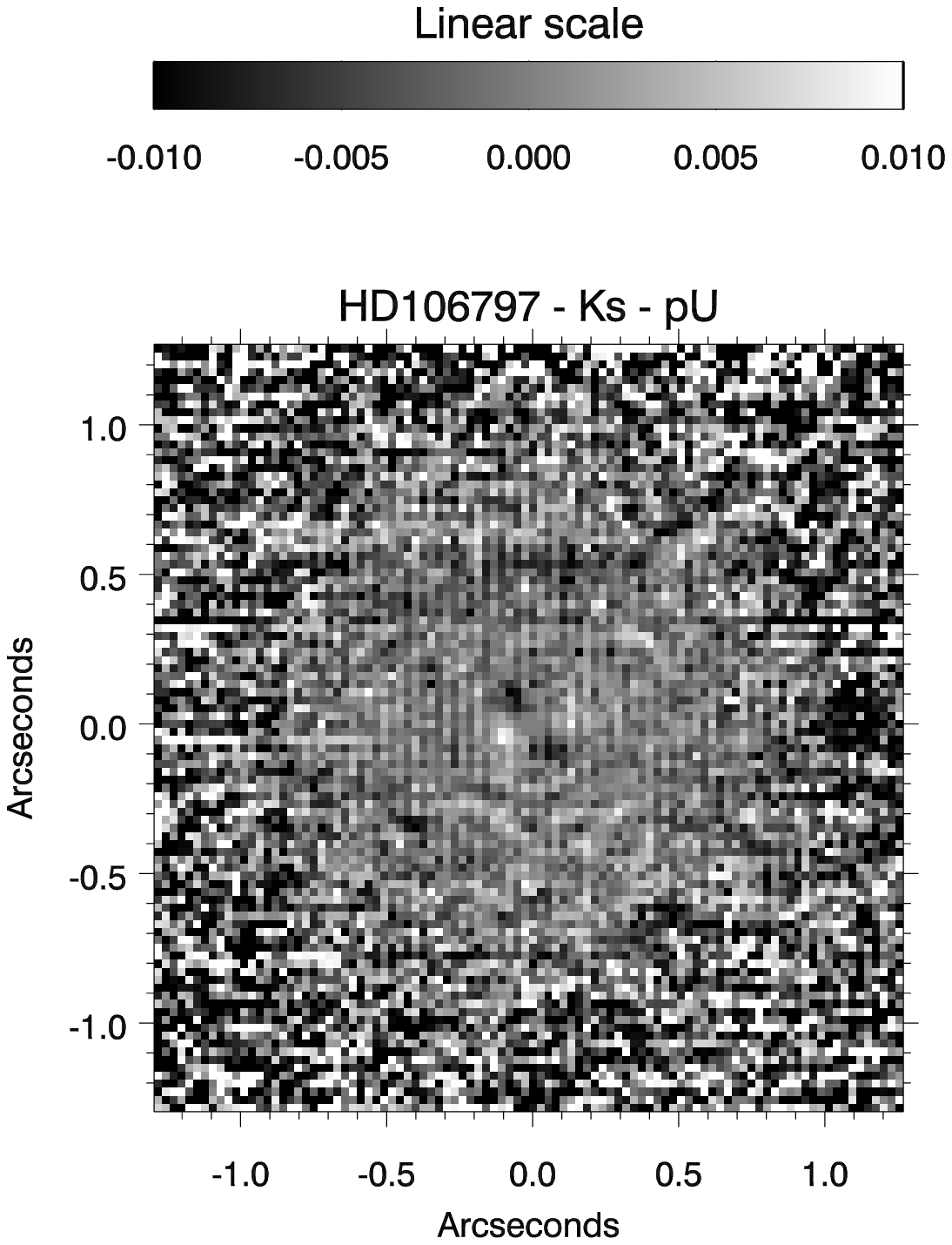}
\caption{Same as Figure~\ref{Ks_images} but for the reference star HD106797. Note that the images have the same stretch. 
\label{reference_images}}
\end{figure*} 

\clearpage
 
\begin{figure*}
\centering
\epsscale{0.9}
\plottwo{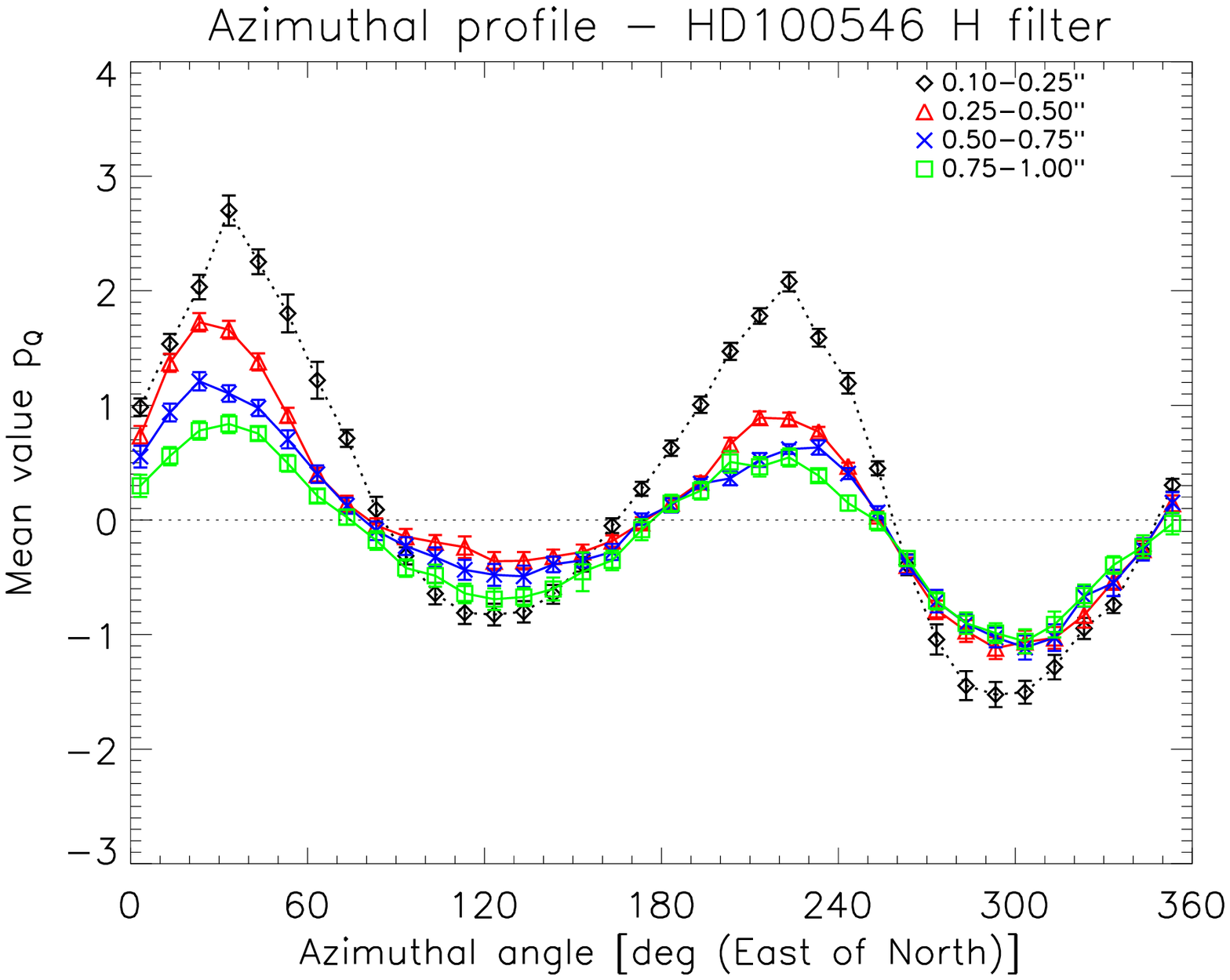}{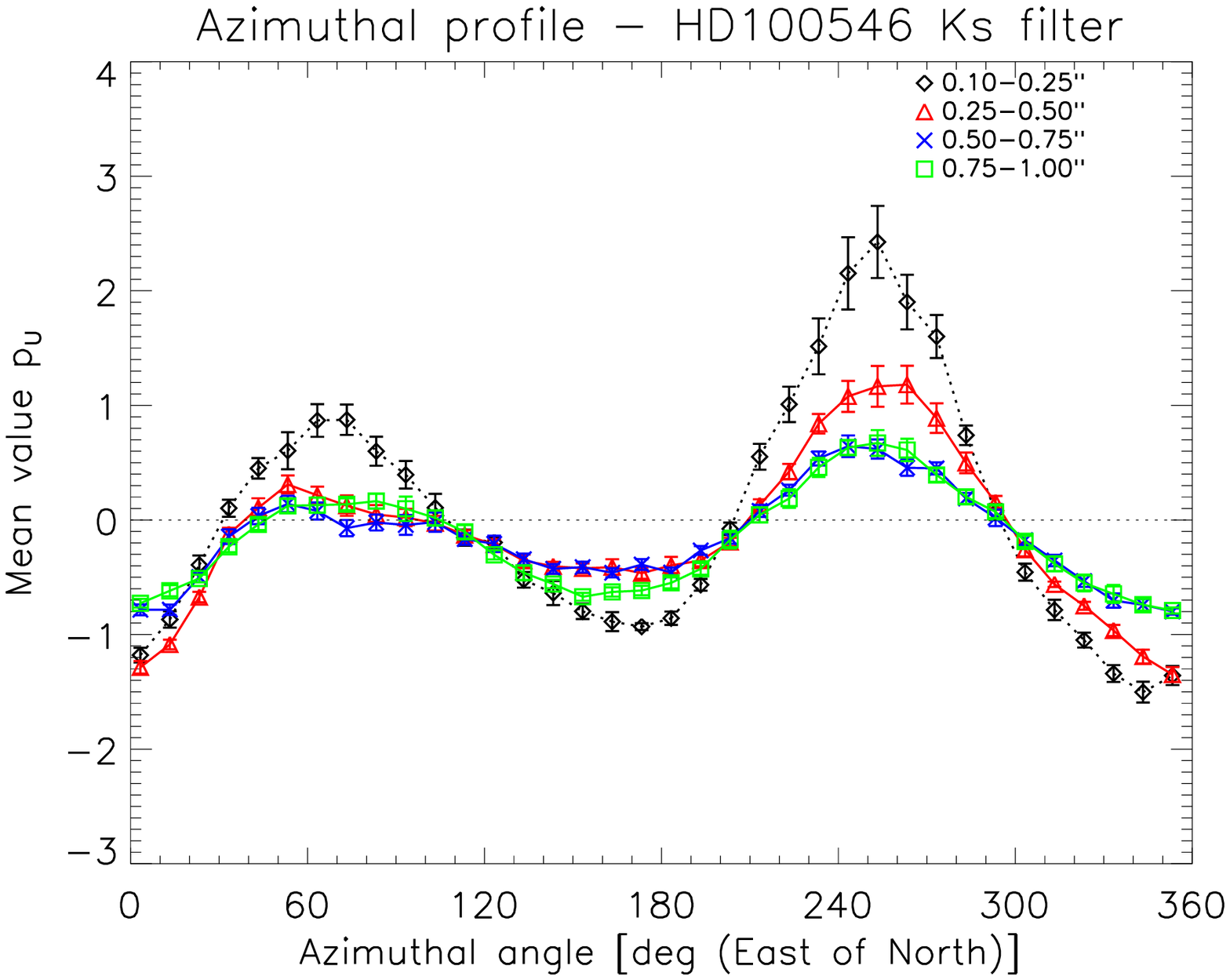}
\plottwo{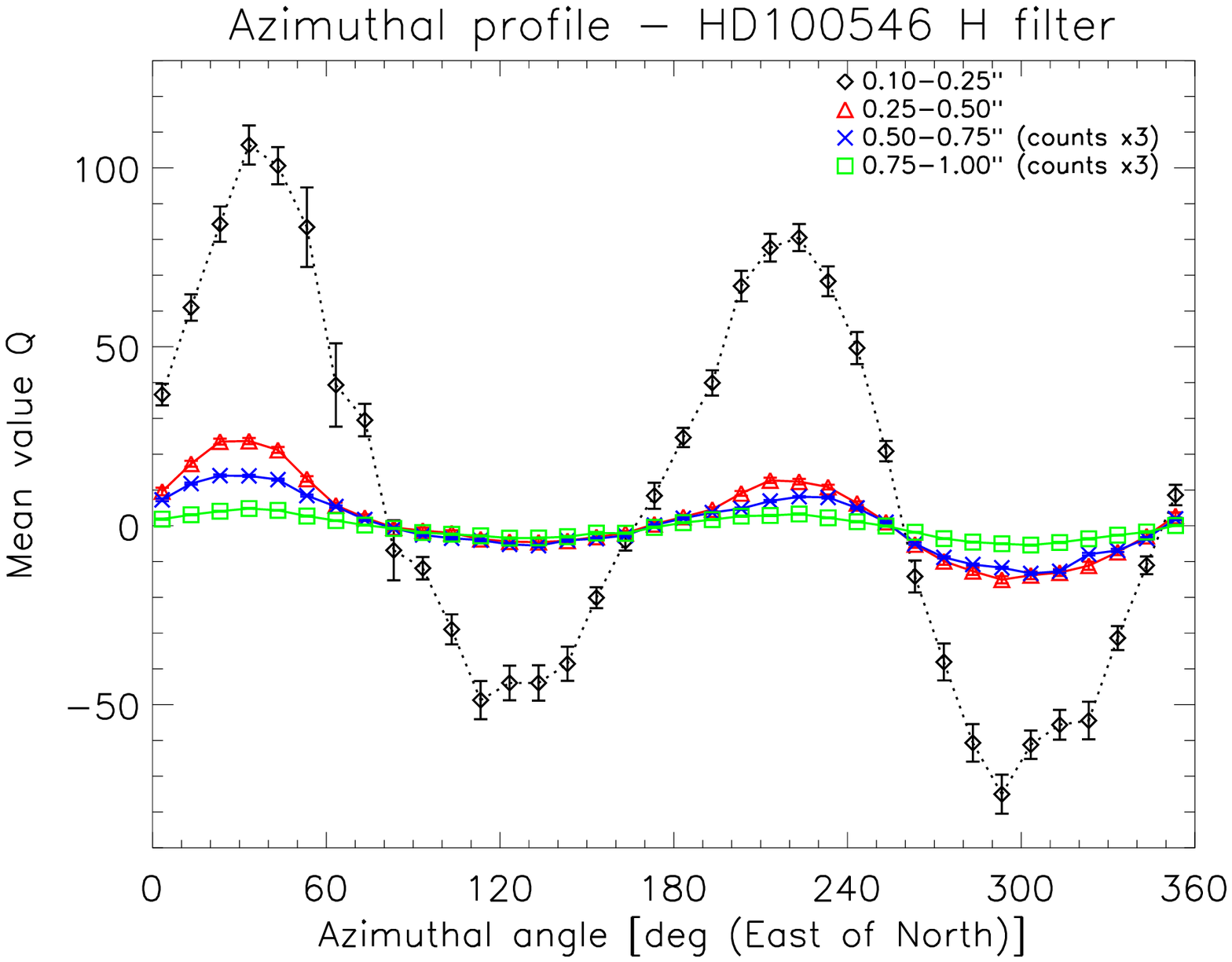}{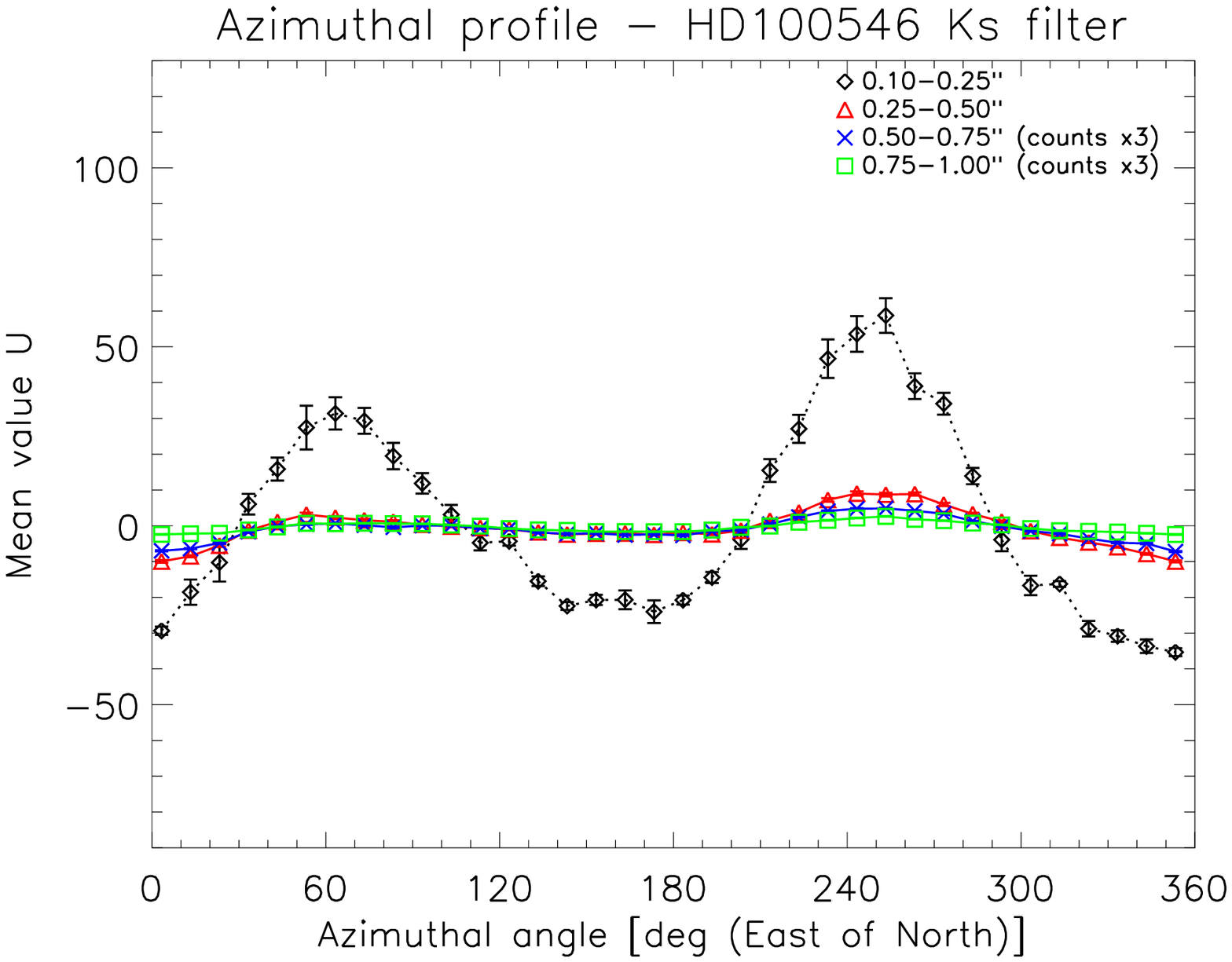}
\caption{Azimuthal profiles of fractional polarization $p_Q$ ($H$ filter) and $p_U$ ($K_s$ filter) in the top row and Stokes parameters $Q$ ($H$ filter) and $U$ ($K_s$ filter) in the bottom row based on Figures~\ref{H_images} and~\ref{Ks_images}. The unit of the y-axis is '\%' in the top row and 'count rate' in the bottom row. Curves for different annuli centered on the star are shown: diamonds (black) = 0.1--0.25$''$, triangles (red) = 0.25--0.5$''$, x (blue) = 0.5--0.75$''$, squares (green) =  0.75--1.0$''$. Each data point shows the mean value in a 10$^\circ$ wedge for a given annulus. In the bottom row the count rates measured for the two outermost annuli have been multiplied by 3 to emphasize their sinusoidal variation. The error bars are the standard deviation of the mean values computed in the individual images at each dither position divided by the square root of the number of images that were combined (see, Table~\ref{observations}).
\label{azimuthal_profile}}
\end{figure*}
 
 \clearpage

\clearpage

\clearpage

\clearpage

\clearpage

\begin{figure*}
\centering
\epsscale{0.9}
\plottwo{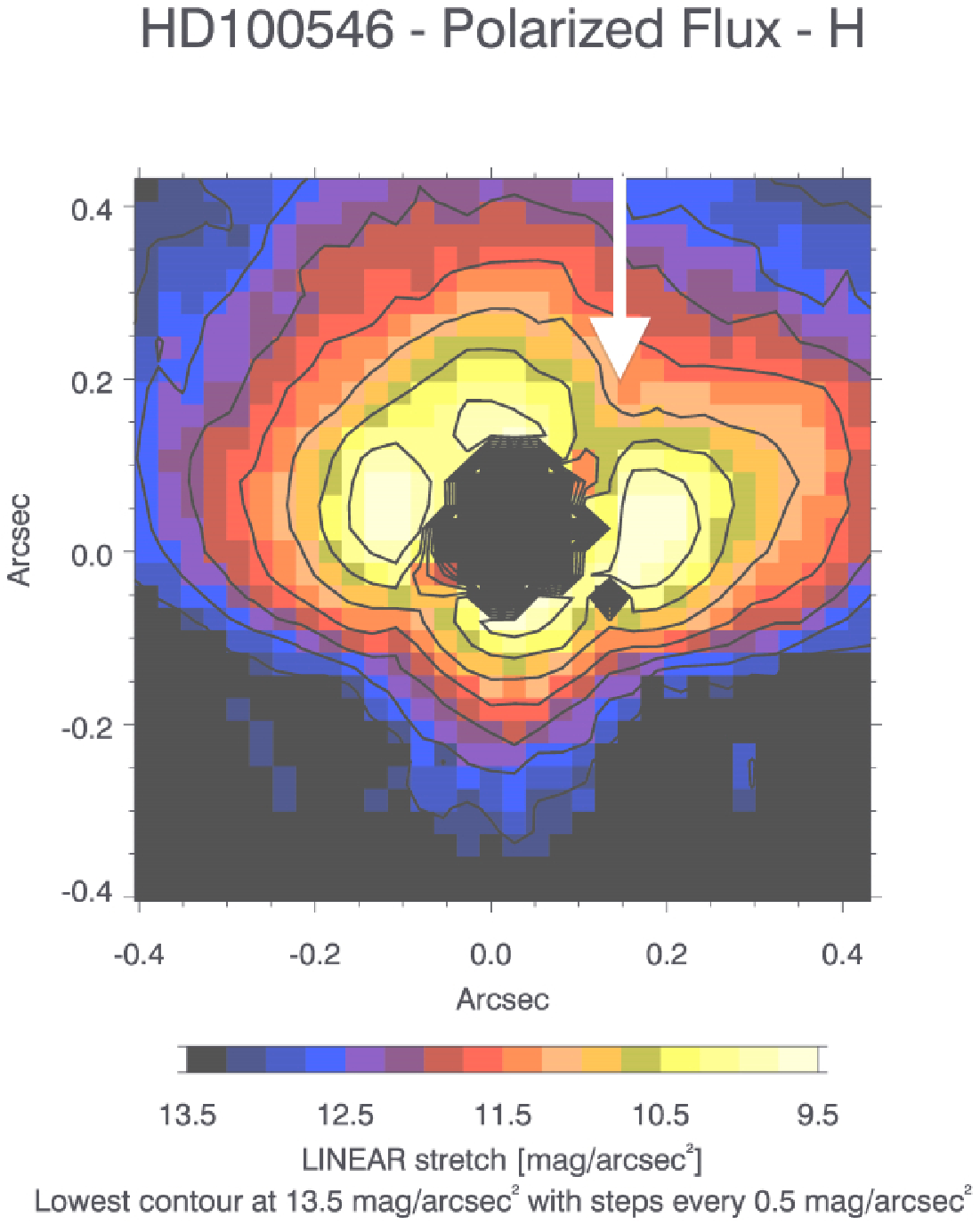}{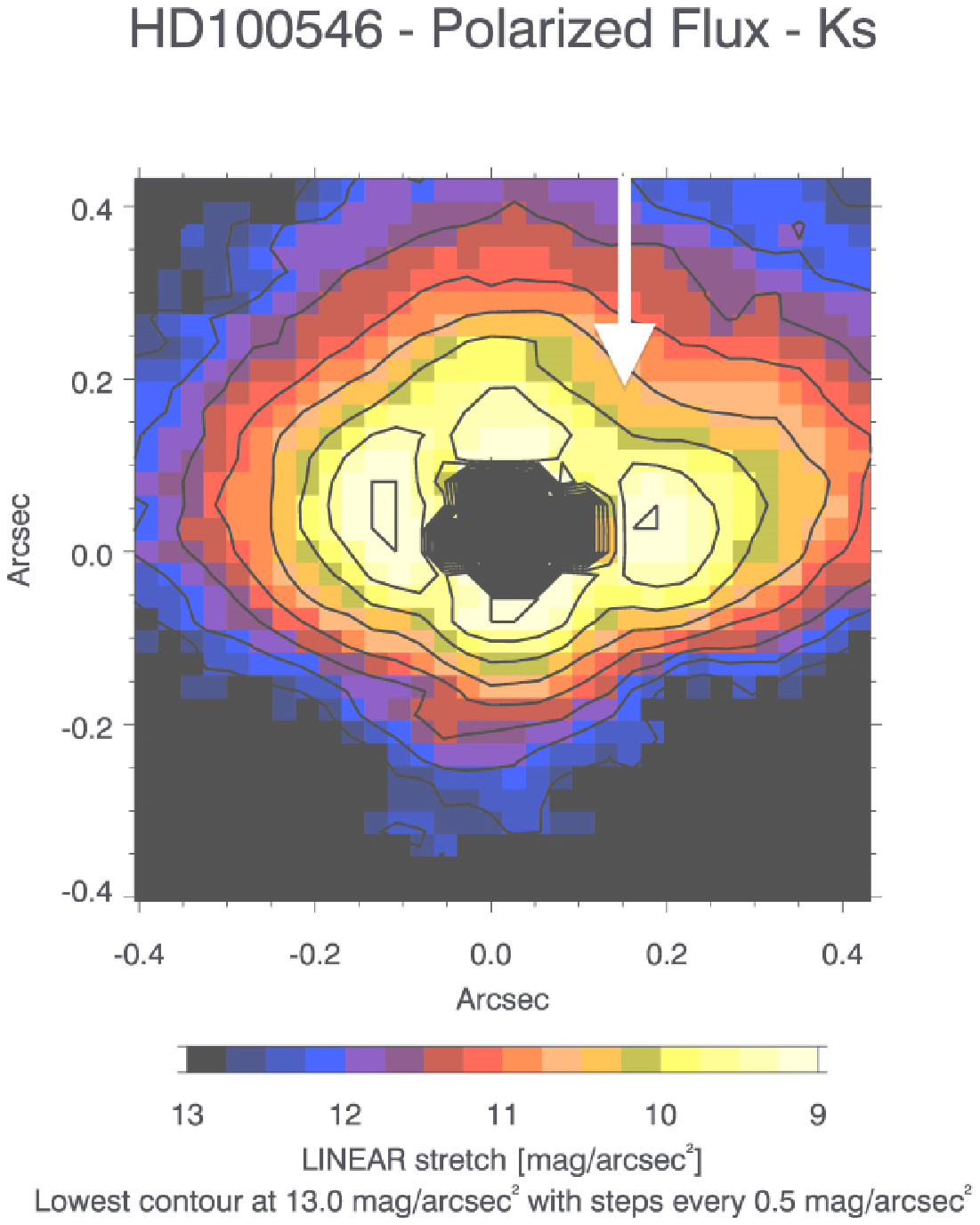}
\plottwo{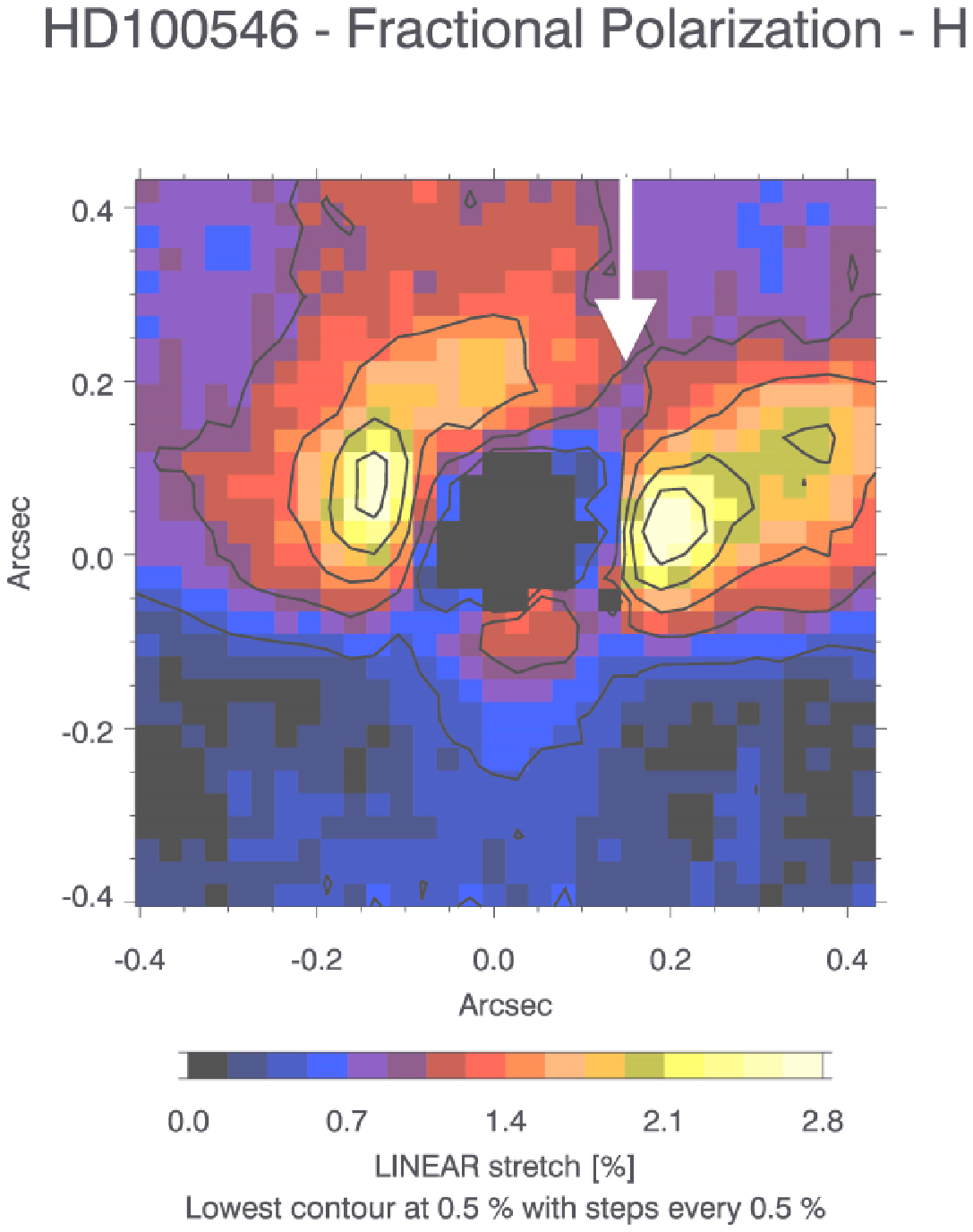}{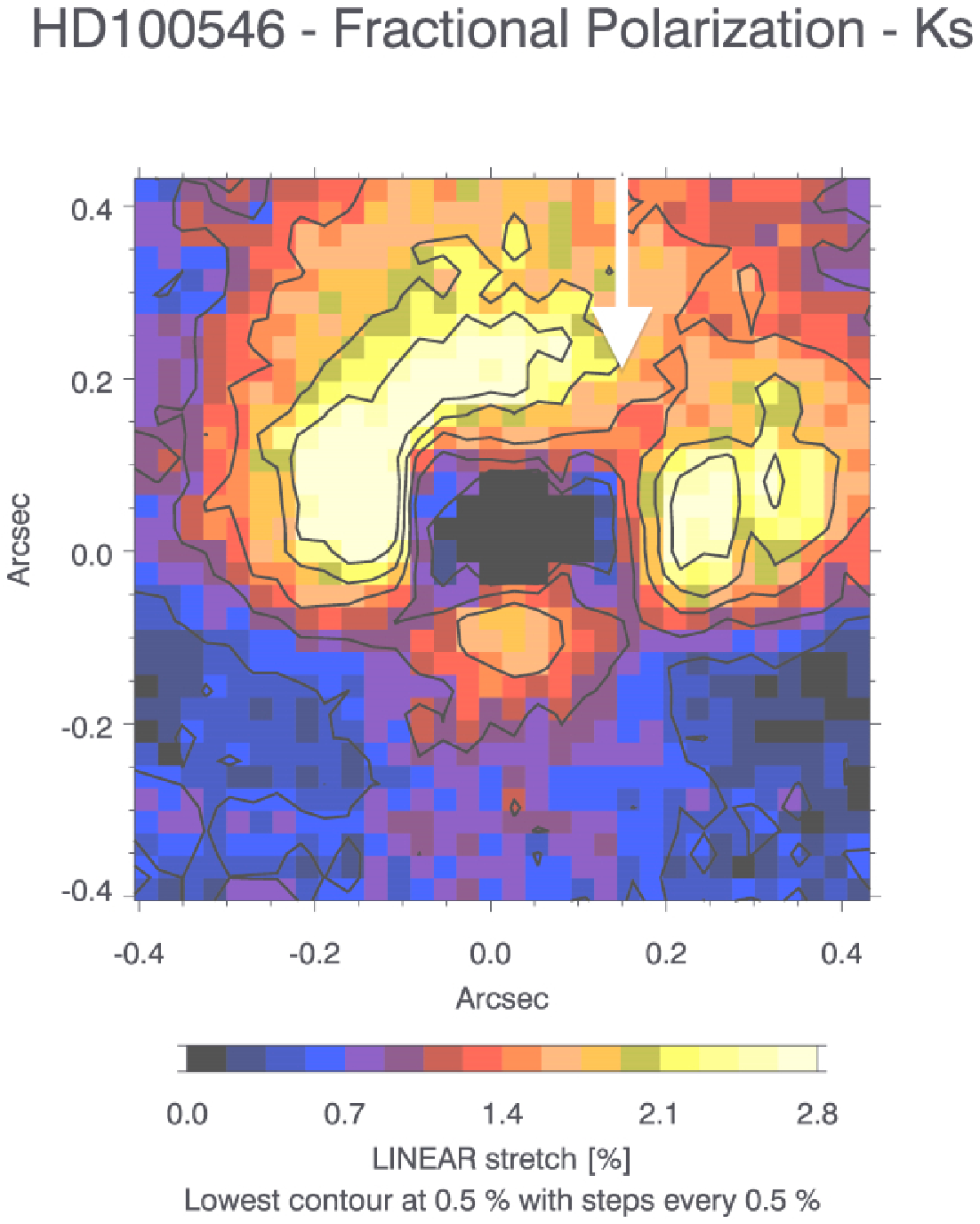}
\caption{Zoom in the inner regions of the images shown in Figure~\ref{P_and_pI_images}.  The images have been rotated clockwise so that the disk major axis runs horizontally. The white arrow indicates the location of the "hole" (see also Figure~\ref{disk_filter}).
\label{disk_hole_zoom}}
\end{figure*}

\clearpage

\begin{figure*}
\centering
\epsscale{.7}
\plotone{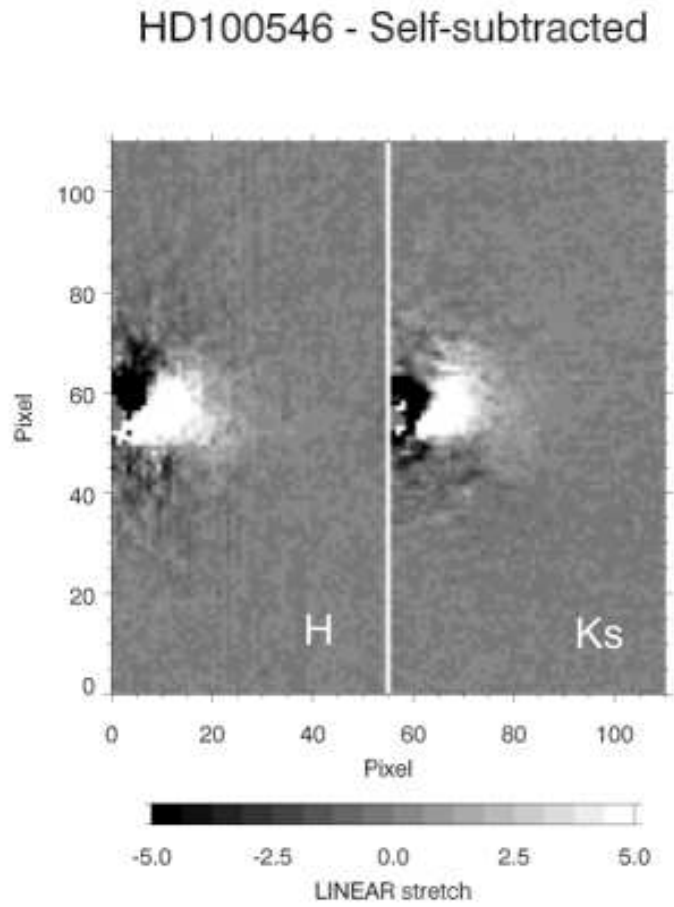}
\caption{Resulting images from subtracting the mirrored left-hand side of the disk from the right hand side of the disk (left panel: $H$ filter; right panel: $K_s$ filter). Before subtraction the images were rotated clockwise so that the disk major axis runs horizontally (same orientation as in Figure~\ref{disk_hole_zoom}).
\label{disk_subtract}}
\end{figure*}

\clearpage

\begin{figure*}
%\centering
\epsscale{.7}
\plotone{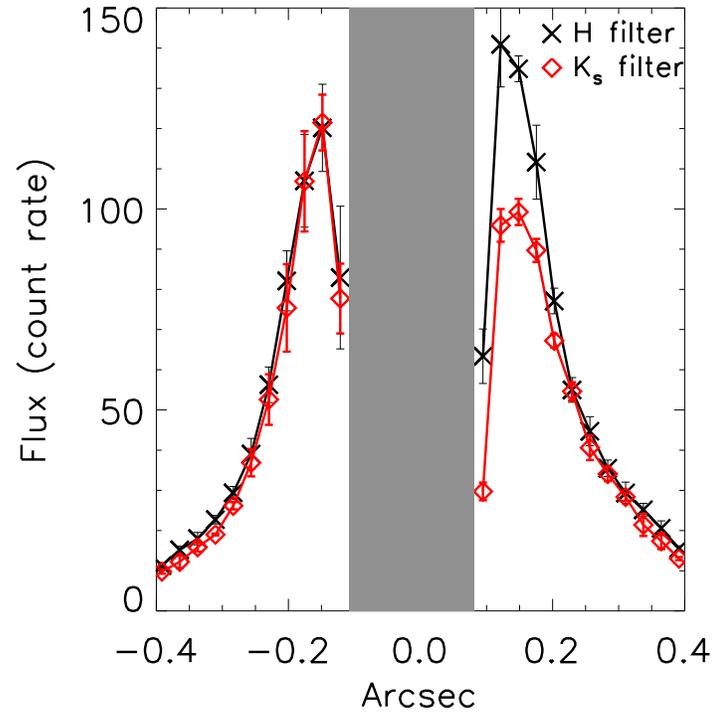}
\caption{Cut along the disk major axis in the polarized flux images shown in Figure~\ref{disk_hole_zoom}. Black 'x' show the results for the $H$ images while red diamonds show the results for the $K_s$ images. Pixels in the image center that were overexposed and no longer in the linear detector regime have been masked out.
\label{inner_rim}}
\end{figure*}

\clearpage

\clearpage

\begin{deluxetable}{lllc}
\centering
\tablecaption{Basic parameters of objects. 
\label{parameters}}           % title of Table
%\tablewidth{0pt}
      % is used to refer this table in the text
\tablehead{
\colhead{Parameter} & \colhead{HD100546}  & \colhead{HD106797} & \colhead{Reference\tablenotemark{a}}
}
\startdata
RA (J2000) & 11$^h$33$^m$25$^s$.44 & 12$^h$17$^m$06$^s$.31 & (1) \\ 
DEC (J2000) & -70$^\circ$11$'$41$''$.24 & -65$^\circ$41$'$34$''$.65  & (1)\\
$J$ & 6.43$\pm 0.02$ mag & 6.00$\pm 0.02$ mag & (2)\\
$H$ & 5.96$\pm 0.03$ mag & 6.03$\pm 0.04$ mag &(2)\\
$K_s$ & 5.42$\pm 0.02$ mag & 6.00$\pm 0.02$ mag & (2)\\
Mass & 2.4$\pm$0.1 M$_\sun$ &  --- & (3)\\
Age &  5...$>$10 Myr & 10-20 Myr & (3),(4),(5),(6)\\
Distance & 97$^{+4}_{-4}$ pc & 96$^{+3}_{-3}$ & (7)\\
Sp. Type & B9Vne & A0V & (8)\\
\enddata
\tablenotetext{a}{References --- (1) \citet{perryman1997}, (2) 2MASS point source catalog \citep{cutri2003}, (3) \citet{vandenancker1997}, (4) \citet{guimaraes2006}, (5) \citet{dezeeuw1999}, (6) \citet{fujiwara2009}, (7) \citet{vanleeuwen2007}, (8) \citet{houk1975}.}
\end{deluxetable}

\clearpage

\begin{deluxetable}{llccccccc}
%\rotate
\tabletypesize{\tiny}
\centering
\tablecaption{Summary of observations.  
\label{observations}}           % title of Table
%\tablewidth{\textwidth}
      % is used to refer this table in the text
\tablehead{
\colhead{Object} & \colhead{Filter} & \colhead{DIT $\times$ NDIT\tablenotemark{a}} & \colhead{Dither}  & \colhead{Position} & \colhead{Airmass} & \colhead{Obs. date} & \colhead{$\langle{EC}\rangle$\tablenotemark{d} [\%]} & \colhead{$\langle \tau_0 \rangle$\tablenotemark{e} [ms]}\\
\colhead{} & \colhead{} & \colhead{} & \colhead{positions\tablenotemark{b}}  & \colhead{angle\tablenotemark{c}} & \colhead{} & \colhead{} & \colhead{mean/min/max} & \colhead{mean/min/max}
}
\startdata
HD 100546 & $H$ & 0.3454 s $\times$ 85 & 11(15) & 155$^\circ$ (141.8$^\circ$) & $\sim$1.67-1.51  & 2006 April 7 & 36/13/47 & 5.4/3.7/7.4 \\
		   & $K_s$ & 0.3454 s $\times$ 85 & 12(13) & 65$^\circ$ (51.8$^\circ$)  & $\sim$1.49-1.43 &  '' & 41/17/60 & 4.5/3.1/7.3 \\
		   & $NB1.64$ & 3 s $\times$ 20 & 4(5) & 65$^\circ$ (51.8$^\circ$) & $\sim$1.43 & " & 49/44/52 & 4.5/4.0/5.0\\
		   & $IB2.16$ & 0.6 s $\times$ 50 & 8(9) & 65$^\circ$ (51.8$^\circ$) & $\sim$1.43-1.45 & " & 44/37/49 & 3.4/2.6/4.3 \\

\\
HD 106797 & $H$ & 0.3454 s $\times$ 85 & 1(2) & 155$^\circ$ (141.8$^\circ$) & $\sim$1.34-1.35 & 2006 April 7 & 30/23/37 & 2.9/2.7/3.1\\
  		    & $K_s$ & 0.3454 s $\times$ 85 & 2(2) & 65$^\circ$ (51.8$^\circ$) & $\sim$1.33-1.34 & " & 40/40/41 & 2.9/2.8/3.1\\
\enddata
\tablenotetext{a}{Detector integration time (DIT) $\times$ number of integrations (NDIT), i.e., total integration time per dither position and per retarder plate position.}
\tablenotetext{b}{Number of dither positions used in final analysis and in parenthesis total number of observed dither position. The difference was disregarded due to poor AO correction. At each dither position NDIT exposures were taken at each of the 4 different retarder plate positions (0.0$^\circ$, -22.5$^\circ$, -45.0$^\circ$, -67.5$^\circ$).}
\tablenotetext{c}{Position angle of the camera on sky (north over east) as taken from the image header and after correction for the offset in the encoder value leading to a false zero point of the retarder plate (in parenthesis). For details see section~\ref{calibration}.}
\tablenotetext{d}{Average, minimum and maximum value of the coherent energy of the PSF. Calculated by the Real Time Computer of the AO system.}
\tablenotetext{e}{Average, minimum and maximum value of the coherence time of the atmosphere. Calculated by the Real Time Computer of the AO system.}
\end{deluxetable}

\clearpage

\begin{deluxetable}{lccc}
\tablecaption{HD100546 disk inclination and position angle from from scattered light imagery.
\label{disk_orientation}}           % title of Table
%\tablewidth{0pt}
      % is used to refer this table in the text
\tablehead{
\colhead{Reference} & \colhead{Inclination\tablenotemark{a}}  & \colhead{PA\tablenotemark{b}}& \colhead{Radius\tablenotemark{c}}
}
\startdata
This work $H$ band &  45.9$^\circ\pm1.7^\circ$ & 136.0$^\circ\pm2.9^\circ$ & 0.2--0.5$''$ \\
This work $K_s$ band &  48.0$^\circ\pm2.0^\circ$& 140.0$^\circ\pm2.4^\circ$ & 0.2--0.5$''$\\
This work mean &  47.0$^\circ\pm2.7^\circ$& 138.0$^\circ\pm3.9^\circ$ & 0.2--0.5$''$\\\\

\citet{ardila2007} & 42.0$^\circ\pm5.0^\circ$& 145.0$^\circ\pm5^\circ$ & 1.6--2.8$''$\\
\citet{pantin2000} & 50.0$^\circ\pm5.0^\circ$& 127.0$^\circ\pm5^\circ$\\
\citet{grady2001} & 49.0$^\circ\pm4.0^\circ$& 127.0$^\circ\pm5^\circ$\\
\citet{augereau2001} & 51.0$^\circ\pm3.0^\circ$& 161.0$^\circ\pm5^\circ$ & 0.7--3.0$''$\\
\enddata
\tablenotetext{a}{Face-on corresponds to 0$^\circ$.}
\tablenotetext{b}{Position angle of disk major axis measured east of north.}
\tablenotetext{c}{Used for isophot fitting.}
%\tablenotetext{d}{Isophot fitting between 1.6--2.8$''$ (HST/ACS data).}
%\tablenotetext{e}{Isophot fitting between 0.7--3.0$''$ (HST/NICMOS data).}

\end{deluxetable}

\clearpage

\appendix
\section{Correction for instrumental polarization}
In the following we describe in detail how the effects of instrumental polarization were quantified and what cross-checks we applied to validate our approach.
\subsection{Quantification of instrumental polarization}
We use the central region around the star as zero polarization calibration source
assuming that its intrinsic polarization is close to zero.
An interstellar polarization component behaves like a static instrumental
polarization offset and this would be corrected with our procedure. 
Assuming $p\approx 0\%$ for the interstellar polarization of HD100546 is a reasonable assumption 
since published polarimetry for the V-band indicate a polarization 
of only $\approx 0.25~\%$ for $\theta \approx 55^\circ$ 
\citep{yudin1998,clarke1999,rodrigues2009}.
This value is composed of some intrinsic polarization component and some 
interstellar polarization $p_{is}$(V). Since HD100546 shows only very little
extinction effects (A$_{\rm V}$=0.28 mag, \cite{vandenancker1997}) we can assume that the
intrinsic polarization is also low in the near-IR for HD100546. 
Of course, a real measurement would be useful but unfortunately we are
not aware of any calibrated near-IR polarimetry for HD100546.  

We assumed that in the final $p_Q$ and $p_U$ images the mean value of the polarization fraction averaged over an annulus between 4--6 pixels centered on the star is zero. In case there are no scattering dust grains  around the star, and hence no polarization signal, the azimuthal average of the fractional polarization should be zero in any case regardless of the radial extend of an annulus. In case there is a disk that is not seen edge-on but with a sufficiently small inclination angle \citep[which for HD100546 we know it is the case, e.g.,][]{augereau2001} we assumed that the substructure in the disk is small enough so that the positive and negative contributions to the polarization fraction in the $p_Q$ and $p_U$ images in a small centro-symmetric annulus around the star would cancel out. Ideally, if the PSF cores were not saturated and the central source would cancel out completely one could use the innermost pixels instead of an annulus. However, since our $H$ and $K_s$ images were saturated in the PSF core (see, section~\ref{observationssection}) we had to rely on those regions that were not saturated and we used the same annulus for all data sets to be consistent. In Table~\ref{instrumental_polarization} we summarize the mean values we found for the polarization fractions in the aforementioned annulus in the different data sets. It turned out that in most cases these values were small, but not zero, and they varied between all filters and both Stokes parameters. We assumed that these offsets were due to instrumental polarization effects and we corrected each individual $p_Q$ and $p_U$ image for these offsets before the final images were median combined. In the final $Q$ and $U$ images these offsets are, however, not additive offsets but factors. Hence, before we did the median combination of the final $Q$ and $U$ images we scaled the individual images with $1+x$, where $x$ is the offset derived in the respective $p_Q$ and $p_U$ images.

\subsection{Validation of approach}
Once we had applied the correction factors, we checked the final results by comparing the observed structures in the different filters for both objects (see section~\ref{butterfly}). Finding strong signals for HD100546 that are similar in all filters and not seeing any consistent and significant structure around the reference source HD106797 gives us confidence that our data reduction technique is robust. %For comparison we show in the Appendix a version of the $H$ filter Stokes $Q$ image \emph{before} it was corrected for instrumental polarization. This was the most extreme case of instrumental effects as can be seen from Table~\ref{instrumental_polarization}. 

An additional check was accomplished by analyzing the azimuthal profiles of the "butterfly" patterns (upper panels of Figure ~\ref{azimuthal_profile}; discussed in detail in section~\ref{butterfly}). Comparing these figures for all filters with similar figures derived from simple generic disk models shows that one would expect that the plotted lines computed for different annuli will always cross each other at $p_Q=p_U=0$ irrespective of disk inclination. While this is the case in our final plots, within the error bars, it was not the case before the correction. 

\subsection{Before-and-After comparison}

\begin{figure}
\centering
\epsscale{1.}
\plottwo{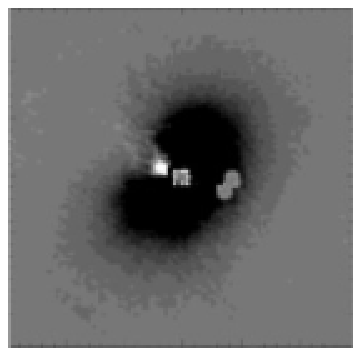}{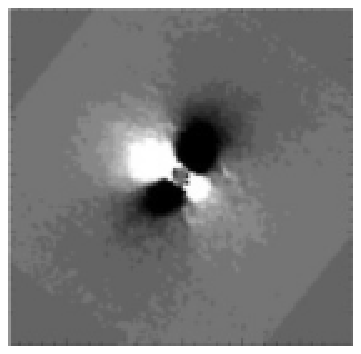}
\caption{Uncorrected (left) and corrected (right) $Q$ image of HD100546 in the $H$ filter.  
\label{instrumental_polarization_image}}
\end{figure}

To demonstrate that instrumental polarization offset has a significant impact and needs to be corrected for we show in Figure~\ref{instrumental_polarization_image} an uncorrected and a corrected version of the $Q$ image of HD100546 in the $H$ filter. As can be seen from Table~\ref{instrumental_polarization} the correction factor here was the largest among all images, thus this image suffered the most contamination by instrumental polarization. To compare these images directly to the final image shown in Figure~\ref{H_images} one has to multiply by $-1$ in order to compensate for the fact that the camera has been rotated by 90$^\circ$ compared to the images in the $K_s$ filter and hence the ordinary and extraordinary beam were switched.

\begin{deluxetable}{lcccc}
\tablecaption{Offsets used to correct for instrumental polarization in $p_Q$ / $p_U$ images in the different filters.
\label{instrumental_polarization}}           % title of Table
%\tablewidth{0pt}
      % is used to refer this table in the text
\tablehead{
\colhead{Object} & \colhead{$H$}  & \colhead{$K_s$} & \colhead{$NB1.64$} & \colhead{$IB2.18$} 
}
\startdata
HD100546 & -0.01275 / -0.00049 & 0.00463 / 0.00450 & -0.00768 / 0.00779  & 0.01033 / 0.00995 \\
HD106797 & -0.00774 / -0.00806 & -0.00134 / 0.00430 & \rm{---}& \rm{---}\\
\enddata
\end{deluxetable}

\newpage
\section{Signal-to-noise estimates}
The main source of uncertainty in our images results from variations in flux in the ordinary and extraordinary beam between the different dither positions. This means that the $Q$, $U$, $p_Q$, and $p_U$ images show slightly different patterns in terms of signal strengths at each dither position. Compared to these variations detector read-out noise, sky noise and photon noise are negligible in the inner regions. Thus, it is valid to base an estimate for the signal-to-noise (S/N) in our images on these variations and compare them to the final images. In addition to the median-combined final images, we also computed the mean images for each Stokes parameter and each filter and the corresponding sigma images which show the standard deviation of each pixel. The mean images and median-combined images were almost identical similar in terms of overall flux levels and surface brightness distribution. Since we are interested in the S/N of the observed extended emission and not in the S/N of individual pixels, we convolved the median images and the sigma images with a Gaussian kernel where the FWHM corresponded to the FWHM of the PSFs in our images. A S/N map can then be computed via
\begin{equation}
{\rm S/N}=\sqrt{N_{\rm Dith}}\cdot\frac{{\rm \bar{I}}}{{\rm I_{\sigma}}}
\end{equation}
where $N_{\rm Dith}$, ${\rm \bar{I}}$, and ${\rm {I_{\sigma}}}$ denote the number of dither position used for the final combination, the median-combined image, and the sigma image, respectively. 

\begin{figure}
\centering
\epsscale{1.}
\plottwo{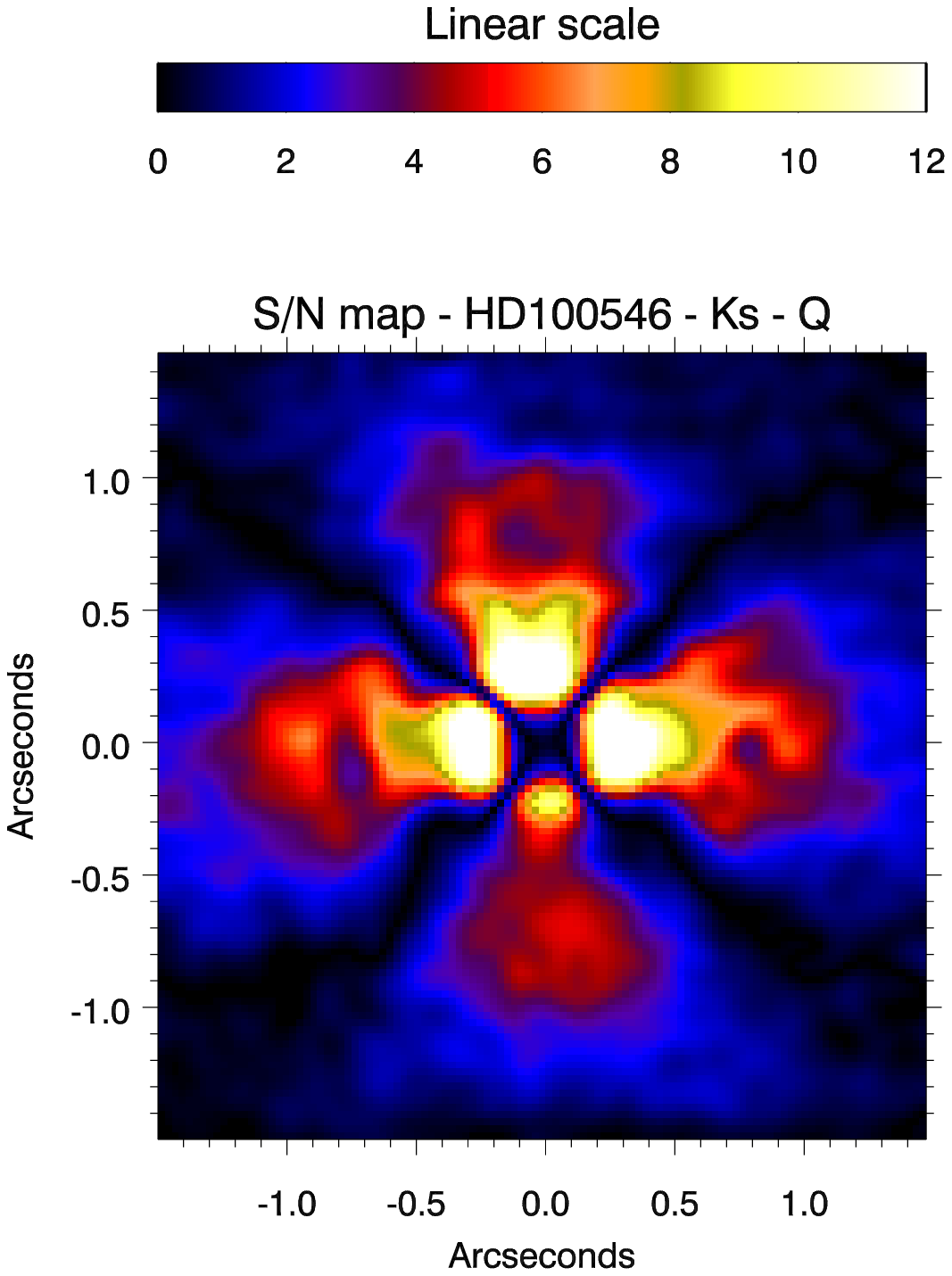}{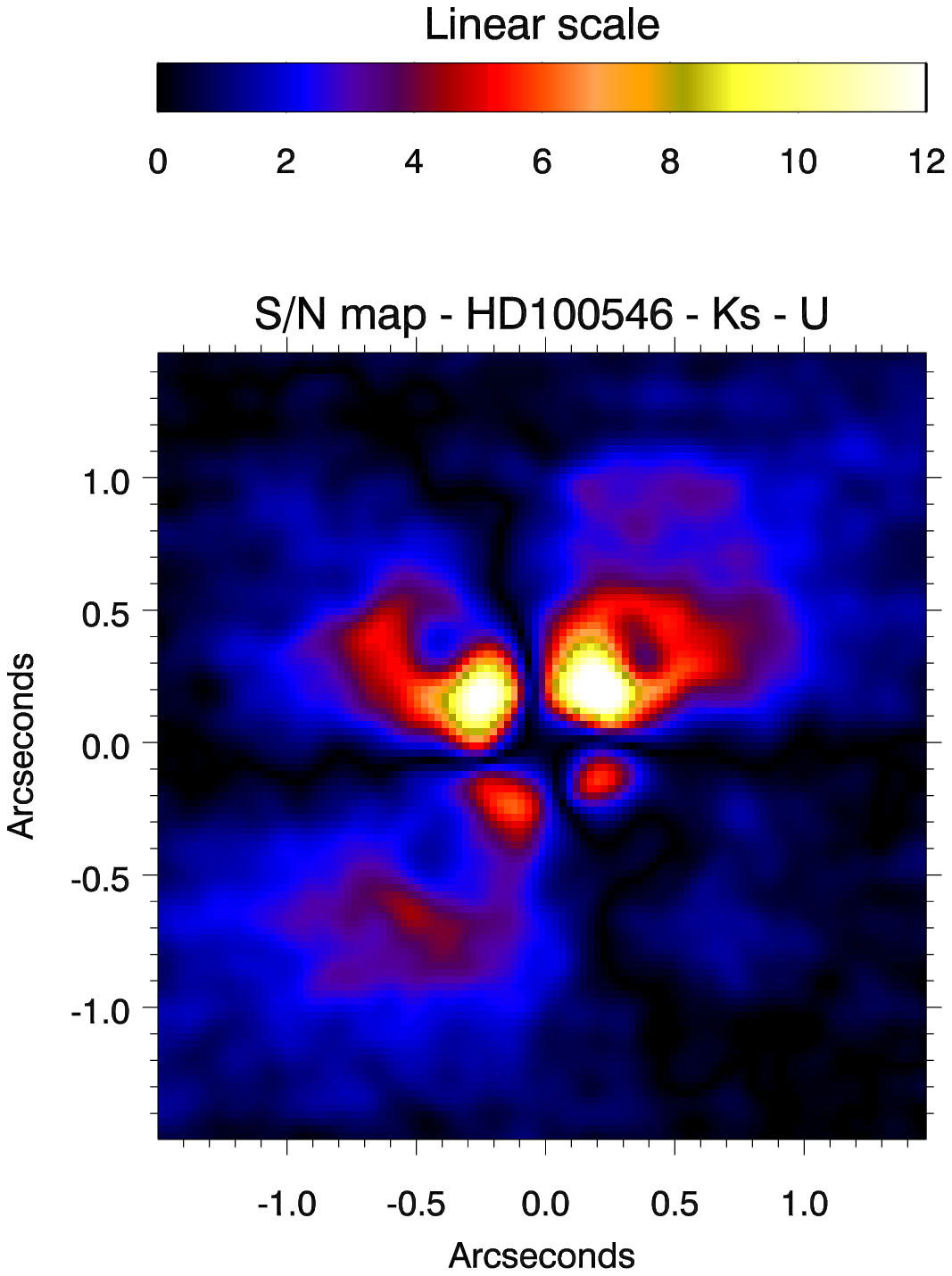}
\caption{S/N maps of the $Q$ (left) and $U$ (right) images for HD100546 obtained in the $K_s$ filter. The disk major axis of the disk runs horizontally. The color scale runs from 0 to 12. See text for more details. 
\label{signaltonoiseimage1}}
\end{figure}

In Figure~\ref{signaltonoiseimage1} we provide an example of these S/N maps and show those for the $Q$ and $U$ image in the $K_s$ filter for HD100546. The orientation of the images is the same as in Figure~\ref{disk_hole_zoom} with the disk major axis running horizontally. To obtain the same orientation as in Figure~\ref{Ks_images} (north up and east to the left) the images have to be rotated by 51.8$^\circ$ counterclockwise. The S/N map clearly resembles the surface brightness distribution of the disk. However, it shows that wherever we detect polarized scattered flux within 1.4$''$ it is detected with a S/N$\gtrsim$3, in particular for the total polarized flux $P$ given by $P=\sqrt{Q^2+U^2}$. 

\begin{figure}
\centering
\epsscale{.5}
\plotone{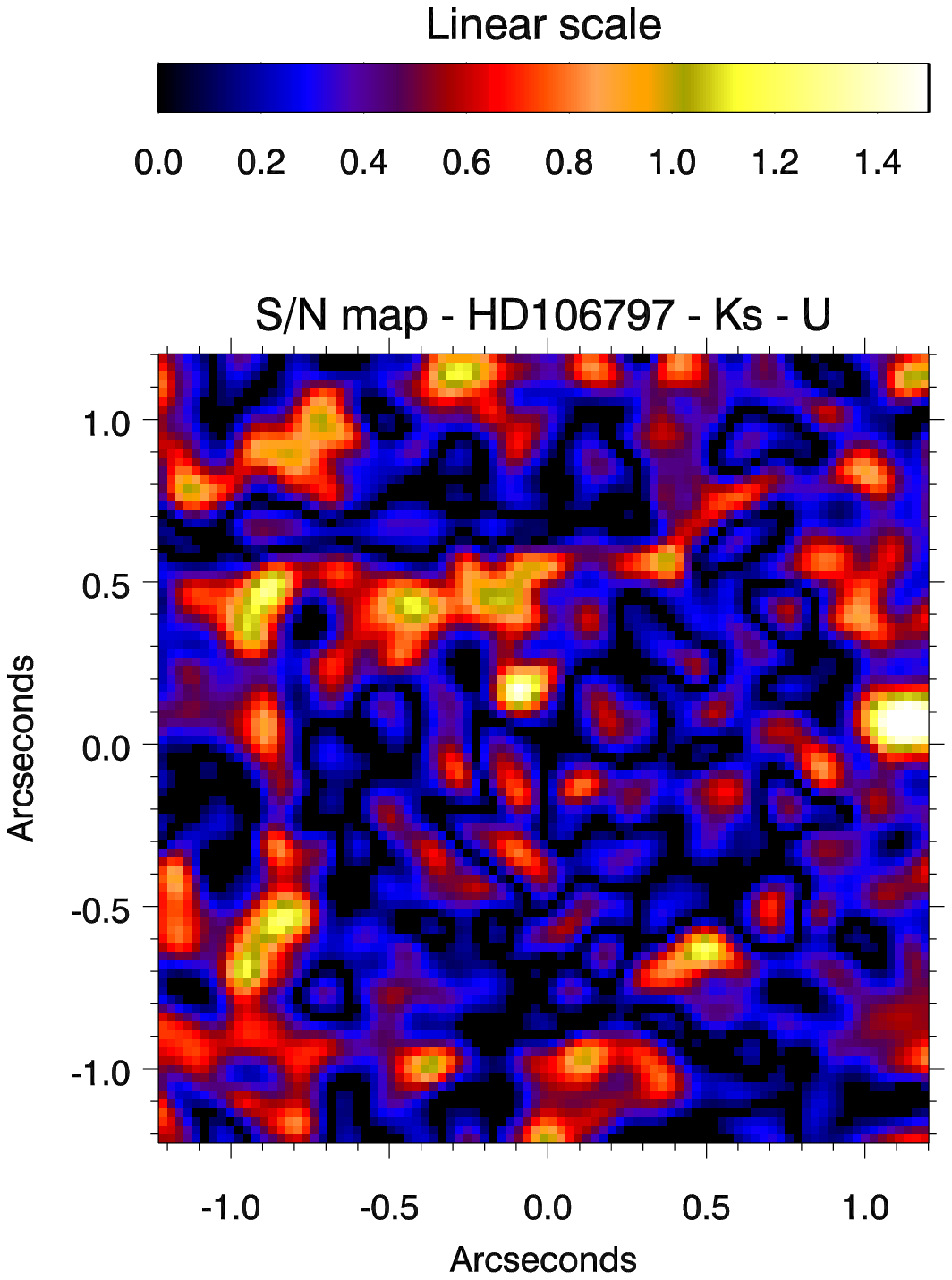}
\caption{S/N map of the $U$ image for HD106797 in the $K_s$ filter. Note that the color scale runs from 0 to roughly 1.5.
\label{signaltonoiseimage2}}
\end{figure}

For comparison, we show the S/N map of the $U$ image in the $K_s$ filter for HD106797 in Figure~\ref{signaltonoiseimage2}. Note the different image stretch compared to the S/N maps of HD100546. Here, no feature is detected with a S/N$>$1.5.

\clearpage

\clearpage

\clearpage

\clearpage

\end{document}